\documentclass[aps,amsmath,amssymb,superscriptaddress,nofootinbib]{revtex4}

\usepackage{soul}
\usepackage{ulem}
\usepackage{graphicx}
\usepackage{hyperref}
\usepackage{color}
\usepackage{epsf}
\usepackage{bm}
 \usepackage{slashed}
\usepackage{hyperref}
\usepackage{enumitem}
\usepackage{lipsum}
\usepackage{bbold}
\usepackage{sidecap}
\usepackage{caption}

\newcommand{\bea}{\begin{eqnarray}}
\newcommand{\eea}{\end{eqnarray}}
\newcommand{\la}{\label}
\newcommand{\be}{\begin{equation}}
\newcommand{\ee}{\end{equation}}

\def\12{\frac{1}{2}}

\def\XXint#1#2#3{{\setbox0=\hbox{$#1{#2#3}{\int}$}
     \vcenter{\hbox{$#2#3$}}\kern-.5\wd0}}

\numberwithin{equation}{section}

\begin{document}

\title{Stretched String with Self-Interaction at High Resolution: \\Spatial Sizes and Saturation}

\author{Yachao Qian and Ismail Zahed}
\address{Department of Physics and Astronomy,
Stony Brook University,  Stony Brook, NY 11794-3800.}

\begin{abstract}
We model the (holographic) QCD Pomeron as  a long and stretched (fixed impact parameter) 
transverse quantum string in flat  $D_\perp=3$ dimensions. 
After discretizing the string in $N$ string bits, we analyze  its length, mass and spatial distribution for large $N$ 
or low-x ($x=1/N$),  and away from its Hagedorn point. The string bit distribution shows sizable asymmetries in the transverse plane that may  translate to azimuthal asymmetries in primordial particle production in the Pomeron kinematics, and the flow moments in minimum bias $pp$ and $pA$ events.  At moderately low-x and relatively small string self-interactions $g_s\approx \alpha_s$ (the gauge coupling), a pre-saturation phase is identified whereby the string transverse area undergoes a  sharp transition from a large 
diffusive growth to a small fixed size area set by few string lengths $l_s$.
For lower values of $x$ the transverse string bit density is shown to increase as $1/x$ before saturating at the Bekenstein bound of one bit per Planck area with the Planck length $l_P/l_s\approx \alpha_s^{2/3}$. We argue that the effects of the AdS$_5$ curvature 
 on the interacting string maybe estimated using an effective transverse dimension between the interacting
 string bits. The result  is a smoother transition  with a transverse string bit density increasing as $1/x^{0.31}$.
\end{abstract}

\date{\today}

\maketitle

\section{\label{sec:introduction}introduction}

Hadron-hadron collisions at high energies but soft momentum transfer are dominated by soft Pomeron
exchange, an effective $0^{++}$ exchange corresponding to the highest Regge trajectory with 
intercept $\alpha_P(0)-1\approx 0.08$~\cite{Donnachie:1992ny}. Reggeon exchanges with spin-isospin quantum numbers
have smaller intercepts and are therefore sub-leading~\cite{Gribov:1972ri,9780511534959}. The growth of the total hadron-hadron cross-section
with the rapidity interval $\chi={\rm ln}(s/s_0)$ is described phenomenologically in the context of
Reggeon field theory. In QCD the re-summation of the soft collinear Bremmstralung contributions
through the BFKL ladders yield a hard Pomeron with a perturbatively small
 intercept and zero slope~\cite{Kuraev:1976ge,Lipatov:1976zz,Sterman:1999yc,Fadin:1975cb,Balitsky:1978ic}. 

Soft Pomerons are altogether non-perturbative. Duality arguments put forth by Veneziano~\cite{Veneziano:1968yb} suggest
that the soft Pomeron is a closed string exchange in the t-channel, with a string world-sheet made of 
planar diagrams like fish-nets~\cite{Greensite:1984sb}. The quantum theory of planar diagrams in the double limit of strong
coupling and large number of colors is tractable in supersymmetric theories using the holographic
principle~\cite{Maldacena:1998im}. Many descriptions of the soft Pomeron in holographic duals to QCD have been suggested 
recently without supersymmetry~\cite{Rho:1999jm,Janik:2000aj,Janik:2000pp,Polchinski:2001tt,Polchinski:2001ju,Polchinski:2002jw,Brower:2006ea,Brower:2007xg,Brower:2010wf,Brower:2011dx,Hatta:2007cs,Hatta:2007he,Albacete:2008ze,Albacete:2008vv,Basar:2012jb,Stoffers:2012zw,Stoffers:2012ai,Stoffers:2013tla,Zahed:2012sg,Shuryak:2013sra}. A simple version  is a stringy exchange in 
AdS$_5$ with a wall with $D_\perp=3$ dimensions,  that reproduces a number of features of 
diffractive scattering, production  and low-x DIS.

The Pomeron as a string exchange in holography can be thought as a chain
of closed but confined gluons, some sort of non-perturbative Weizacker-Williams field tying two 
colorless dipoles separated by  a large rapidity interval $\chi$. In this spirit, lepton on proton scattering 
in DIS at low-x can be described through a holographic string exchange with the identification 
$\chi\approx {\rm ln}(1/x)$. In the proton rest frame, the leptonic dipole of size $1/Q$ acts as a small 
probe dipole scattering off the larger dipole composing the proton at a fixed impact parameter $b$. DIS experiments are always
averaged over this impact parameter when measuring  gluonic densities in structure functions. 
However, the dominant contribution in the averaging stems from large $b$~\cite{Stoffers:2012zw,Stoffers:2012ai,Stoffers:2013tla}.
More  exclusive experiments could be done in future electron-Ion-Colliders  to unravel the impact parameter dependence
at low-x as well.

Low-x physics translates to a large $N=1/x$ resolution of the holographic string as we detail below.
This is achieved for long strings by discretizing the transverse Polyakov scalar action in $N$ string bits
and  initially ignoring  the stringy interactions (free string) and the curvature of AdS$_5$. 
String bits have been identified with wee (gluonic) partons by Thorn
\cite{Bergman:1997ki,Karliner:1988hd}. The slow logarithmic growth of the free string  transverse area 
translates to an anomalously large  transverse string bit density at low-x. Repulsive string 
interactions can cause the transverse density to conform with the maximum Bekenstein bound for a black-hole
as argued by Susskind for wee gravitons~\cite{Polchinski:2001ju,Susskind:1994hb,Susskind:1994vu}. However,  
such a growth appears to be at odd with the Froissart bound~\cite{Froissart:1961ux}.

A high string bit density at low-x points towards a liquid of string bits, a priori resolving the string. However, the underlying
presence of the string is still paramount to maintain the (Gribov) diffusion of the string bits in the transverse plane. Recall that
the diffusion constant ${\bf D}= l_s^2/2$ is dimensionfull  and ties with the squared string length.  Also, a highly resolved string
provides an optimal  desccription of  low-x saturation in QCD as wee partons reaching the Bekenstein
bound~\cite{Bekenstein:1973ur,Bekenstein:1972tm,Bekenstein:1974ax,Hawking:1974rv,Hawking:1974sw,Cornalba:2010vk,Cornalba:2009ax,Cornalba:2008sp}. 
In this work we will show that the bound is reached in two stages in flat $D_\perp=3$. 
First a dilute pre-saturation stage where the string transverse
area undergoes a first order transition from a large diffusive growth to a small but fixed size set by the string scale for relatively
weak string self-interactions.  Second a dense saturation stage at very low-x whereby the transverse string bit density saturates the Bekenstein bound of one bit per transverse Planck area. To assess the role of the AdS curvature on our results we suggest the use
of an effective transverse dimension for the string bit interactions. The result is a smoothening of the transition to the Bekenstein bound.

Our pre-saturation condition is overall consistent with the saturation condition following from the stringy dipole-dipole cross section analysis derived by Stoffers and one of us~\cite{Stoffers:2012zw,Stoffers:2012ai,Stoffers:2013tla}. In some ways, our stringy description of saturation can be regarded as the dual of the weak coupling description of gluon saturation in QCD based on the color glass condensate~\cite{McLerran:2001sr,TapiaTakaki:2010zz,Iancu:2003xm,Iancu:2002xk,Navelet:2002zz,Iancu:2001yq,Levin:2001eq,McLerran:1993ka,Gelis:2010nm,Marquet:2005hu,Iancu:2003uh,Iancu:2009nd,Gelis:2012ri} and is variant in the impact parameter space~\cite{Tribedy:2011aa,Tribedy:2011yn,Tribedy:2010ab}. The exponential rise of the string density of states with its mass provides the most efficient way of scrambling information and reaching the Bekenstein bound and thus the saturation point as we will show below.

This paper consists of a number of new results:  1/ A detailed numerical spatial shape 
analysis of  an open and free string in flat $D_\perp=3$ dimensions for increasing 
resolution; 2/ A variational analysis of the effects of two-body interactions on the string shape
as a function of the resolution; 3/ A contraction of the string to a black-hole-like configuration 
under attraction and an expansion of the string  shape under repulsion; 4/ A physical
interpretation of the contracted string at high resolution with saturation in DIS dipole-dipole
scattering in curved $D_\perp(\lambda)<3$; 5/ A prompt transverse azimuthal asymmetry
in dipole-dipole scattering.

In section 2 we detail the discretized version of the transverse scalar string in flat $D_\perp$ dimensions. 
We analyze numerically its geometrical distributions for different resolutions.  Self-string interactions both attractive
and repulsive are introduced and discussed in the mean-field approximation in section 3. Using a Gaussian
variational approach we re-assess the geometrical properties of the transverse string at various resolution in
section 4. Detailed numerical sampling of the string using the variational analysis are given in section 5.
At low-x a pre-saturation stage with fixed and small string geometry, followed by saturation when the Bekenstein bound
is reached are discussed in section 6. Details about the azimuthal deformation of the string distribution are given
in section 7 in terms of standard flow moments  in the diffusive and pre-saturation phases. Our conclusions are in section 8.


\section{\label{}Discretized Free Transverse String}

Scattering of dipoles in the pomeron kinematics with a large rapidity interval $\chi={\ln}(s/s_0)$ and fixed impact parameter $b$ 
is dominated by a closed t-channel string exchange. In leading order in $\chi$, the exchange amplitude can be shown to be that 
of a free transverse string at fixed Unruh temperature $T=a/2\pi$ with the mean world-sheet acceleration $a=\chi/b$~\cite{Basar:2012jb, Stoffers:2012zw, Stoffers:2012ai, Stoffers:2013tla, Qian:2014jna}. For long strings the Unruh temperature is low. These strings will be referred to as cold strings.
With this in mind, the free  transverse string with fixed end-points in flat $D_\perp$ dimensions is characterized by

\be
S_\perp = \frac{\sigma_T}{2} \int d \tau \int_0^\pi d \sigma \ \ \left[  \left( \dot{x}_\perp  \right)^2  + \left( {x'}_\perp \right)^2    \right]
\label{XS1}
\ee
with the end-point condition 
\be
x_\perp^i (\sigma=0, \tau) = 0 \ \ \ \ \ \ x_\perp^i (\sigma=\pi, \tau) = \bold{b}^i
\ee
The string tension is $\sigma_T=1/(2\pi \alpha^\prime)$ with $\alpha^\prime=l_s^2$. For simplicity, we will set $2l_s\equiv1$ throughout
and restore it by inspection when needed. The purpose of the present work is to show how the concept of saturation at low-x emerges
from the string description and identify its key parameters in QCD through holography.
 We will also study the general geometrical structure of the transverse string, in particular its spatial size and deformation in the cold or pomeron regime
both for a free and interacting string. Initial geometrical string deformations maybe the source
of large prompt azimuthal deformations in the inelastic channels and for high multiplicity events. 

The transverse free string (\ref{XS1})  can be thought as a collection of $N$  string bits connected by identical
strings~\cite{Karliner:1988hd,Bergman:1997ki} and discretized as follows

\be 
\mathcal{L}_\perp  =   \frac{1}{N}    \sum_{k=0}^{N}  \left( \dot{x}_\perp^i (k) \right)^2 - \frac{1}{N}     \sum_{k=1}^N \left(  \frac{x_\perp^i(k) - x_\perp^i (k-1)}{\frac{\pi}{N}}    \right)^2 
\ee
with $S_\perp=\int d\tau{\cal L}_\perp$. For $N \rightarrow \infty$ the (\ref{XS1}) is recovered.
Using the mode decompostion for the amplitudes  $x_\perp^i$ 
\be
x_\perp^i (k , \tau) = \bold{b}^i \frac{k}{N }  +   \sum_{n=1}^{N-1}  X_{n}^i (\tau) \sin \left( \frac{n k}{N } \pi\right)   \ \ \ \ \ \ \ \ \    (k=0,1, \cdots, N)
\label{N1}
\ee
 and their conjugate momenta

\be
p_\perp^i (k, \tau) = \frac{\partial\mathcal{L}}{\partial\dot{x}_\perp^i} =   \frac{2}{N}   \dot{x}_\perp^i =  \frac{2}{N}   \sum_{n=1}^{N-1} \dot{X}_{n}^i (\tau) \sin \left( \frac{n k}{N } \pi\right) \equiv \frac{2}{N}  \sum_{n=1}^{N-1}  P_{n}^i (\tau) \sin \left( \frac{n k}{N } \pi\right)
\ee
allow us to write the Hamiltonian

\bea
\mathcal{H}_\perp  &=&   \frac{N}{4}   \sum_{k=0}^{N} \left( p^i_k \right)^2 + \frac{1}{N}    \sum_{k=1}^N  \left(  \frac{x^i_k - x^i_{ k-1}}{\frac{\pi}{N}}    \right)^2 \nonumber\\
&=&   \frac{1}{2 } \sum_{n=1}^{N-1}    \left(         P _{n}^i (\tau) P_{n }^i (\tau)   + 
\Omega_n^2  {X}_{n}^i (\tau)  {X}_{n }^i (\tau) \right)  + \frac{  b^2}{  \pi^2 } 
\label{HAR}
\eea
with  free harmonic oscillators of frequencies 

\be
\Omega_n = \frac{2 N}{\pi} \sin  \left(  \frac{n \pi}{2 N} \right)
\ee

Each oscillator in (\ref{HAR}) carries a small mass $m_N=2/N$ and a large compressibility $k_N=4/(\pi^2m_N)$.
 The ground state of this dangling  N-string bit Hamiltonian  is a product of Gaussians~\cite{Karliner:1988hd}

\be
\Psi[X]=\prod_{n, i} \Psi (X_n^i) = \prod_{n, i} \left( \frac{\Omega_n}{\pi}  \right)^{\frac{1}{4}} \exp\left[  - \frac{\Omega_n}{2} (X_n^i)^2   \right]
\ee
leading to the ground state energy
\be
\left< \mathcal{H}_\perp \right>   =   \frac{D_\perp}{2 } \sum_{n=1}^{N-1}    \Omega_n     + \frac{  b^2}{  \pi^2 } =  \frac{D_\perp}{2 } \frac{N}{\pi} \left[ \cot\left(  \frac{\pi}{4 N} \right)  -1  \right]+ \frac{  b^2}{  \pi^2 } 
\la{hga}
\ee
The string transverse squared size is

\be 
R_\perp^2  =  \frac{1}{N} \sum_{k= 0}^N  \left< \left(  x_k^i  - \bold{b}^i \frac{k}{N} \right)^2 \right> \nonumber\\
= \frac{D_\perp}{4}\sum_{n=1}^{N-1} \frac{1}{\Omega_n}
\label{RT0}
\ee 
while its transverse squared mass is

\be
M_\perp^2 = \frac{1}{2} \left< \mathcal{H}_\perp \right>   =  \frac{D_\perp}{4} \sum_{n=1}^{N-1}    \Omega_n     + \frac{  b^2}{  2 \pi^2 } =  \frac{D_\perp}{4}\frac{N}{\pi} \left[ \cot\left(  \frac{\pi}{4 N} \right)  -1  \right]  + \frac{  b^2}{  2 \pi^2 }
\label{H0}
\ee
We note that back to the continuum $\Omega_n\rightarrow n$ with the ground state wave functions

\be
\Psi (X_n^i) = \left( \frac{n}{\pi}  \right)^{\frac{1}{4}} \exp\left[  - \frac{n}{2} (X_n^i)^2   \right]
\ee
so that
 
\be
\left< \mathcal{H}_\perp \right>   \approx   \frac{2 D_\perp}{\pi^2  }  N^2 + \frac{  b^2}{  \pi^2 } 
\ee
The transverse squared  radius of the string diverges logarithmically
\be 
R_\perp^2   \approx  \frac{D_\perp}{4} \ln (N)
\label{RT}
\ee 
 while its effectve squared mass diverges quadratically
\be\la{mperpanalytic}
M_\perp^2   \approx   \frac{  D_\perp}{\pi^2  }  N^2 + \frac{  b^2}{ 2 \pi^2 } 
\ee
 with the number $N$ of string bits.

A simple interpretation of $N$ in relation to the holographic Pomeron follows from the 
diffusive equation for the tachyonic mode of the closed string exchange in~\cite{Stoffers:2012zw},

\be
\left( \partial_\chi - \frac{D_\perp}{12}\right) \bold{K} =  \frac{\alpha'}{2} \Delta_\perp^2 \bold{K} \equiv  \frac{1}{8} \Delta_\perp^2 \bold{K}
\ee
where $ \bold{K}$ is the quantum propagator for long closed strings in flat $D_\perp+2$ space.
The last equality follows after setting $2l_s=1$ in our current conventions.
Thus the transverse diffusive size of the Pomeron  is

\be
R_\perp^2 = \frac{{D_\perp}}{4} \chi \equiv \frac{D_\perp}{4} \ln \left( \frac{Q^2}{s_0} \left(\frac 1x -1\right)\right) 
\ee
where the last equality uses the DIS kinematics~~\cite{Stoffers:2012zw}. Thus, the identification

\be
R_\perp^2\approx \frac{{ D_\perp}}{4} \ln \left( \frac{1}{x  } \right) 
\ee
for small $x$, which leads to 

\be
N \equiv \frac{1}{x}
\ee
as the string resolution as suggested earlier. The curvature of AdS$_5$  causes the leading Pomeron intercept $D_\perp/12\rightarrow D_\perp(\lambda)/12$  in leading order in $\lambda=g_{YM}^2N_c$ with $D_\perp(\lambda)<3$~\cite{Stoffers:2012zw,Stoffers:2012ai,Stoffers:2013tla}. The string diffusion is reduced to a diffusion in a smaller effective dimension. We will return to this point below.

\begin{figure}[!htb]
\minipage{0.48\textwidth}
\includegraphics[height=42mm]{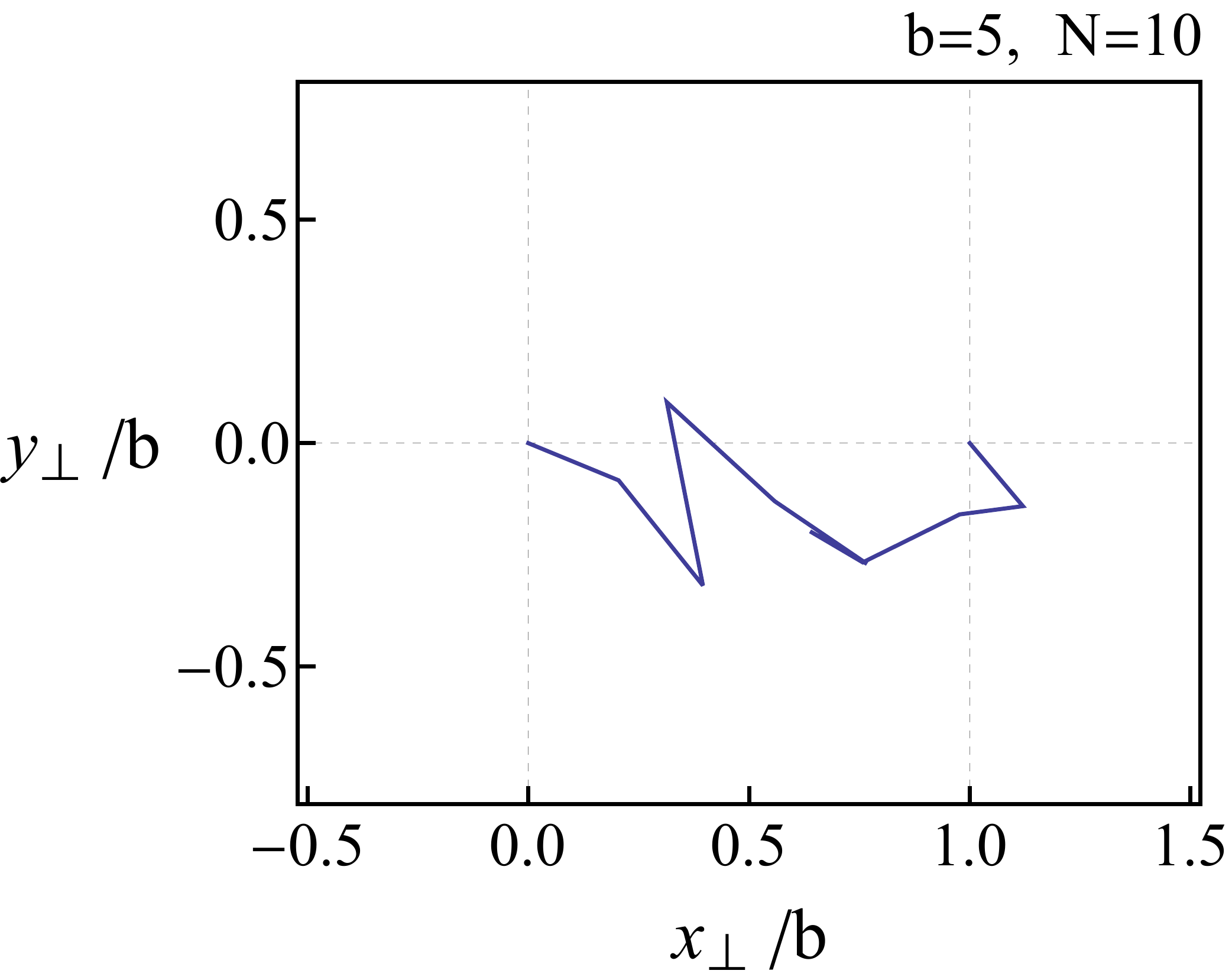}
\endminipage\hfill
\minipage{0.48\textwidth}
\includegraphics[height=42mm]{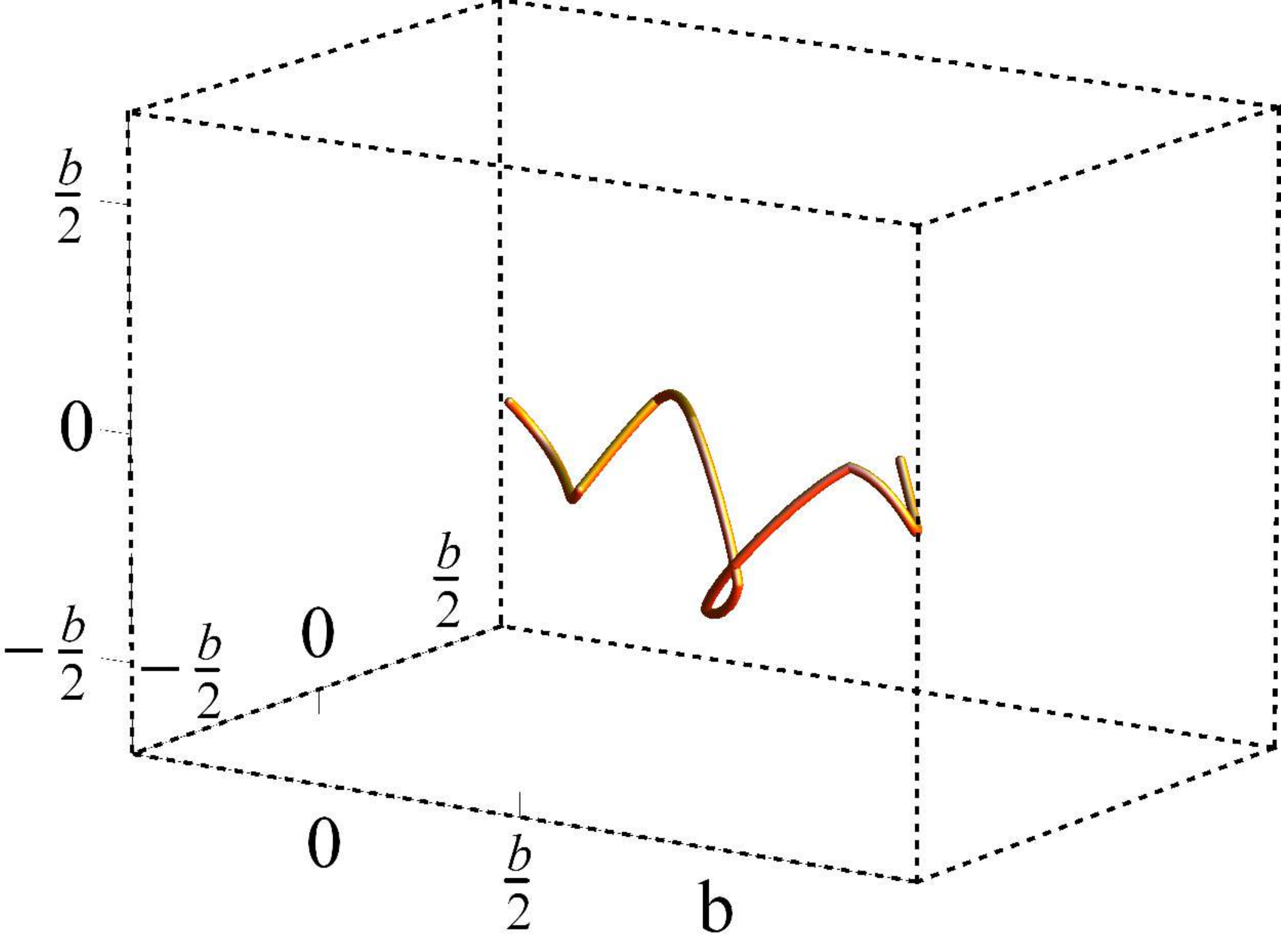}
\endminipage
  \caption{  Free transverse string shape at a resolution $x=1/10$ and $b=5$.}\label{StringN=10}
\end{figure}

\begin{figure}[!htb]
\minipage{0.48\textwidth}
\includegraphics[height=42mm]{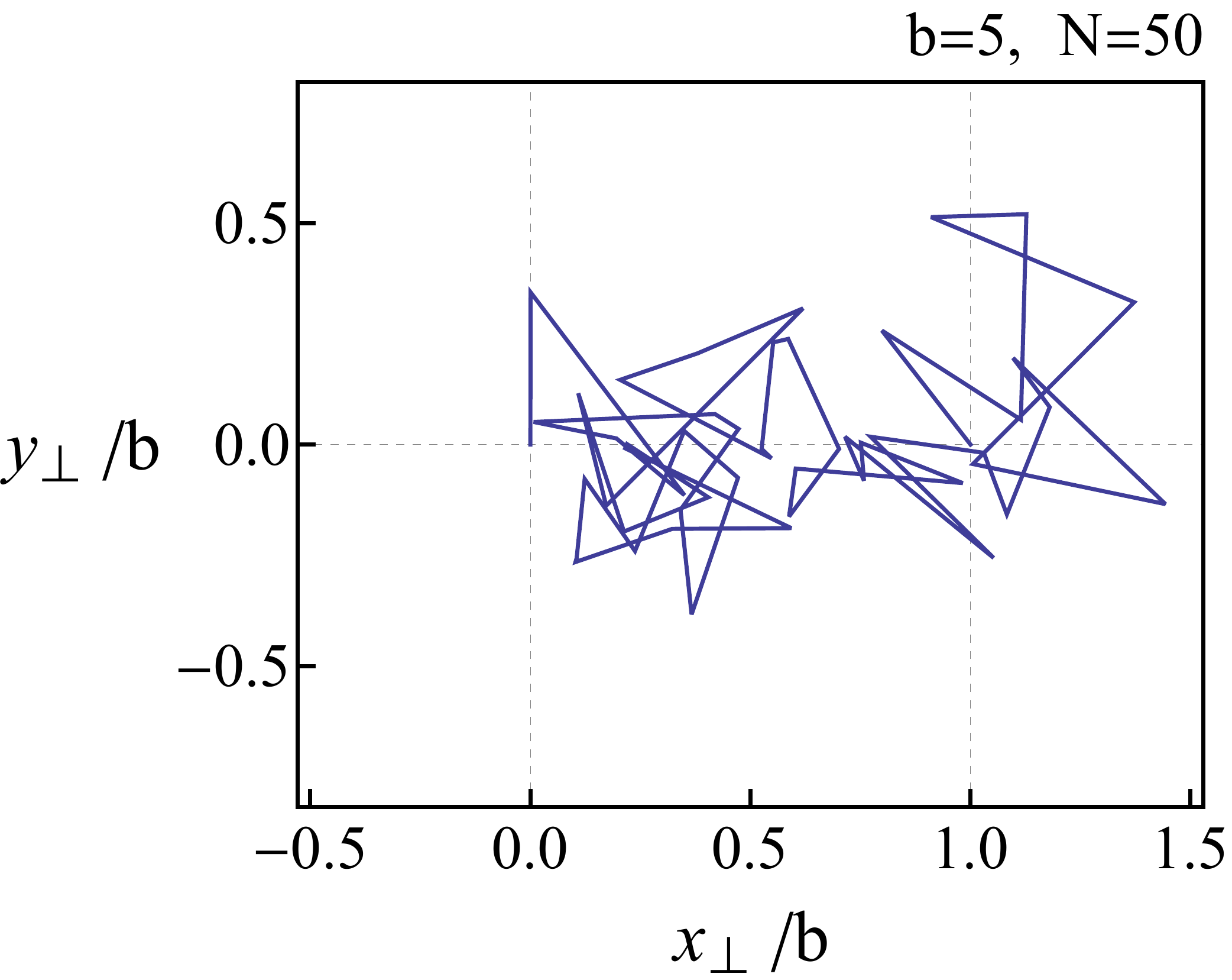}
\endminipage\hfill
\minipage{0.48\textwidth}
\includegraphics[height=42mm]{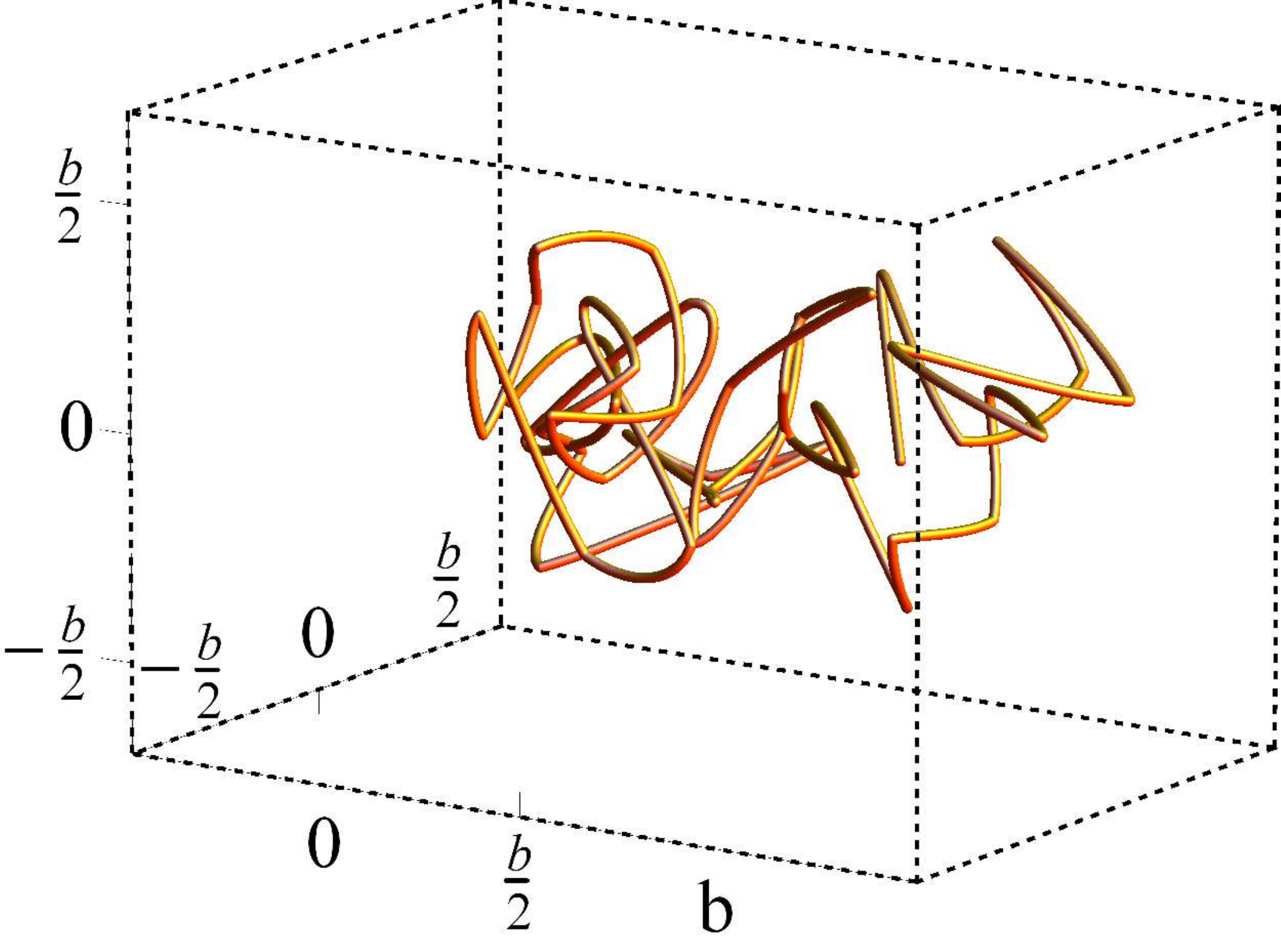}
\endminipage
  \caption{ Free transverse string shape at a resolution of $x=1/50$ and $b=5$.}\label{StringN=50}
\end{figure}

\begin{figure}[!htb]
\minipage{0.48\textwidth}
\includegraphics[height=42mm]{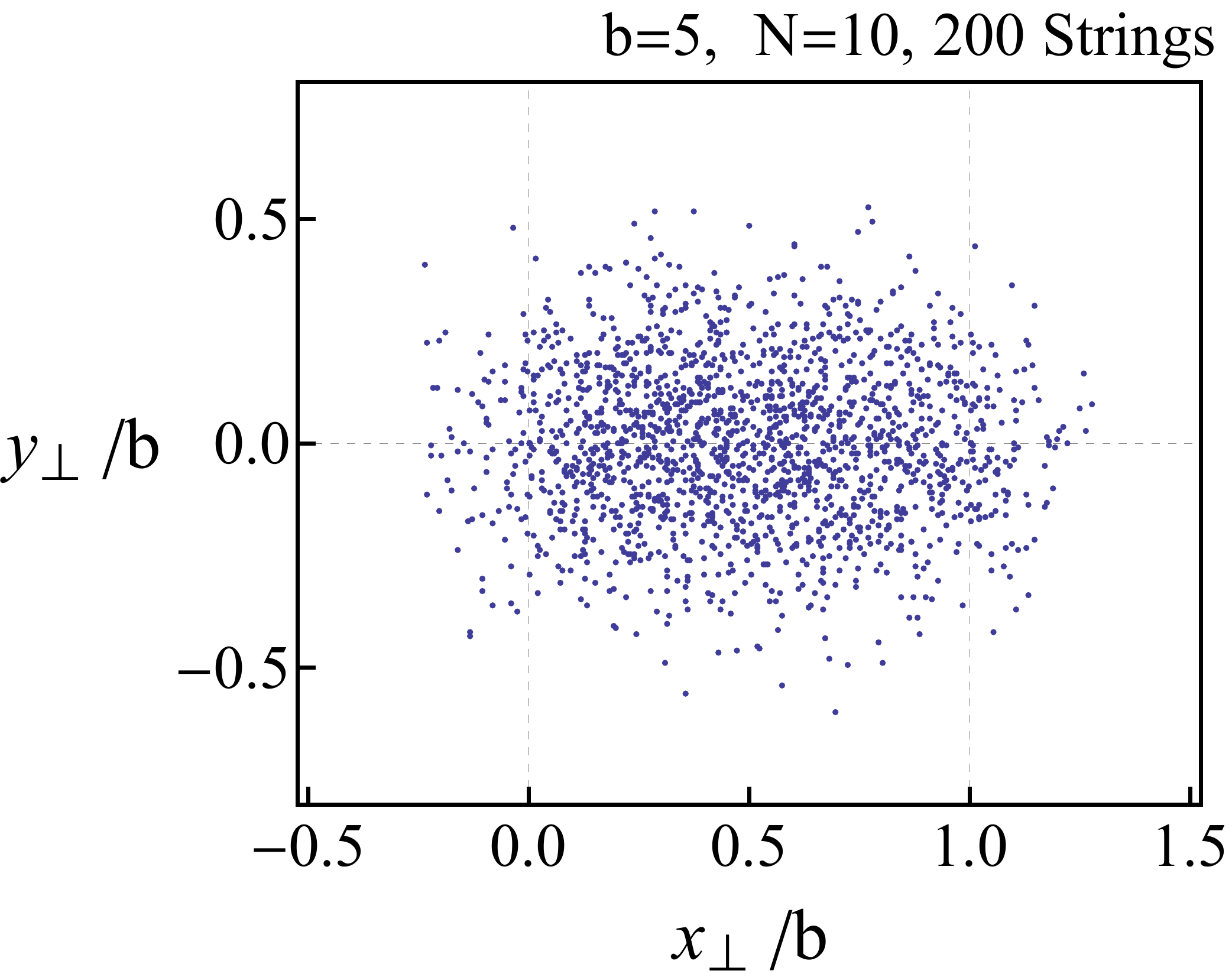}
\endminipage\hfill
\minipage{0.48\textwidth}
\includegraphics[height=42mm]{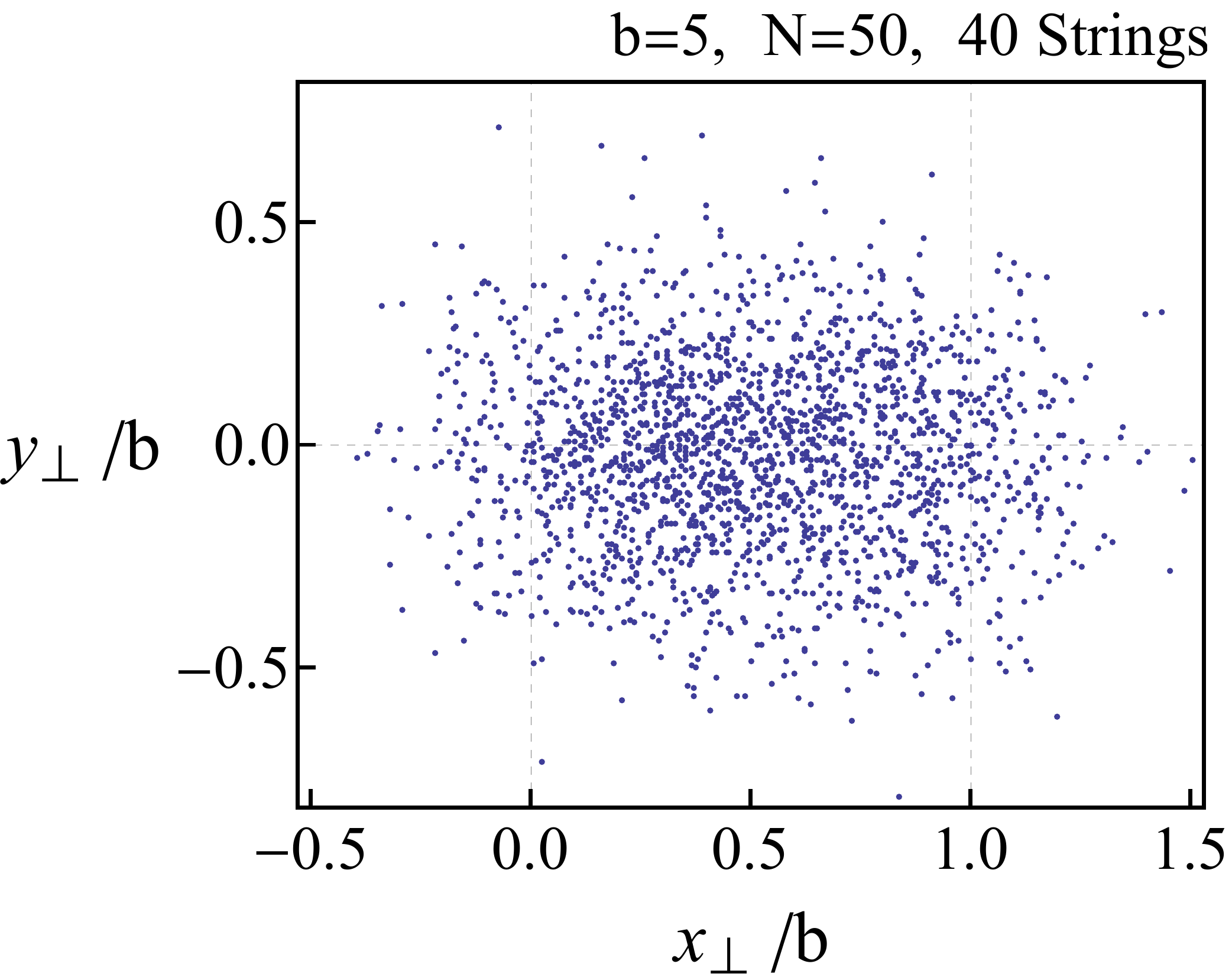}
\endminipage
  \caption{ Transverse string bit distributions:  at $x=1/10$ sampling 200 strings (${\bf Left}$) and at $x=1/50$ 
  sampling 40 strings (${\bf Right}$).}\label{RandomStringShape}
\end{figure}

 
 In its ground state, each of the discretized string bit coordinates $X_n^i$ is normally distributed with probability $|\Psi(X_n^i)|^2$.
 This gives rise to a random walk of the string bits along the chain in the transverse direction with fixed end-points. This is also true
 for the continuum. In Fig.~\ref{StringN=10} and Fig.~\ref{StringN=50} we show the string shape for a fixed distance ${\bf b}=5$ 
 for two distinct resolutions $N=10$ and $N=50$ respectively. The left figure is the string projected in the transverse plane, while the right
 figure is the string in $D_\perp=3$ dimensions. Fig.~\ref{RandomStringShape} (left) show the string bits in the transverse plane for an ensemble
 of 200 strings at a resolution of $N=10$ with fixed $b=5$. Fig.~\ref{RandomStringShape} (right) shows the same for 40 strings at a 
 higher resolution $N=50$.

\section{\label{ } Self-interacting string in the Mean-Field Approximation}


Attractive string self-interaction will cause the string to  shrink transversely, while
repulsive self-interactions will cause the transverse string  to grow outward, in a way pushing the string bits out.
While string bits are held by  confinement which is harmonic in our discretized case, self-string interactions are not well-known. We now postulate that for a sufficiently high resolution or large $N$ we may average the inter-bit interactions in the string using two-body self-interactions

\be
V =-\frac{g^2}{2}   \sum_{k \neq k'} \int \frac{d^{D_\perp+1}p}{(2 \pi)^{D_{\perp}+1}} \,\frac{  M(\vec{x}_k )   M(\vec{y}_{k} )}{p^2 + m^2} \exp \left( i \vec{p} \cdot (\vec{x}_k - \vec{y}_{k})  \right)
\label{V1}
\ee
where $M(\vec{x}_k)$ is the mass of the discrete point at $\vec{x}_k$.  Here $m$ is a finite mass in units of the string length that characterizes the
range of the interaction.  Most of our numerical analyses to follow will be for $m=0$. Results at finite $m$ may be mapped on $m=0$ through 
a pertinent re-scaling of the bare coupling $g\rightarrow g(m)$.  
Note that the static interaction involves the virtual exchange in $D_\perp+1$ as the holographic
set up is in $D_\perp+2$. The effect of the curvature of AdS$_5$ will be assessed phenomenologically below.

In holographic QCD $m$ is typically  the mass of the graviton in bulk which is dual to the glueball mass on the boundary.  In the large number of colors limit, the value of $m$ is large. However, for a finite number of colors and flavors mixing between the glueballs and the flavor scalars lead to  a much lighter $m$~\cite{Kalaydzhyan:2014tfa,Liu:2014fda,Liu:2014qrt}. Also, in a dense but cold gluon medium the glueball mass
maybe lighter.  In our case we will consider  $m$ a parameter that could be re-absorbed by redefining $g$.
Throughout we will discuss in detail the attractive self-interactions or $g^2>0$.  The repulsive case and results will only be quoted. 
Note that our analysis of the string ground state is quantum so that self-interactions do not result in a string collapse thanks to the quantum
uncertainty principle.

For large $N$,  the bit coordinates $x_k$ and $x_{k'}$ are approximately independent. They are normally distributed with a probability distribution

\be
\rho( \vec{x}_k ) = \left( \frac{1}{\Sigma_k \sqrt{2 \pi}} \right)^{D_\perp} \exp\left( - \frac{(\vec{x}_k - \vec{b} \frac{k}{N} )^2}{2 \Sigma_k^2}   \right)
\label{PDF1}
\ee
The squared variance is

\be
 \Sigma_k^2  =   \sum_{n=1}^{N-1} \frac{\sin^2 \left(  \frac{n k}{N} \pi \right)}{2 \omega_n   }    \approx  \sum_{n=1}^{N-1} \frac{1}{4 \omega_n   }    =  \frac{R_\perp^2}{D_\perp}
\ee
We note that the normal frequencies $\omega_n$ differ from the free frequencies $\Omega_n$. They are defined variationally 
below. (\ref{V1}) is a highly simplified two-body interaction as higher-order many-body interactions are also possible. We just note
that $g\approx 1/N_c$ justifying the dominance of the two-body interactions.

Using (\ref{PDF1}) we may define  the bit mass distribution on the string in the mean-field type approximation as

\be
M(\vec{x}_k ) \approx \frac{M_\perp}{N+1} \rho( \vec{x}_k )
\label{M1}
\ee
Inserting (\ref{PDF1}-\ref{M1}) into (\ref{V1})  yield

\be
V=-  \frac{g^2 }{2} \sum_{k \neq k'}  \left(  \frac{M_\perp  }{N+1} \right)^2   \int\frac{d^{D_\perp+1} p}{(2 \pi)^{D_\perp+1}} \,\, \rho(\vec{x}_k) \rho(\vec{y}_{k'}) \frac{e^{i \vec{p} \cdot \left( \vec{x}_k - \vec{y}_k \right)}}{p^2 +m^2}  
\label{V2}
\ee
In the large $N$ limit, we may average (\ref{V2}) over $x_k$ and $y_k$ to obtain in the mean-field approximation

\bea
\overline{V} &\equiv &  -  \frac{g^2 }{2} \sum_{k \neq k'}  \left(  \frac{M_\perp  }{N+1} \right)^2    \int d^{D_\perp} x_k  \int d^{D_\perp} y_k \int\frac{d^{D_\perp+1} p}{(2 \pi)^{D_\perp+1}} \,\, \rho(\vec{x}_k) \rho(\vec{y}_{k'}) \frac{e^{i \vec{p} \cdot \left( \vec{x}_k - \vec{y}_k\right)}}{p^2 +m^2}  \nonumber\\
&=&  -  \frac{g^2 }{2} \sum_{k \neq k'}   \left(  \frac{M_\perp  }{N+1} \right)^2 \int\frac{d^{D_\perp+1} p}{(2 \pi)^{D_\perp+1}} \,\, \frac{1}{p^2 + m^2} \exp \left[ - \frac{p^2}{2} \left(  \Sigma_k^2 + \Sigma_{k'}^2 \right) +  i \vec{p} \cdot \vec{b} \frac{(k - k')}{N}  \right]  
\eea
Thus

\bea
\overline{V} &\approx&  -  \frac{g^2}{2}  M_\perp^2 \int\frac{d^{D_\perp+1} p}{(2 \pi)^{D_\perp+1}} \,\, \frac{1}{p^2 + m^2}\int_0^1 d k \int_0^1 d k' \exp \left[ -  p^2  \frac{R_\perp^2}{D_\perp} +  i \vec{p} \cdot \vec{b}  (k - k')   \right] \nonumber\\
&=&  -  \frac{g^2}{2}  M_\perp^2 \int\frac{d^{D_\perp+1} p}{(2 \pi)^{D_\perp+1}} \,\, \frac{1}{p^2 + m^2}  \exp \left( -  p^2  \frac{R_\perp^2}{D_\perp} \right)   \frac{4 \sin^2 \left(  \frac{\vec{p} \cdot \vec{b}}{2} \right)}{\left( \vec{p} \cdot \vec{b}  \right)^2}
\label{VV2}
\eea
For $m=0$ and $\vec{b} \rightarrow 0$, (\ref{V2}) simplifies

\be
\overline{V}\approx  -{\bf C}(D_\perp)\,g^2\,\frac{M^2_\perp}{R_\perp^{D_\perp-1}}
\label{NEWTON}
\ee
with $ {\bf C}\equiv  ( 1 /  \sqrt{16 \pi D_\perp} )      \left(  D_\perp / 4 \pi \right)^{\frac{D_\perp}{2}} \Gamma(D_\perp/2 -1/2)/ \Gamma(D_\perp/2 + 1/2 )  $. In this limit, the self-interactions between the string bits
reduce to a Newtonian potential acting as a mean-field approximation. The Newtonian constant is identified as $ G_N=g^2 D_\perp^{(D_\perp -1)/2 }/ (8 \pi)$ through the bottom-up holographic setting in $D_\perp+2$ dimensions~\cite{Stoffers:2012zw}. Thus,

\be
  g^2= 8 \pi D_\perp^{\frac{1- D_\perp }{2}} \,l_P^{D\perp}=  8 \pi D_\perp^{\frac{1- D_\perp }{2}} \, g_s^2 l_s^{D_\perp} \equiv
   2^{3 - D_\perp} \pi D_\perp^{\frac{1- D_\perp }{2}}\,  g_s^2 
  \label{GPLANCK}
\ee
where in the last equality we reset $2l_s\equiv 1$ as per our current conventions.
Recall that the curvature effects
of AdS$_5$, which we are ignoring so far,  amounts to an effective $D_\perp\rightarrow D_\perp(\lambda)$ in leading order on the
transverse string propagator as we noted earlier. This observation will be used below to estimate the curvature corrections to the
current analysis.

\section{\label{subattraction}Variational Analysis}

For small perturbative interactions, we can modify the transverse Hamiltonian through

\be\la{hamiltonianpostulate}
\mathcal{H}_\perp  = 2 M_\perp^2 + 2 M_\perp \delta (2 M_\perp)\equiv 2 M_\perp^2 + 2M_\perp\, \overline{V}
\ee 
$\mathcal{H}_\perp$  in Eq.~\ref{hamiltonianpostulate} is difficult to analyze analytically in the presence of $\overline V$.  
We follow Thorn and Ogerman~\cite{Bergman:1997ki} and analyze it variationally by using a trial Gaussian distribution for each string bit

\be
\Psi (X_n^i) = \left( \frac{\omega_n}{\pi}  \right)^{\frac{1}{4}} \exp\left[  - \frac{\omega_n}{2} (X_n^i)^2   \right]
\label{VAR1}
\ee
where the set of normal modes $\omega_n$ will be defined below by minimizing the energy of the string in the presence of $V$.
In terms of (\ref{VAR1}) the scalar part is

\be
\mathcal{H}_\perp^0  =   \frac{D_\perp}{4 } \sum_{n=1}^{N-1}  \left(      \omega_n   + 
 \frac{\Omega_n^2}{\omega_n} \right)  + \frac{  b^2}{  \pi^2 } 
\ee
and reduces to (\ref{H0})  when $\omega_n=\Omega_n$ for $V=0$. 
The effective mass of the string is
\be
M_\perp^2 [\omega_n]= \frac{1}{2}\mathcal{H}_\perp^0   =  \frac{D_\perp}{8} \sum_{n=1}^{N-1}  \left(      \omega_n   + 
 \frac{\Omega_n^2}{\omega_n} \right)  + \frac{  b^2}{ 2 \pi^2 } 
 \label{MVAR}
\ee
The squared effective transverse radius is

\be 
R_\perp^2[\omega_n]  =  \frac{1}{N} \sum_{k= 0}^N  \left< \left(  x_k^i  - \bold{b}^i \frac{k}{N} \right)^2 \right>  
=  \frac{D_\perp}{4}\sum_{n=1}^{N-1} \frac{1}{\omega_n}
\label{RVAR}
\ee 
 
With our conventions the total string energy is $E[\omega_n]= 2 M_\perp^2 + 2 M_\perp \overline{V}$ depends on the set of variational parameters
$\omega_n$ which are fixed through the minimum

\bea
\frac{\delta E}{\delta \omega_n} &=&   \frac{\delta M_\perp^2}{\delta \omega_n} \left[  2 -   \frac{3g^2 }{2} M_\perp  \int\frac{d^{D_\perp+1} p}{(2 \pi)^{D_\perp+1}} \,\, \frac{1}{p^2 + m^2}  \exp \left( -  p^2  \frac{R_\perp^2}{D_\perp} \right)   \frac{4 \sin^2 \left(  \frac{\vec{p} \cdot {b}}{2} \right)}{\left( \vec{p} \cdot \vec{b}  \right)^2}  \right] \nonumber\\
&& +  \frac{1}{D_\perp}   \frac{\delta R_\perp^2}{\delta \omega_n}   g^2   M_\perp^3 \int\frac{d^{D_\perp+1} p}{(2 \pi)^{D_\perp+1}} \,\, \frac{p^2}{p^2 + m^2}  \exp \left( -  p^2  \frac{R_\perp^2}{D_\perp} \right)   \frac{4 \sin^2 \left(  \frac{\vec{p} \cdot {b}}{2} \right)}{\left( \vec{p} \cdot \vec{b}  \right)^2}=0
\label{VAR2}
\eea
The mass and size variations can be made explicit

\be
\frac{\delta M_\perp^2}{\delta \omega_n} =  \frac{D_\perp}{8}   \left(    1  -  \frac{\Omega_n^2}{\omega_n^2} \right) \ \ \ \ \ \ \ \ \ \ 
 \frac{1}{D_\perp}   \frac{\delta R_\perp^2}{\delta \omega_n} = -   \frac{1}{4 \omega_n^2   } 
\label{S1}
\ee
Inserting (\ref{S1}) into (\ref{VAR2}) and rearranging yield

\bea 
 \omega_n^2 =   \Omega_n^2  +      \frac{  \frac{g^2 M_\perp^3}{D_\perp}  \int\frac{d^{D_\perp+1} p}{(2 \pi)^{D_\perp+1}} \,\, \frac{p^2}{p^2 + m^2}  \exp \left( -  p^2  \frac{R_\perp^2}{D_\perp} \right)   \frac{4 \sin^2 \left(  \frac{\vec{p} \cdot {b}}{2} \right)}{\left( \vec{p} \cdot \vec{b}  \right)^2}}{ 1 -   \frac{3 g^2}{4} M_\perp \int\frac{d^{D_\perp+1} p}{(2 \pi)^{D_\perp+1}} \,\, \frac{1}{p^2 + m^2}  \exp \left( -  p^2  \frac{R_\perp^2}{D_\perp} \right)   \frac{4 \sin^2 \left(  \frac{\vec{p} \cdot {b}}{2} \right)}{\left( \vec{p} \cdot \vec{b}  \right)^2}}
 \label{OVAR}
\eea
where both $M^2_\perp[\omega_n]$ and $R_\perp^2[\omega_n]$ depend on the variational parameters through (\ref{MVAR}-\ref{RVAR}).
(\ref{OVAR}) define a highly non-linear set of equations for the variational parameters $\omega_n$ defining the Gaussian ansatz (\ref{VAR1}).
The generic solution is of the form $\omega_n = \sqrt{\Omega_n^2 + \eta^2}$ with

\bea 
\eta^2 =          \frac{   \frac{ g^2 M_\perp^3 }{ D_\perp}  \int_0^\infty d p \int_0^\pi d\phi \,\, \left(\sin \phi\right)^{D_\perp -1}   \frac{p^{D_\perp + 2}}{p^2 + m^2}  \exp \left( -  p^2  \frac{R_\perp^2}{D_\perp} \right)   \frac{4 \sin^2 \left(  \frac{p b \cos \phi}{2} \right)}{p^2 b^2 \cos^2 \phi}}{2^{D_\perp} \pi^{\frac{D_\perp+2}{2} } \Gamma(\frac{D_\perp}{2}) -   \frac{3 g^2}{4} M_\perp   \int_0^\infty d p \int_0^\pi d\phi \,\, \left(\sin \phi\right)^{D_\perp -1}  \frac{p^{D_\perp}}{p^2 + m^2}  \exp \left( -  p^2  \frac{R_\perp^2}{D_\perp} \right)   \frac{4 \sin^2 \left(  \frac{p b \cos \phi}{2} \right)}{p^2 b^2 \cos^2 \phi}}
\label{E1}
\eea
to be determined numerically.  Note that for $b =0$, (\ref{E1}) simplifies as

\bea 
\eta^2 =    \frac{M_\perp^2}{R_\perp^2}      \frac{  \frac{1 }{2}  \frac{ g^2 M_\perp }{    R_\perp^{D_\perp - 1}}\exp \left( \frac{m^2 R_\perp^2}{D_\perp}  \right)  \Gamma\left( \frac{3 + D_\perp}{2}  \right) \left(  \frac{m^2 R_\perp^2}{D_\perp} \right)^\frac{D_\perp + 1}{2} \Gamma \left( - \frac{1 + D_\perp}{2} , \frac{m^2 R^2}{D_\perp}  \right) }{2^{D_\perp} \pi^{\frac{D_\perp+1}{2} } \left( D_\perp \right)^{\frac{1 - D_\perp }{2}} \Gamma(\frac{D_\perp + 1}{2}) -   \frac{3    }{8 } \frac{g^2 M_\perp}{R_\perp^{D_\perp -1}}\exp \left( \frac{m^2 R_\perp^2}{D_\perp}  \right)  \Gamma\left( \frac{1 + D_\perp}{2}  \right) \left(  \frac{m^2 R_\perp^2}{D_\perp} \right)^\frac{D_\perp - 1}{2} \Gamma \left( - \frac{  D_\perp - 1}{2} , \frac{m^2 R^2}{D_\perp}  \right)}\nonumber\\
\label{E2}
\eea

For $m=0$, Eq.~\ref{E2} further simplifies as
\bea 
\eta^2 =   \frac{M_\perp^2}{R_\perp^2}      \frac{  \frac{1 }{2}  \frac{ g^2 M_\perp }{    R_\perp^{D_\perp -1}}  }{ 2^{D_\perp} \pi^{\frac{D_\perp+1}{2} } \left( D_\perp \right)^{\frac{1 - D_\perp }{2}}  -   \frac{3    }{4 (D_\perp -1) }   \frac{g^2 M_\perp}{R_\perp^{D_\perp -1}}  }
\label{E3}
\eea

Note that for $b\neq 0$, (\ref{E1}) simplifies as

\bea 
\eta^2 =  \frac{M_\perp^2}{b^2}        \frac{  \frac{4}{D_\perp}  \frac{   g^2 M_\perp  }{   b^{D_\perp -1}} {\bf I}[0]  }{2^{D_\perp} \pi^{\frac{D_\perp+2}{2} } \Gamma(\frac{D_\perp + 1}{2}) - 3   \frac{  g^2 M_\perp}{ b^{D_\perp-1}} {\bf I}[2]  }
\label{E4}
\eea
where
\be
{\bf I}[a] \equiv   \int_0^\infty d x \int_0^\pi d\phi \,\, \left(\sin \phi\right)^{D_\perp -3}   \tan^2\phi \,\,  \frac{x^{D_\perp -a }}{x^2 + m^2 b^2}  \exp \left( -  \frac{x^2}{D_\perp }  \frac{R_\perp^2}{b^2 } \right)     \sin^2 \left(  \frac{x  \cos \phi}{2} \right)
\ee

 The repulsion will cause the string bits to expand. A rerun of the  precedent arguments  yields now  $\omega_n = \sqrt{\Omega_n^2 - \eta^2}$ with 

\bea 
\eta^2 =  \frac{M_\perp^2}{b^2}        \frac{  \frac{4}{D_\perp}  \frac{   \tilde{g}^2 M_\perp  }{   b^{D_\perp -1}} {\bf I}[0]  }{2^{D_\perp} \pi^{\frac{D_\perp+2}{2} } \Gamma(\frac{D_\perp + 1}{2})+ 3   \frac{  \tilde{g}^2 M_\perp}{ b^{D_\perp-1}} {\bf I}[2]  }
\label{Erep}
\eea
The variational analysis will be now carried numerically for both the attractive and repulsive string interaction in the mean-field
approximation.

\section{Numerical Results}

\begin{figure}[!htb]
\minipage{0.33\textwidth}
\includegraphics[width=42mm]{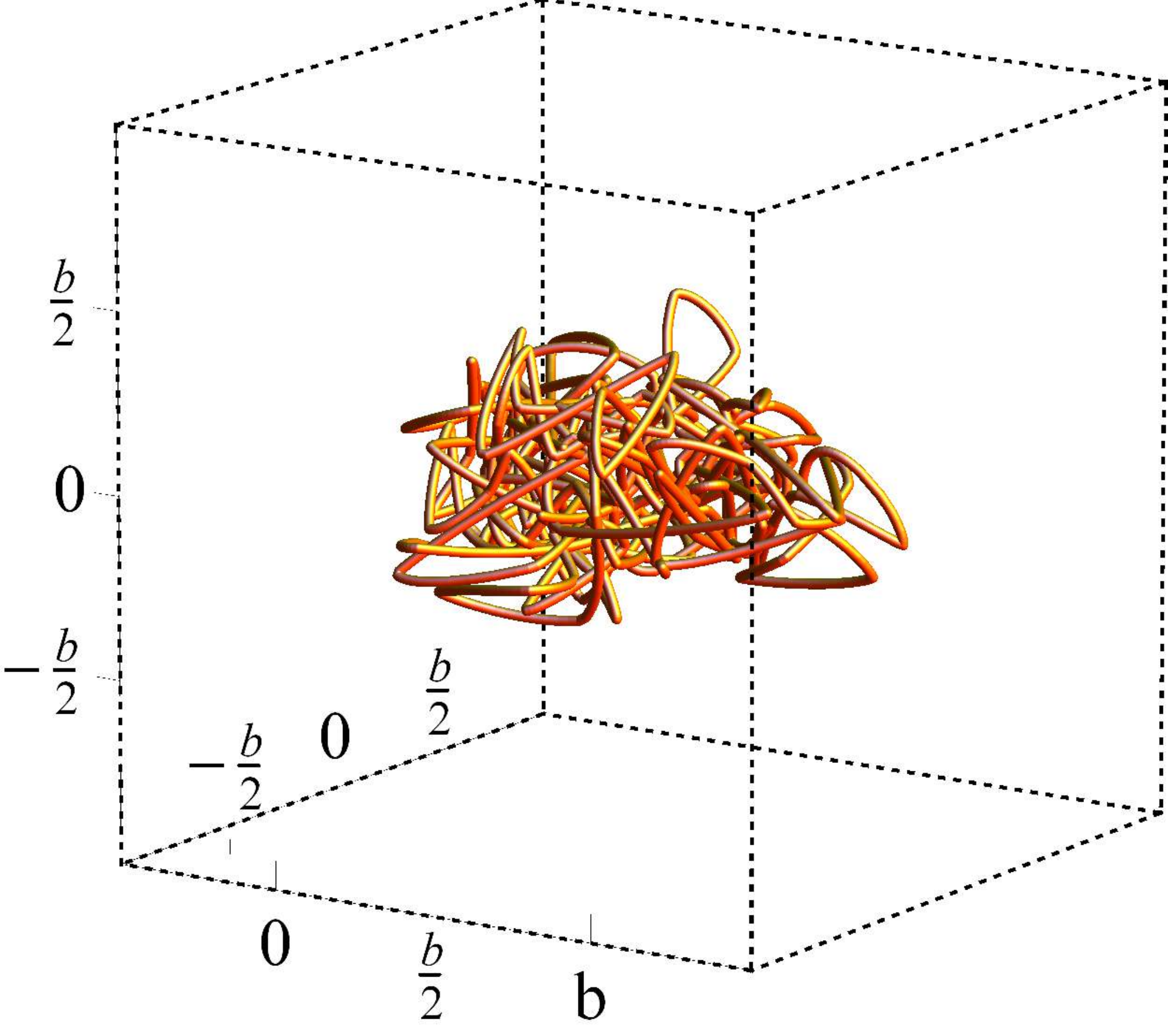}
\endminipage\hfill
\minipage{0.33\textwidth}
\includegraphics[width=42mm]{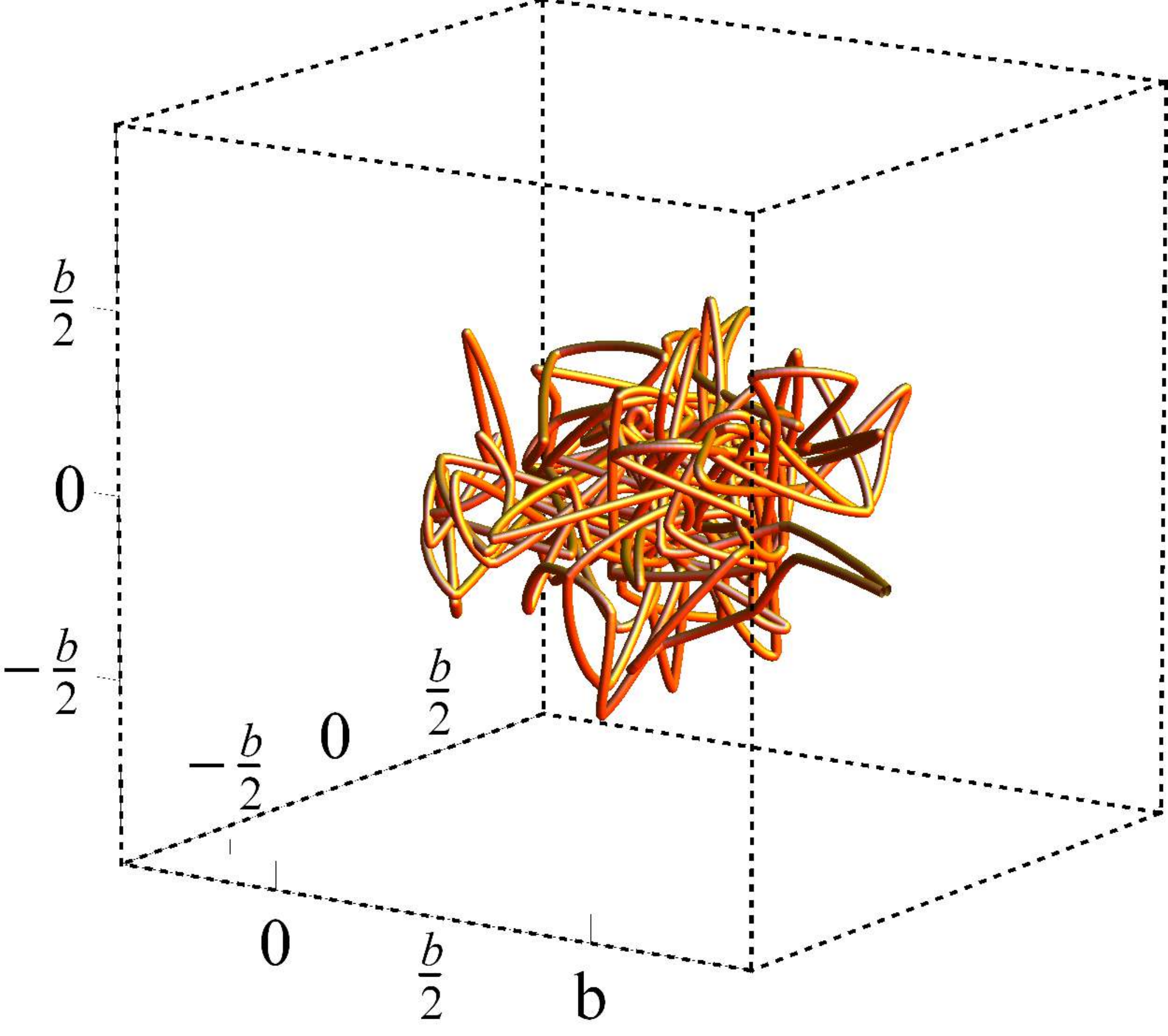}
\endminipage
\minipage{0.33\textwidth}
\includegraphics[width=42mm]{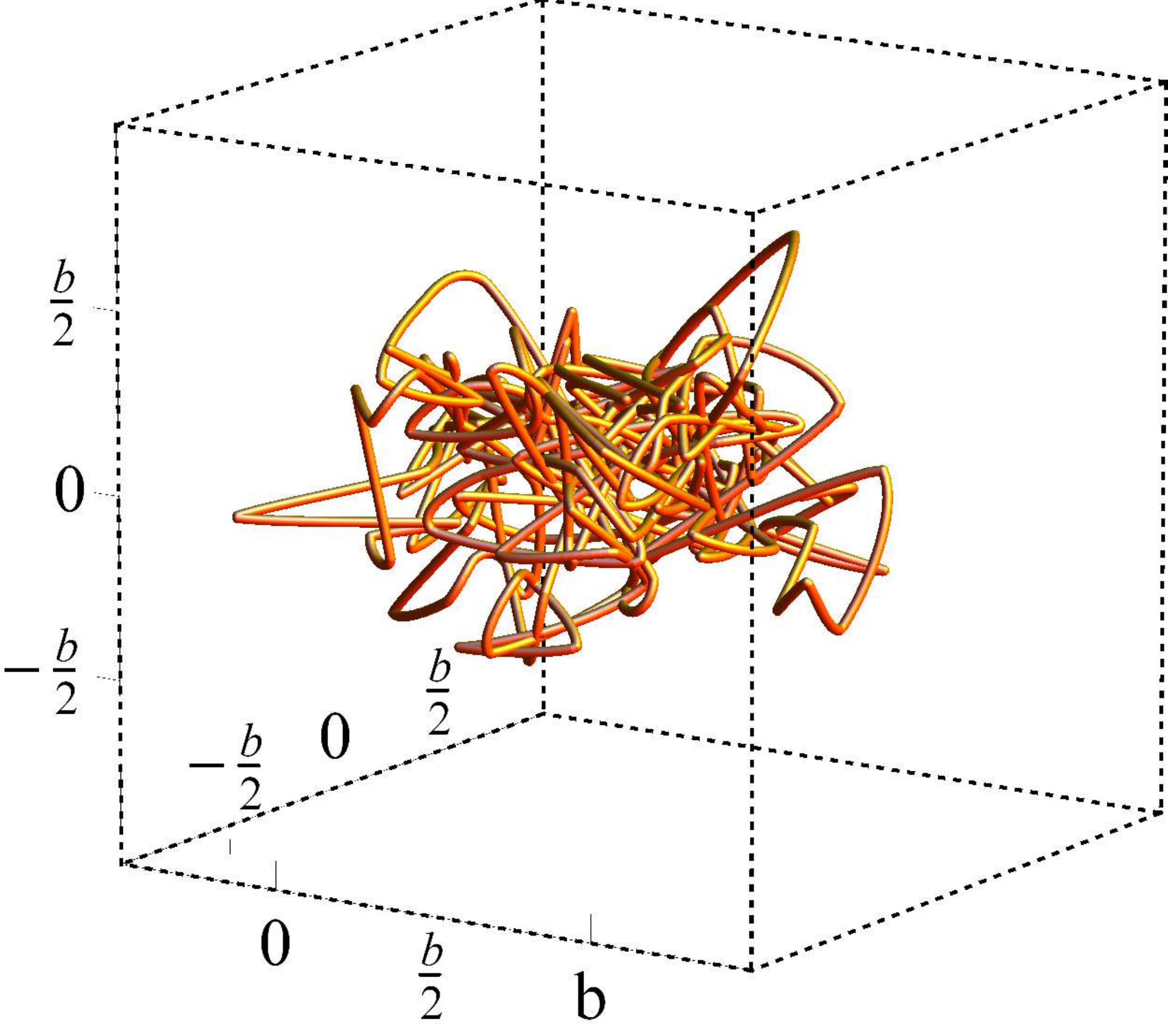}
\endminipage
 \caption{Attractive interaction: $g = 0.3$~($\bold{Left}$). \,\, No interaction:  $g=\tilde{g} = 0$~($\bold{Center}$). \,\,  Repulsive interaction: $\tilde{g} = 0.3$~($\bold{Right}$). }\label{S603} 
\end{figure}

\begin{figure}[!htb]
\minipage{0.33\textwidth}
\includegraphics[width=50mm]{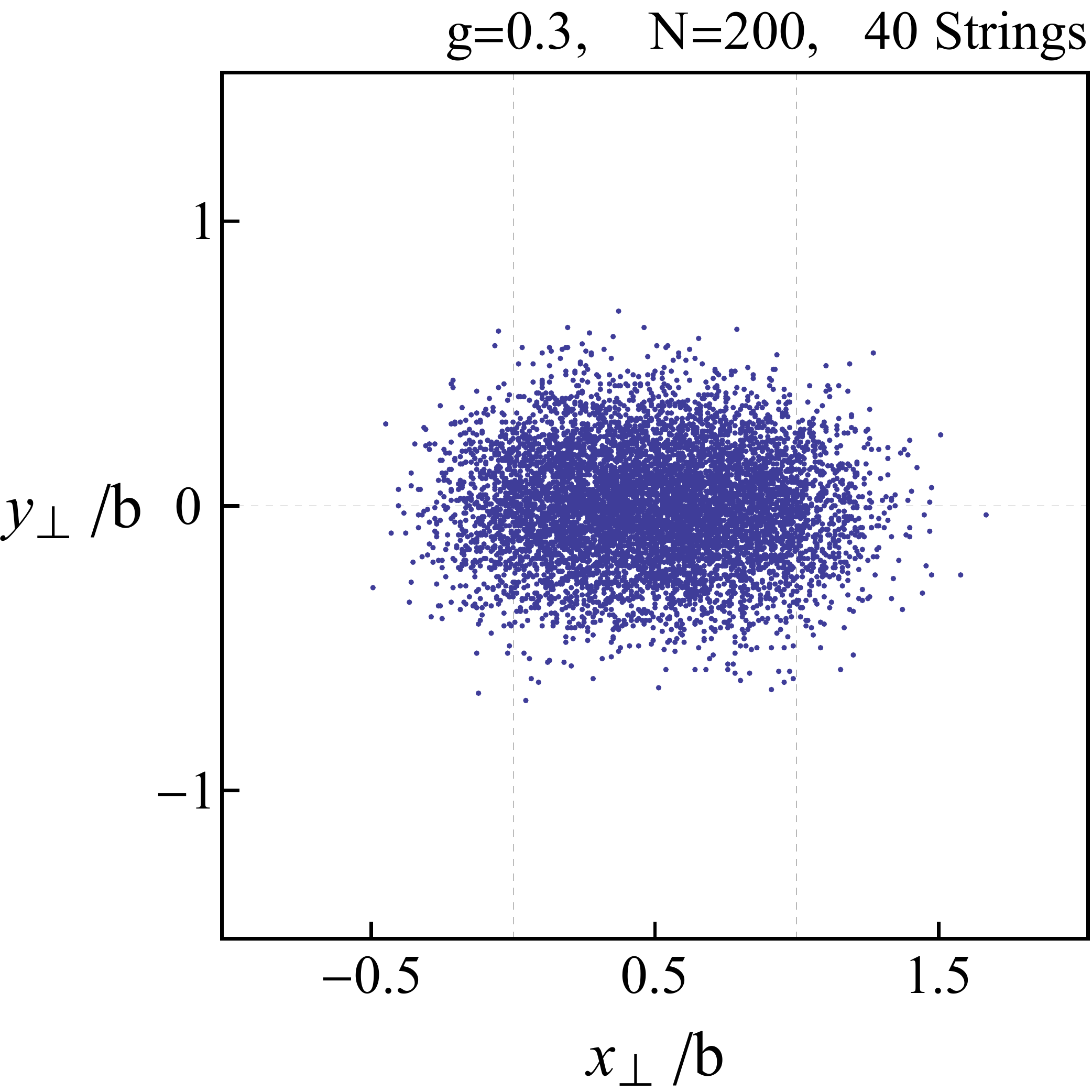}
\endminipage\hfill
\minipage{0.33\textwidth}
\includegraphics[width=50mm]{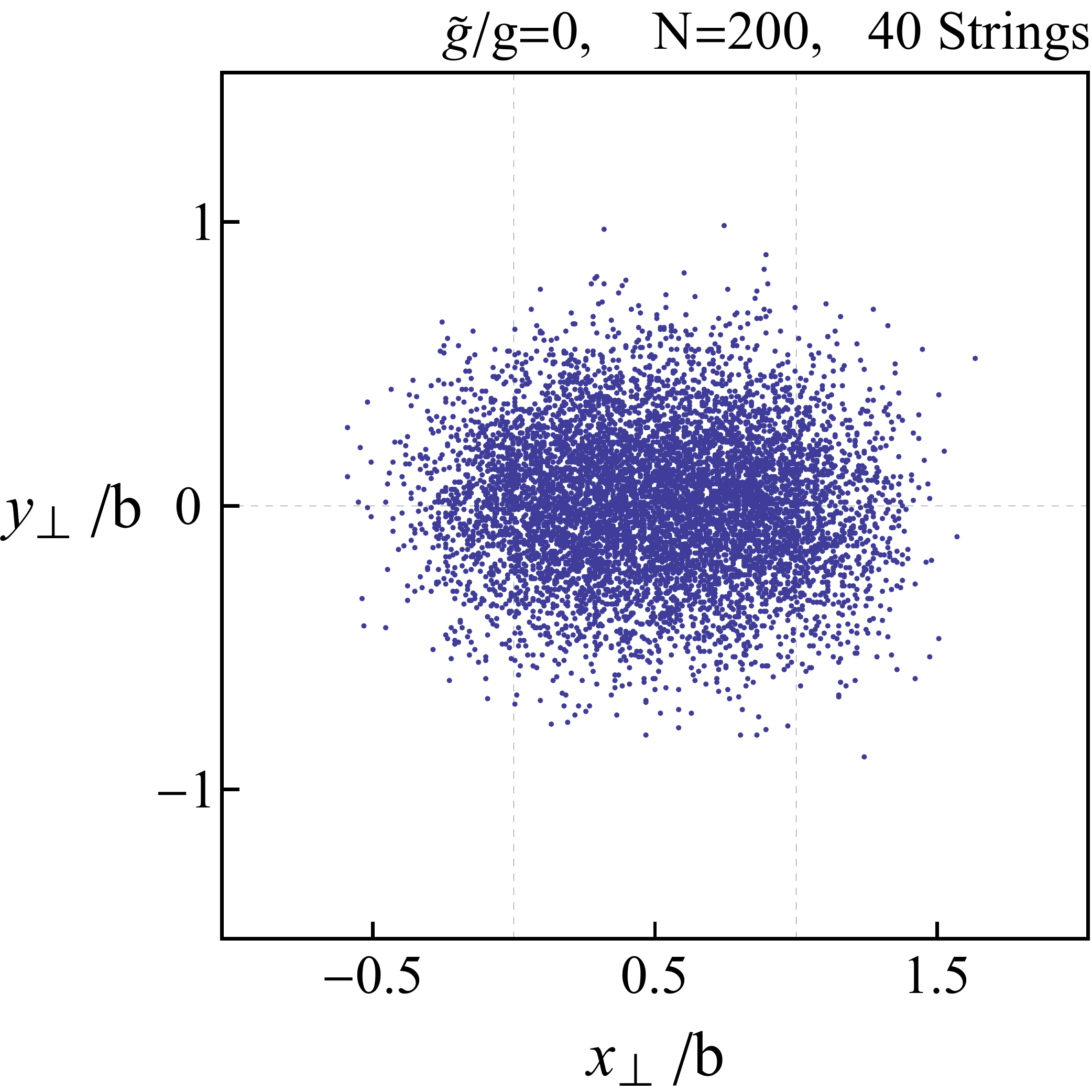}
\endminipage
\minipage{0.33\textwidth}
\includegraphics[width=50mm]{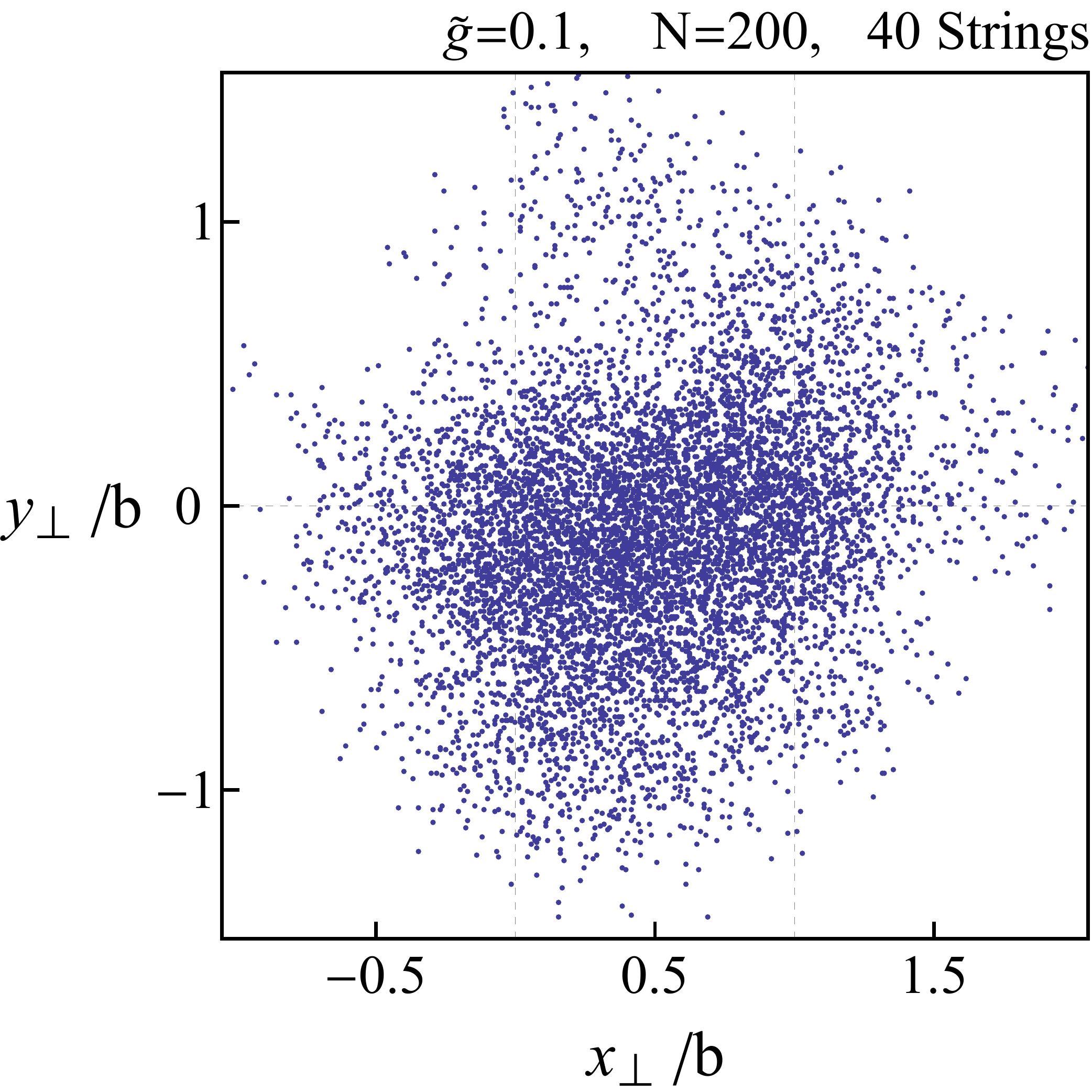}
\endminipage
\caption{Attractive interaction: $g = 0.3$~($\bold{Left}$). \,\, No interaction:  $g=\tilde{g} = 0$~($\bold{Center}$). \,\,  Repulsive interaction: $\tilde{g} = 0.1$~($\bold{Right}$). }\label{transversedisg}
\end{figure}

The Gaussian variation ansatz (\ref{VAR1}) can be used to define a variational probability distribution 
$|\Psi(X_n^i)|^2$ for the string amplitudes $X_n^i$ in the normal mode decomposition (\ref{N1}). Each
string bit undergoes a Gaussian random walk which is free for $g=0$ but constrained by the interaction
through $\omega_n$ for $g\neq 0$.   In Fig.~\ref{S603}  we show the spatial geometry of our discretized
strings in $D_\perp=3$ with a resolution $N=200$ for the attractive interaction $g=0.3$, no interaction and repulsive interaction $\tilde{g}=0.3$.  The string
is stretched with $b=5$. In Fig.~\ref{transversedisg}  we show the transverse distribution of the string bits for
an ensemble consisting of 40 stretched strings. The string bits are the dots and we have left out the 
string connection for a better visualization. The resolution is $N=1/x=200$. The attractive configurations
are denser along $b$, while the repulsive configurations are spread out of $b$.

\begin{figure}[!htb]
\minipage{0.48\textwidth}
\includegraphics[width=56mm]{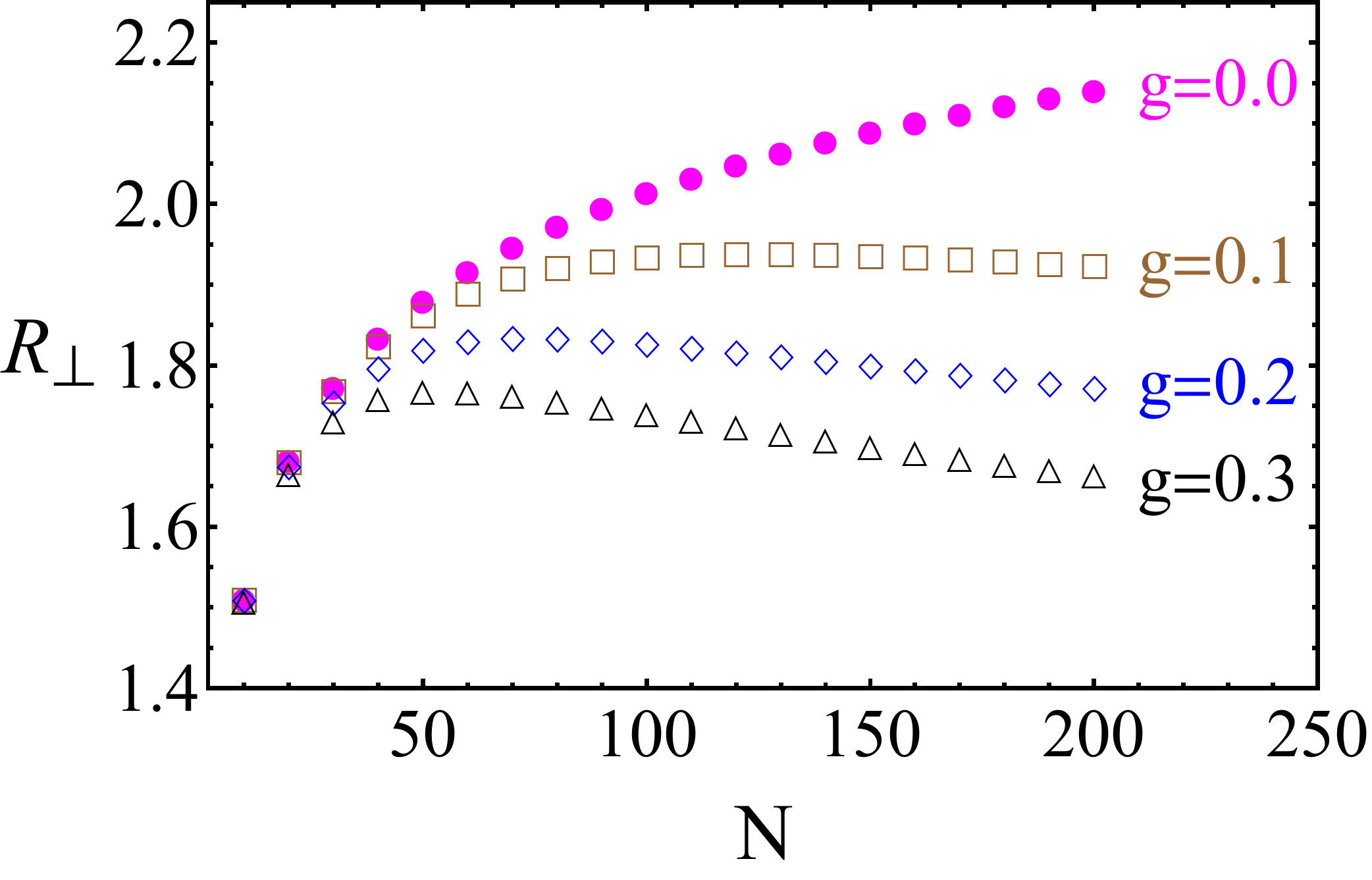}
\endminipage\hfill
\minipage{0.48\textwidth}
\includegraphics[width=56mm]{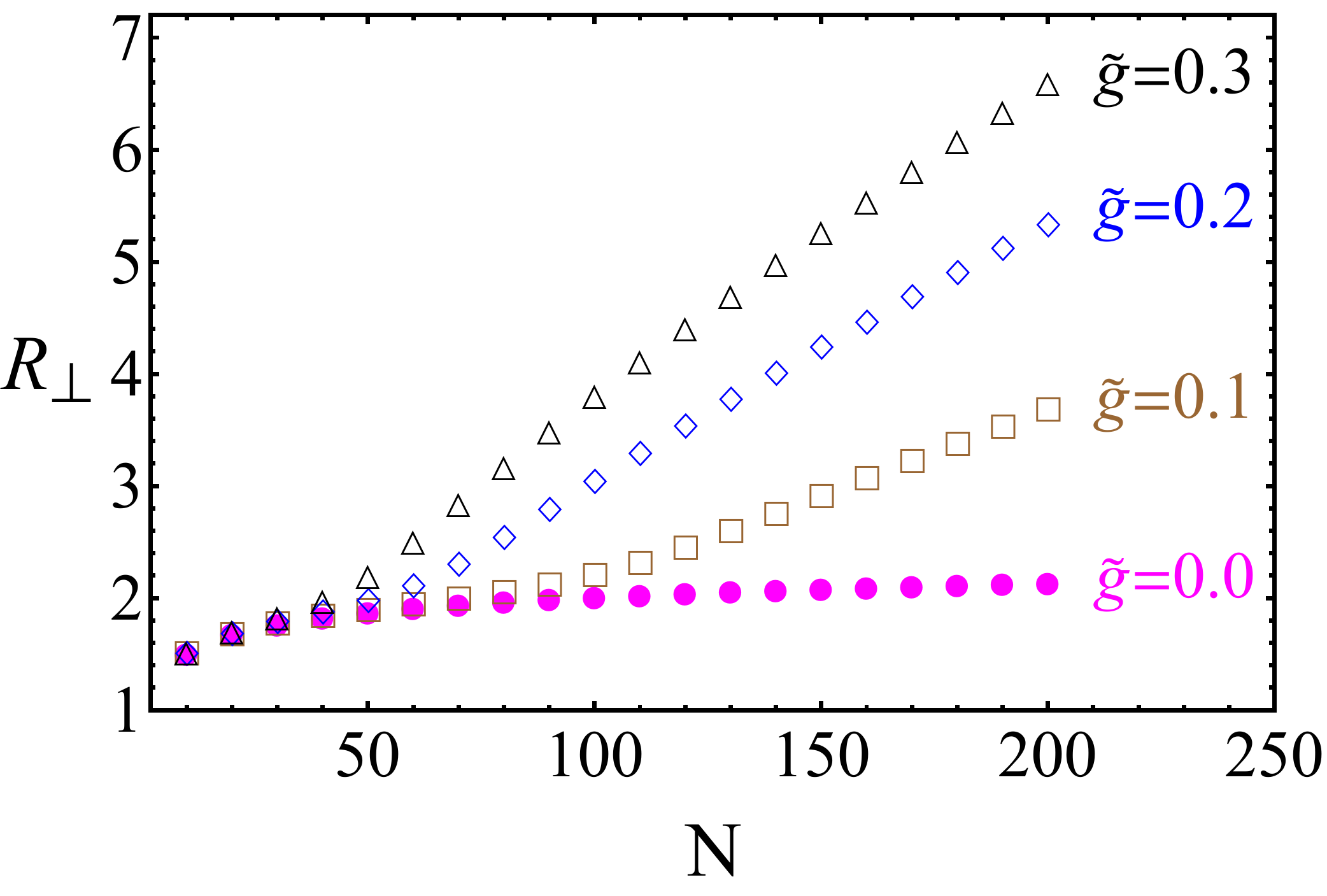}
\endminipage
 \caption{Attractive interaction: $g = 0.1, 0.2, 0.3$~($\bold{Left}$). \,\,   Repulsive interaction: $\tilde{g} = 0.1, 0.2, 0.3$~($\bold{Right}$). } \label{radius}
\end{figure}

\begin{figure}[!htb]
\minipage{0.48\textwidth}
\includegraphics[width=56mm]{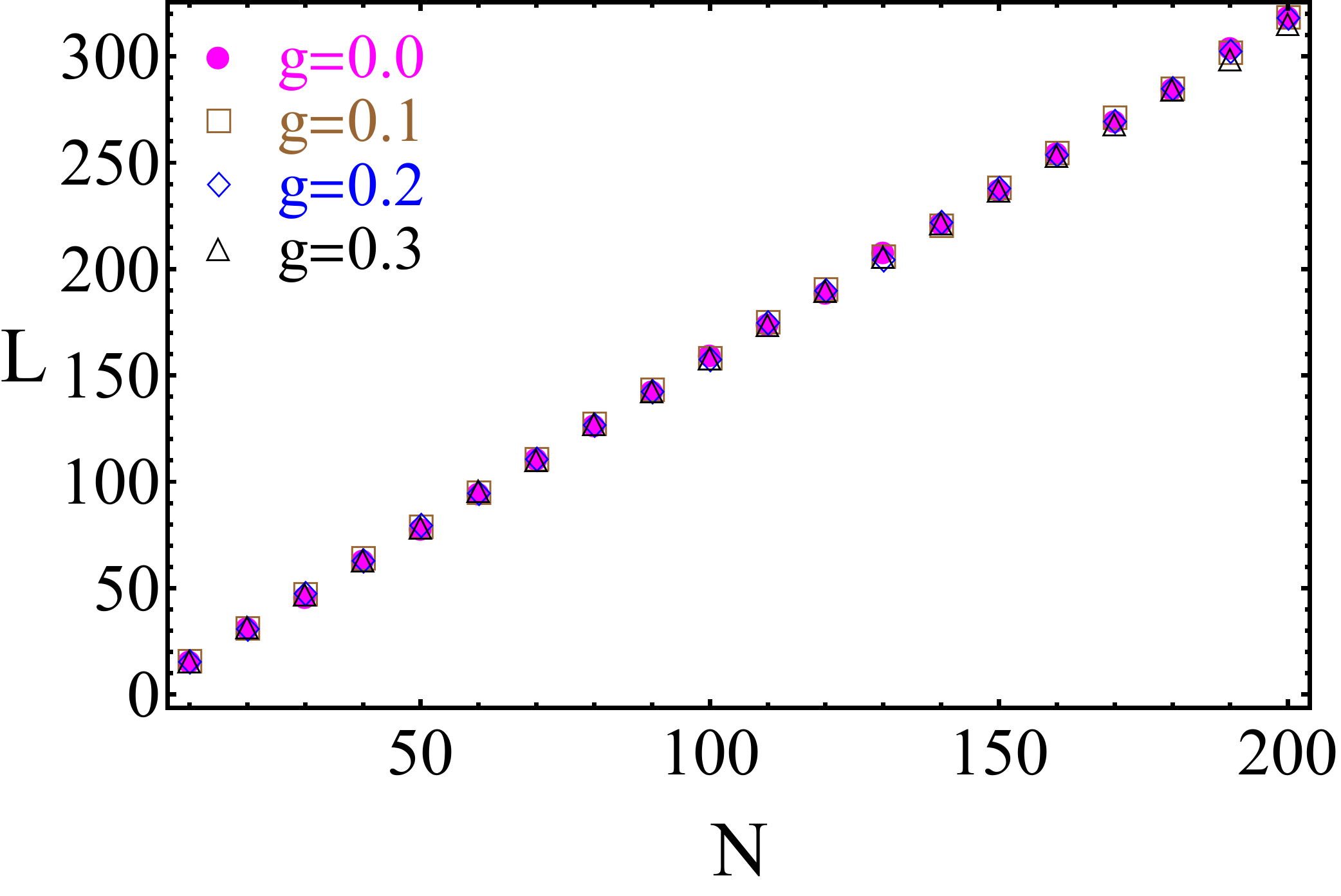} 
\endminipage\hfill
\minipage{0.48\textwidth}
\includegraphics[width=56mm]{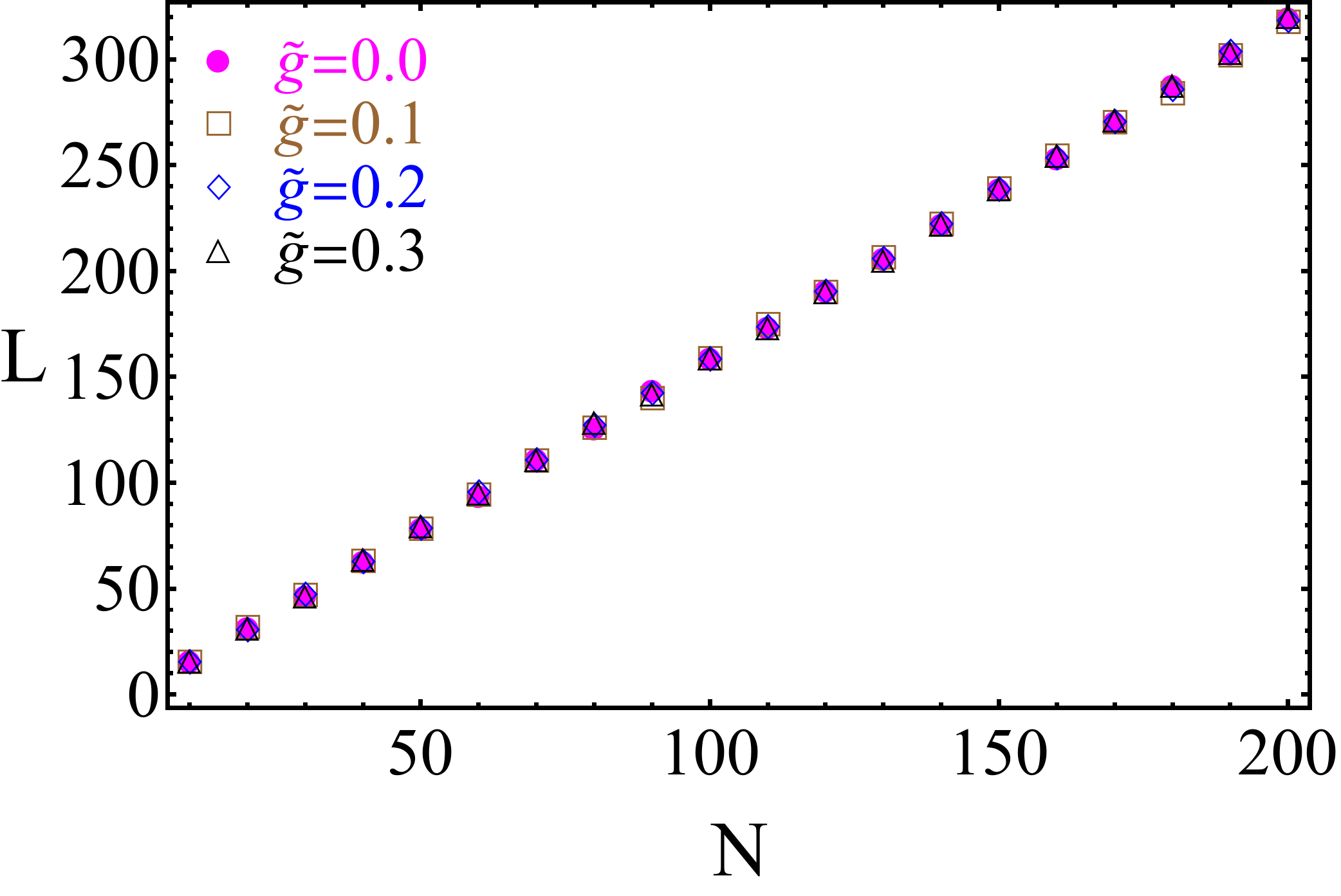}
\endminipage
 \caption{Attractive interaction: $g =0.1, 0.2, 0.3$~($\bold{Left}$). \,\,   Repulsive interaction: $\tilde{g} = 0.1, 0.2, 0.3$~($\bold{Right}$). } \label{length}
\end{figure}

\begin{figure}[!htb]
\minipage{0.48\textwidth}
\includegraphics[width=56mm]{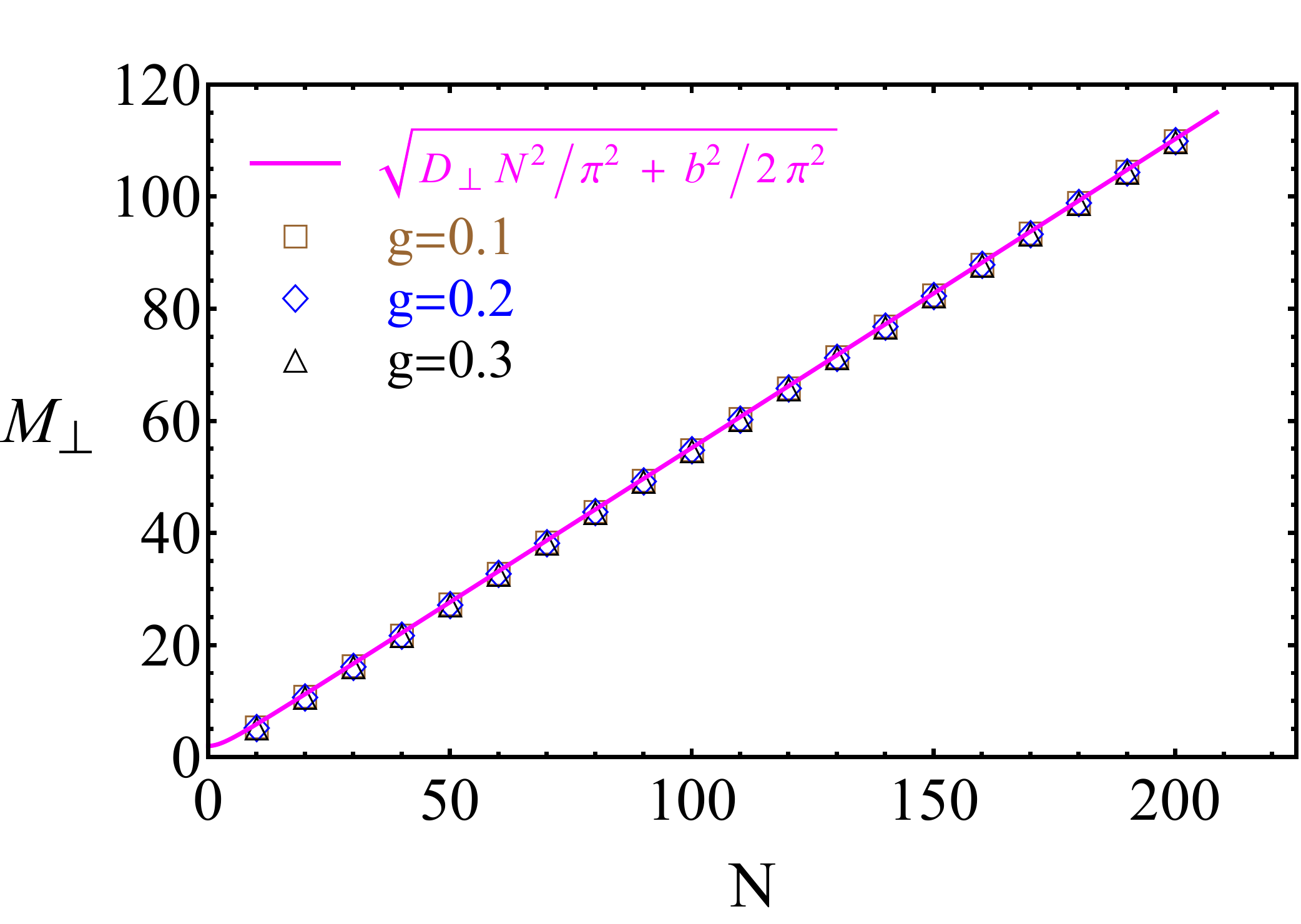}
\endminipage\hfill
\minipage{0.48\textwidth}
\includegraphics[width=56mm]{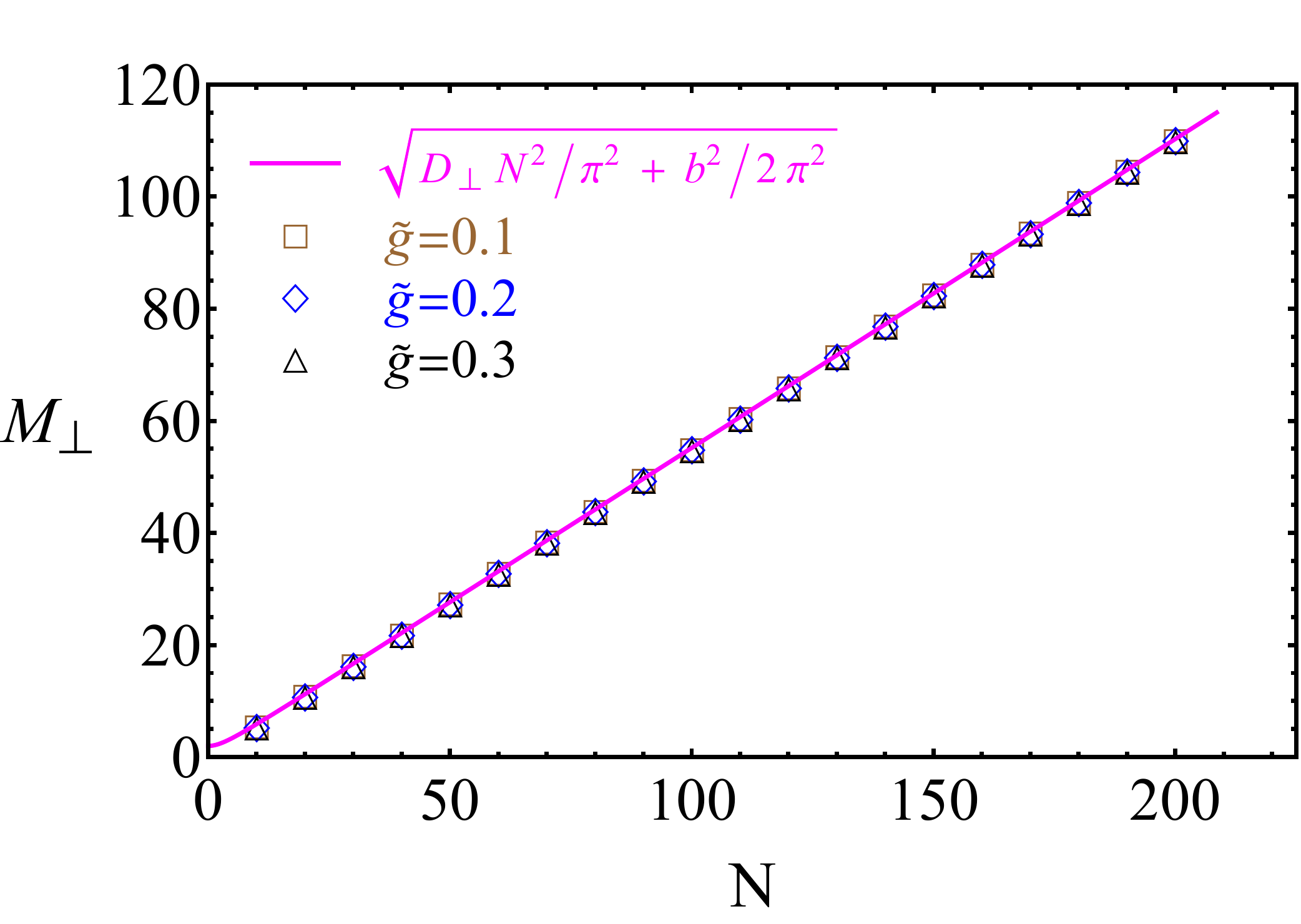}
\endminipage
 \caption{Solid line is the analytical result in (\ref{mperpanalytic}). Numeric datas are for attractive interaction: $g = 0.1, 0.2, 0.3$~($\bold{Left}$) and repulsive interaction: $\tilde{g} = 0.1, 0.2, 0.3$~($\bold{Right}$).  } \label{Mass}
\end{figure}

In Fig.~\ref{radius} we show the growth of the transverse radius as measured by (\ref{RVAR}) versus the resolution
$N$ for different strengths of the attractive forces (left) and repulsive forces (right). For comparison, we also show the 
full length of the string

\be\label{LTO}
L \equiv \sum_{k=1}^{N} \sqrt{  \left(  x_k^i - x_{k-1}^i \right)^2}
\ee
The analogue change of the total length of the
string with the resolution as defined in (\ref{LTO}) and the mass of the string as defined in (\ref{MVAR})  are also shown in Fig.~\ref{length} and Fig.~\ref{Mass}. While the length and mass scale  linearly with $N$ whatever the interaction, the transverse size of the string bit distribution shows sensitivity to $N$. 
For $g=\tilde{g}=0$ the transverse
radius grows logarithmically as expected. As the attraction is switched on, the transverse radius asymptotes a constant about the string length. 
In contrast, as the repulsion is switched on, the transverse radius asymptotes a linear rise with the resolution as also noted
in~\cite{Bergman:1997ki} in their non-relativistic string bit models with a variety of repulsive string interactions of different ranges. This supports our
earlier earlier observation that at large resolution $N$ the mean-field approximation is generic.

\begin{figure}[!htb]
\minipage{0.48\textwidth}
\includegraphics[width=61mm]{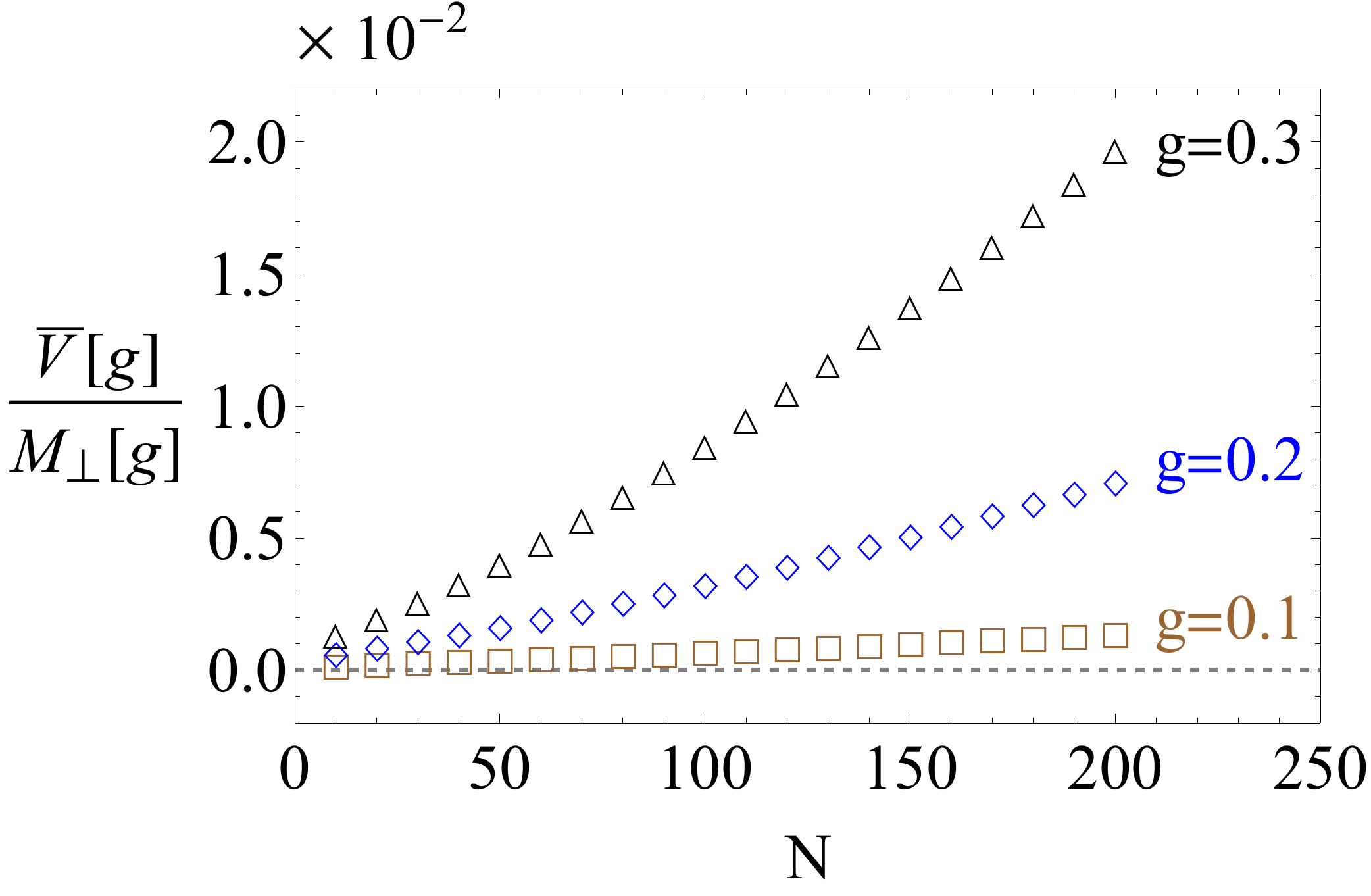}
\endminipage\hfill
\minipage{0.48\textwidth}
\includegraphics[width=61mm]{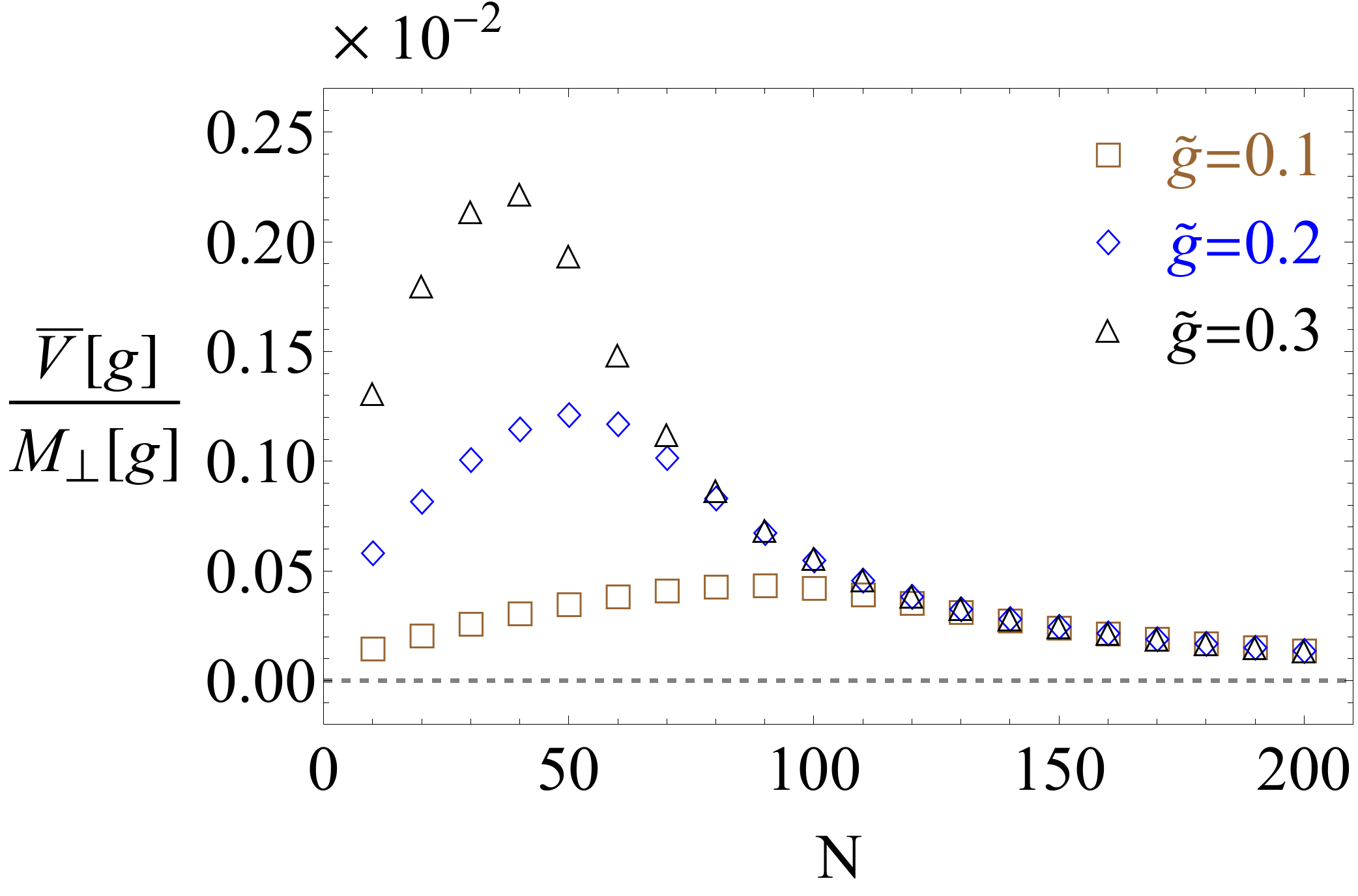}
\endminipage
 \caption{Attractive interaction: $g = 0.1, 0.2, 0.3$~($\bold{Left}$). \,\,   Repulsive interaction: $\tilde{g} = 0.1, 0.2, 0.3$~($\bold{Right}$). } \label{Validity}
\end{figure}

We  note that all attractive string self-interactions
result in transverse area that are less than or equal to the Froissart bound. In contrast, all repulsive string self-interactions result in a transverse area
that upsets the Froissart bound at asymptotic $N$ or asymptotically low-x. Thus our observation that saturation of the Bekenstein bound
by the the string bits or wee gluons, follows from weakly attractive string-self interactions in conformity with the Froissart bound. 
We also note that our treatment of the interaction assumes weak self-interactions, or the smallness of the ratio

\be
\frac{2 M_\perp \overline{V}}{ \mathcal{H}_\perp^0} = \frac{\overline{V} [g]}{ M_\perp [g]}
\ee
We show in~Fig.~\ref{Validity} that this is indeed the case.

\section{Saturation}

At low-x or large $N$ and $b=0$, 
the transverse string density is high ${\bf n}_\perp(0,N)\approx N/R_\perp^{D_\perp}$ as $R_\perp$ shrinks under the
effect of attractive self-interactions. For $0<g<0.3$ our numerical results yield $1.2<R_\perp<1.6$ in units where $2l_s=1$, i.e. 
$2.4<R_\perp/l_s<3.2$. To understand the effects of the self-interaction on the string size configuration, we re-write schematically 
the squared mass $E=2M_\perp^2+2M_\perp V$  of the self-interacting string in terms of $N$ and $R_\perp$ dropping all numerical factors

\be
E\approx N^2\frac{1}{R_\perp^2}+N^2 \frac{R_\perp^2}{D^2_\perp{{\rm ln}^2N}}-N^2\frac{g^2N}{R_\perp^{D_\perp-1}}
\label{EE}
\ee
The first contribution in (\ref{EE}) follows from the kinetic contribution in (\ref{HAR}) using the estimate $N^2p_i^2$ and the
uncertainty principle $p_i\approx 1/R_\perp$. The second contribution in (\ref{EE}) follows from the harmonic potential in
(\ref{EE}) using the estimate $N^2(\Delta x_i)^2$ with typically 
$\Delta x_i\approx R_\perp/(D_\perp {\rm ln}N)$ in the diffusive regime. The third 
contribution in (\ref{EE}) is the potential contribution $2M_\perp\,\overline{V}$ to the squared mass after using $M_\perp\approx N$.
Note that for $g=0$ the minimum of (\ref{EE}) yields the diffusive result $R^2_\perp\approx D_\perp{{\rm ln}N}$ whatever
$D_\perp$. For finite $g$, the minimum of (\ref{EE}) depends on the dimensionality $D_\perp$.  A similar relation to (\ref{EE})
was found to hold for classical strings at hi high temperature by Damour and Veneziano~\cite{Damour:1999aw} using different arguments.

\subsection{Flat Space: $D_\perp=3$}

For our case $D_\perp=3$ so the minimum of (\ref{EE}) occurs for

\be
R_\perp\approx \left(1-g^2N\right)^{1/4}\,\sqrt{{\rm ln}N}
\label{EE1}
\ee

\begin{figure}[!htb]
\minipage{0.48\textwidth}
\includegraphics[width=61mm]{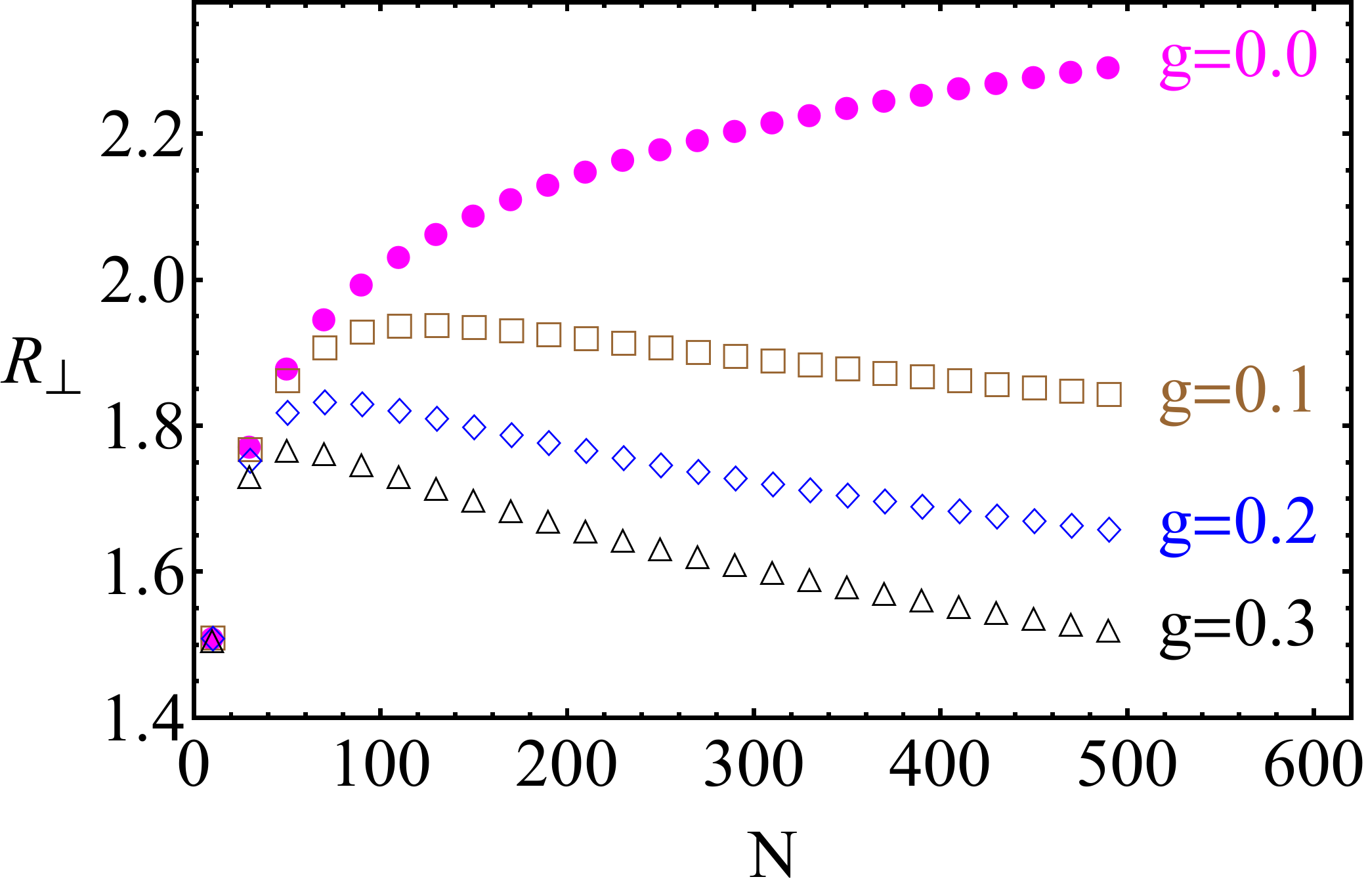}
\endminipage\hfill
\minipage{0.48\textwidth}
\includegraphics[width=61mm]{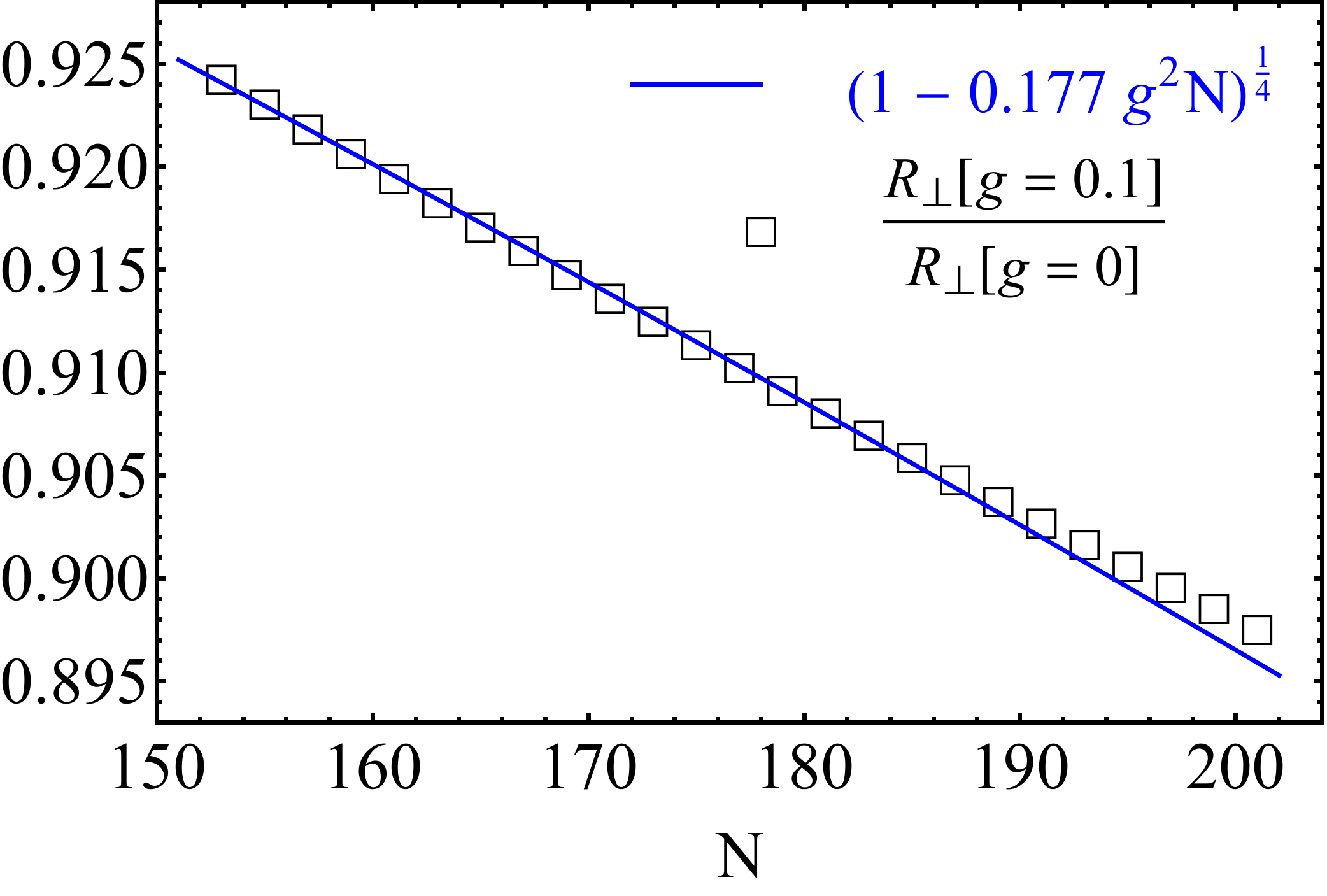}
\endminipage
  \caption{   $R_\perp[g]$ for $g = 0.1, 0.2, 0.3$~$(\bold{Left})$.   \,\,\,\,  $R_\perp[g=0.1]/R_\perp[g=0] $  for large $N$~$(\bold{Right})$.    }\label{largeNfit}
\end{figure}

\begin{figure}[!htb]
\includegraphics[width=71mm]{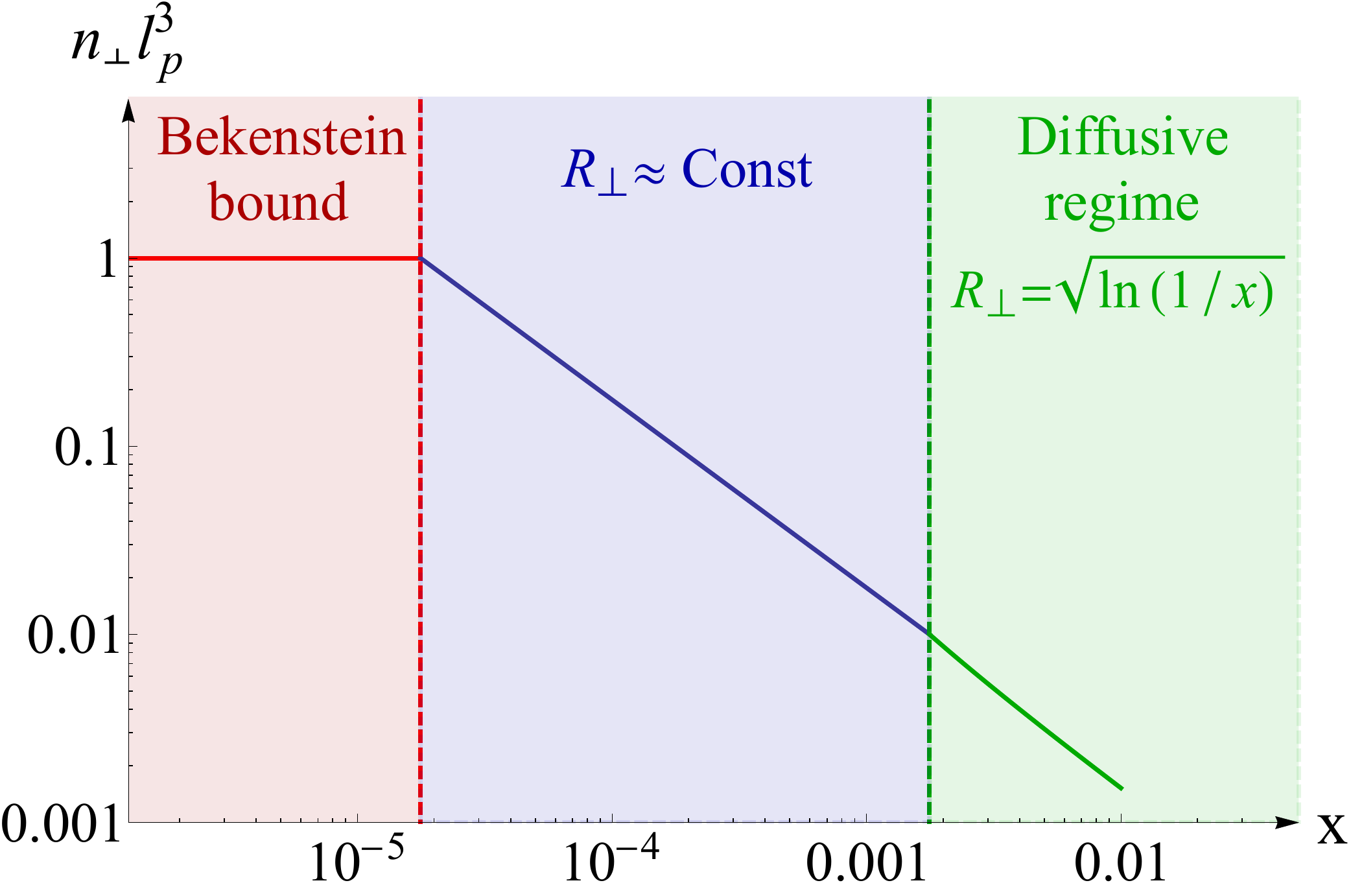}
  \caption{Saturation (red), pre-saturation (blue) and diffusive (green) regimes for a transverse string with decreasing
  resolution in $D_\perp=3$. See text.}
  \label{PhasePlot}
\end{figure}

For a relatively small attraction $g^2 N\approx 1$, $R_\perp$  in (\ref{EE1}) undergoes a numerical change from
an increasingly large and diffusive string to a small and fixed size string of about few string lengths. 
Fig.~\ref{largeNfit}~(left) shows that for $g=0.1$ the transverse size flattens out at about $R_\perp \sim 1.85\equiv 3.7l_s\approx 0.3$ fm in the range $200<N<600$.  For fixed $b$ the transverse area is ellipsoidal with a transverse bit density

\be
{\bf n}_\perp(b, N)\approx \frac N{b \left( 2 R_\perp \right)^{D_\perp-1}}\equiv \frac{N}{N_c} \frac{1}{g^2 l_s^{D_\perp}} \frac{g^2 N_c}{\frac{b}{l_s} \left( 2 \frac{R_\perp}{l_s} \right)^{D_\perp-1}} 
\label{nT}
\ee
The critical resolution $x_c=1/N_c$  at which this change  takes place can be  read from Fig.~\ref{largeNfit}~(right)

\be
x_c = \frac{1}{N_c} \approx  0.1^2 \times 0.177 = 1.77 \times 10^{-3}
\ee
for $g=0.1$ with $D_\perp=3$ and $b=5\equiv 10\, l_s\approx 1$ fm.  We identify the onset $x\approx x_c$ as the pre-saturation phase of the
string at high resolution whereby its transverse area contracts to the string scale under weak self-attraction.  However, the transverse string
bit density is still dilute at this resolution since

\be
{\bf n}_{\perp}(5,x)\approx  \frac{x_c}{x} \frac{1}{\frac{\pi}{3} l_p^{3}} \frac{1}{10 \times \left( 4  \times 1.85 \right)^{2} \times 0.177}  =
 \frac{x_c}{x} \frac{0.01}{  l_p^{3}} \equiv   \frac{x_s}{x} \frac{1}{  l_p^{3}}
\label{nT2}
\ee
or $n_\perp(5,x_c)\approx 0.01/l_P^3$.  
Recall that the Planck length $l_P^{D_\perp}=g_s^2l_s^{D_\perp}$ and that $g^2=(\pi/3)g_s^2$  from (\ref{GPLANCK}).  At
the saturation point or 
$x\equiv x_s=0.01x_c\approx 10^{-5}$ the transverse density saturates the Bekenstein bound of one bit per
transverse Planck area or $n_\perp(5,x_s)\approx 1/l_P^3$. We identify this point with the saturation scale or black hole regime.
A schematic  rendering of the pre-saturation and saturation phases in the low-x regime for $b=5$ are shown in~Fig.~\ref{PhasePlot}.

\subsection{Curved Space: $D_\perp(\lambda)<3$}

\begin{figure}[!htb]
\minipage{0.48\textwidth}
\includegraphics[width=61mm]{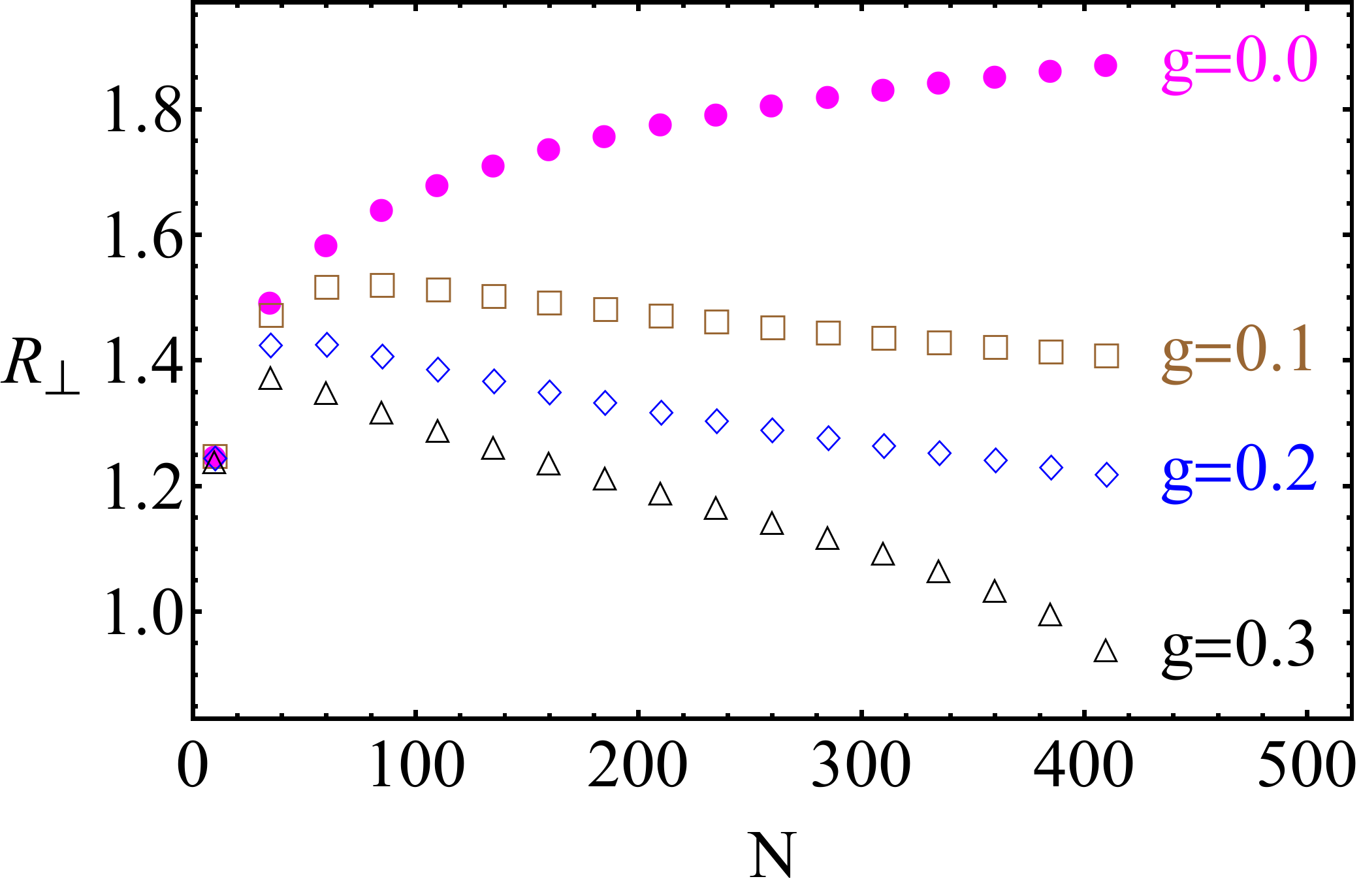}
\endminipage\hfill
\minipage{0.48\textwidth}
\includegraphics[width=61mm]{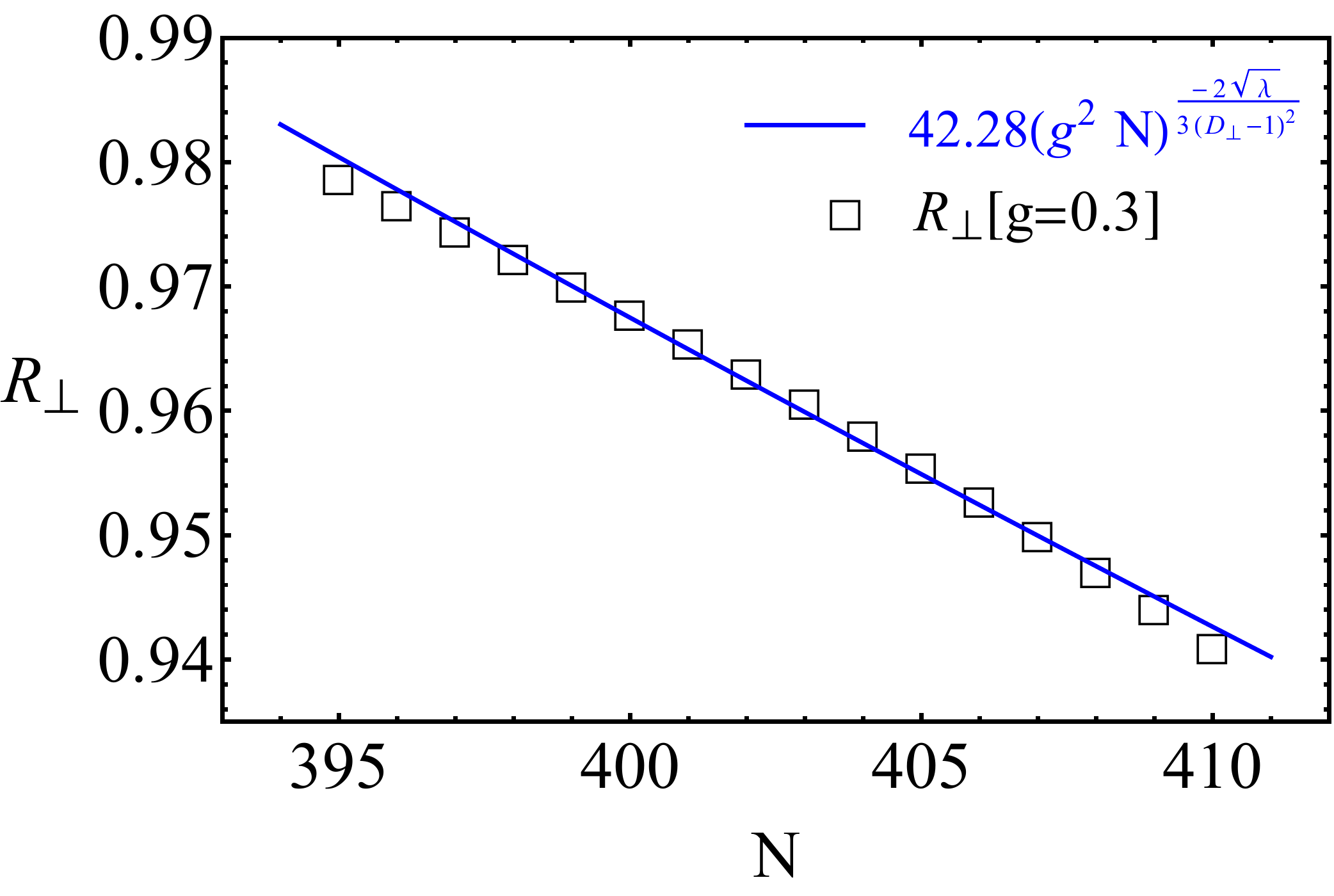}
\endminipage
  \caption{  $R_\perp[g]$ for $g = 0.1, 0.2, 0.3$~$(\bold{Left})$.   \,\,\,\,\,\,\,\,\,\,  $R_\perp[g=0.3]$ and Eq.~\ref{RNEW} $(\bold{Right})$.    }\label{largeNfitlambda}
\end{figure}

\begin{figure}[!htb]
\includegraphics[width=49mm]{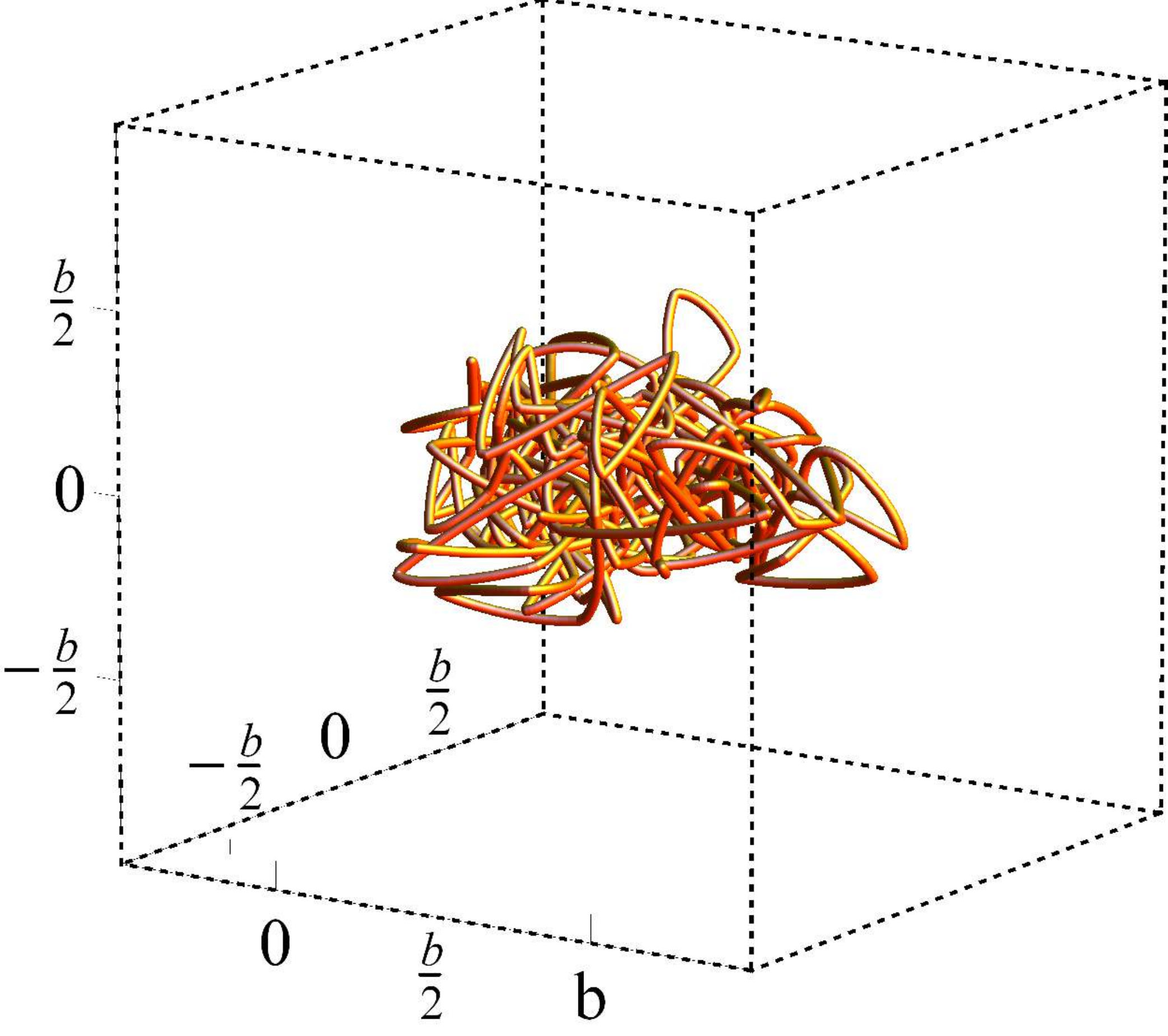}
  \caption{3D configuration of the string with $N=200$ and $g=0.3$ using $D_\perp(\lambda)$ in the interaction. See text.}
  \label{3DStringN200g=03Lambda40}
\end{figure}

An exact  spatial analysis of the transverse string in curved AdS$_5$ space is beyond the scope of this work. 
In this section we will attempt to give simple estimates of the effects of the curvature of AdS$_5$ on some of 
our previous results.  For that we first note that an aspect of the curved geometry on the Pomeron
is to cause the string transverse degrees of freedom to effectively feel a reduced transverse spatial dimension 
~\cite{Stoffers:2012zw,Stoffers:2012ai,Stoffers:2013tla,Shuryak:2013sra}

\be
D_\perp\rightarrow D_\perp(\lambda)=D_\perp\left(1-\frac{3(D_\perp-1)^2}{2D_\perp\sqrt{\lambda}}+{\cal O}\left(\frac 1\lambda\right)
\right)
\label{DEFFXL}
\ee
with $\lambda=g^2_{YM}N_c$.  Indeed, (\ref{DEFFXL}) causes the Pomeron intercept to move from $D_\perp/12=0.25$
to $D_\perp(\lambda \approx 40)\approx 0.17$ closer to the empirical  interceptt of $0.08$~\cite{Donnachie:1992ny}.  
A phenomenological way to implement this effect is to
add warping factors on the oscillators in (\ref{XS1}) as we detail in the Appendix and repeat the numerical analysis.
A simpler estimate follows from the 
substitution (\ref{DEFFXL}) in the interacting part of our variational analysis. Indeed,
the schematic estimate (\ref{EE}) shows that the first contribution reflects on the uncertainty principle
which probes short distances and thus is
not sensitive to the curvature of AdS$_5$. The second diffusive contribution is sensitive through $D_\perp$
but will turn out to be
subleading as we will show below. The third contribution is long ranged and senses the curvature of AdS$_5$.
Thus

\be
E\rightarrow  N^2\frac{1}{R_\perp^2}+N^2 \frac{R_\perp^2}{D_\perp^2(\lambda)\,{{\rm ln}^2N}}-N^2\frac{g^2N}{R_\perp^{D_\perp(\lambda)-1}}
\label{EEXL}
\ee
For very small values of $g$ the first two contributions in (\ref{EEXL}) are dominant and the string
transverse size grows diffusively. The minimization of the first two dominant contributions in this regime 
yields $R^2_\perp\approx D_\perp(\lambda)\,{\rm ln}N$.
This is consistent with the growth of the Pomeron in curved AdS$_5$ 
noted in~\cite{Stoffers:2012zw,Stoffers:2012ai,Stoffers:2013tla,Shuryak:2013sra}.
However, for

\be
g^2 > \frac{1}{N} \left( \ln N \right)^{\frac{D_\perp(\lambda) - 3}{2}}
\label{CONDX}
\ee
the string size shrinks and the transverse string size follows from balancing the first term 
with the last term due to the interaction. The balance
between the self-interaction and the uncertainty principle, yields a  continuously
decreasing transverse string size

\be
R_\perp\approx \left(\frac 1{g^2N}\right)^{\frac{2\sqrt{\lambda}}{3(D_\perp-1)^2}}
\label{RNEW}
\ee
in units of the string length.  A typical configuration of the string with $N=200$ 
using the string interaction (\ref{V2}-\ref{VV2}) with the effective substitution $D_\perp\rightarrow D_\perp(\lambda)$ is
displayed in~Fig.~\ref{3DStringN200g=03Lambda40}. For $\lambda=40$, $D_\perp = 3$ and $g = 0.3$, 
the scaling regime (\ref{RNEW}) is observed to take place for our string samplings for
$N_c\approx 400$ as shown in Fig.~\ref{largeNfitlambda}. As before, we identify the critical resolution 
$x_c=1/N_c\approx 0.0025$ with the onset of the scaling regime (\ref{RNEW}). 

\begin{figure}[!htb]
\includegraphics[width=71mm]{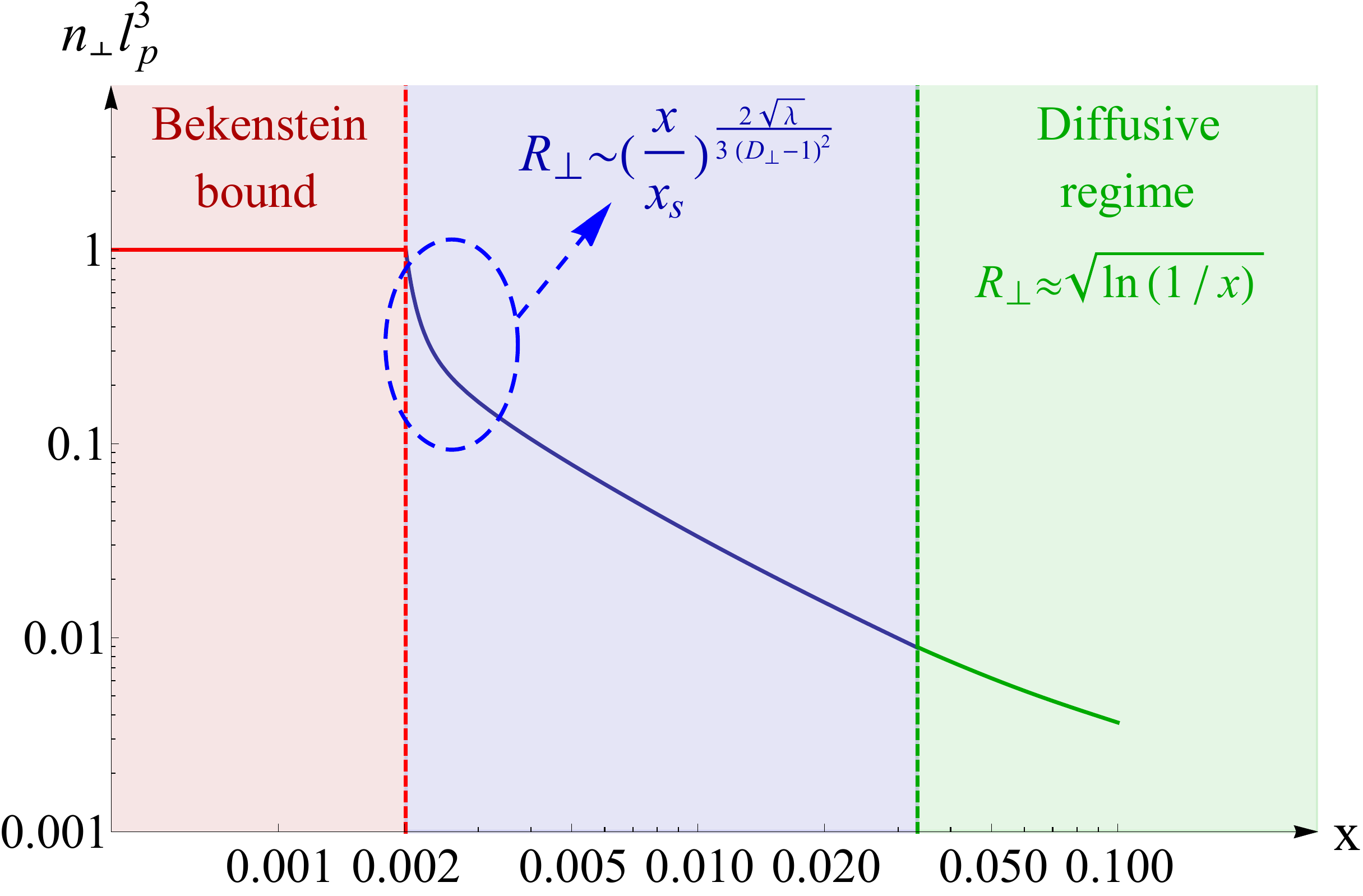}
  \caption{Saturation (red), pre-saturation (blue) and diffusive (green) regimes for a transverse string with decreasing
  resolution in $D_\perp(\lambda)$. See text.}
  \label{PhasePlotLambda40}
\end{figure}

The transverse density for fixed impact parameter $b$ is now

\be
{\bf n}_\perp(b,x=1/N)\approx \frac N{b \left(2 R_\perp\right)^{{D_\perp} - 1}}
\approx  \left( {\frac{(0.3)^2\times 400}{x/x_c}}\right)^{\frac{2\sqrt{\lambda} } {3(D_\perp-1)}+ 1}
  \left(\frac{l_s}{b} \right)  \left( \frac{1}{2 \times 42.28} \right)^{D_\perp -1}   \,
\frac 1{l_p^{{D_\perp}}}
\label{nTX}
\ee
For a typical impact parameter of $b=5l_s$, it saturates the Bekenstein bound
for $x\equiv x_s\approx 0.6 x_c$ ($1/x_s=N_s\approx 633$)

\be
{\bf n}_\perp(5,x_s\approx 0.0016  )\approx \frac 1{l_p^{{D_\perp}}}
\ee
In Fig.~\ref{PhasePlotLambda40} we give a schematic rendering of the diffusive (green, pre-saturation (blue) and saturation
(red) regimes foliowing from the effective $D_\perp\rightarrow D_\perp(\lambda)$ substitution.

\subsection{Stringy Saturation}

In flat $D_\perp=3$ the transverse string size distribution remains diffusive or logarithmic in $N$ for small self-attractive interactions in the range
$0<g^2N<1$. However for $g^2N\approx 1-1/ {\rm ln}^2 N $ the transverse string size shrinks to a fixed size comparable to the string length. The change sets in at weak coupling with $g^2\approx 1/N$, for which the transverse density  at $b=0$ is now

\be
{\bf n}_\perp(0,x=1/N)\approx \frac N{R_\perp^{D_\perp}}
\rightarrow {(g^2N)} \frac  1{l_p^{D_\perp}}
\label{nT}
\ee
after restoring the string length. The first transition occurs in a very narrow range of $g$ and thus appears to be first order by our analysis  in
(\ref{EE}-\ref{EE1}). It is a pre-saturation transition where the string size shrinks away from its diffusive growth and remains about fixed at a relatively dilute transverse string bit density. At much higher resolution or low-x a saturation transition takes place when the transverse string bit distribution reaches the Bekenstein bound of one string bit per Planck scale.
This maybe intuitively understood by noting that low-x follows from large boosts a situation analogous to falling matter on a black-hole.
For completeness, we note that self-repulsive strings increase in sizes following the substitution $g^2\rightarrow -g^2$ in (\ref{nT}). 

Using the estimates for the AdS curvature through the substitution (\ref{DEFFXL}) yields 

\be
{\bf n}_\perp(0,x=1/N)\approx \frac N{R_\perp^{{D_\perp}}} \longrightarrow \left(g^2 N  \right)^{\frac{2 \sqrt{\lambda} D_\perp}{3 (D_\perp - 1)^2}} \frac  1{l_p^{D_\perp}}
\label{nTT}
\ee
instead of (\ref{nT}). 
(\ref{nTT}) reaches more smoothly the Bekenstein bound as the string self-interaction satisfies $g^2N\approx 1$.
Alternatively, the effective density using the effective dimension $D_\perp(\lambda)$

\be
\tilde{\bf n}_\perp(0,x=1/N) \approx \frac N{R_\perp^{{D_\perp}(\lambda)}}
\rightarrow \left( {g^2N}\right)^{\frac{2\sqrt{\lambda}D_\perp(\lambda)}{3(D_\perp-1)^2}+\frac{D_\perp(\lambda)}{D_\perp}}
\,\left(\frac 1x\right)^{\frac{3(D_\perp-1)^2}{2D_\perp\sqrt{\lambda}}}    \,
\frac 1{l_p^{{D_\perp}(\lambda)}}
\label{nTX}
\ee
 is seen to increase beyond the Bekenstein bound as the string self-interaction reaches $g^2N\approx 1$. There is no black-hole
 to saturate in fractional  dimension.

\subsection{Relation to Saturation in DIS}

The present observations on stringy saturation are consistent with the arguments presented in~\cite{Stoffers:2012zw,Stoffers:2012ai,Stoffers:2013tla} whereby the stringy but eikonalized dipole-dipole cross section  was found to saturate in the impact parameter space when $g_s^2l_s^{D_\perp}n_\perp\equiv l_P^{D_\perp}n_\perp\approx 1$ (see their Eq. 47). 
 Although the relationship between the string coupling and the gauge coupling depends on the holographic extension of QCD used, for the generic model of AdS$_5$ with a wall $g_s\approx C\,g^2_{YM}/4\pi\equiv C \,\alpha_s$ ($C=1$ for AdS$_5$
without a wall). Our numerical analysis puts $g_s\approx  0.1-0.3$.

The 3-dimensional 
density $n_\perp$ was physically interpreted in~\cite{Stoffers:2012zw,Stoffers:2012ai,Stoffers:2013tla}
as the number of wee dipoles per unit transverse 2-dimensional space per unit dipole size $z$ along 
the holographic direction. The latter enforces  hyperbolic evolution of the dipole size through the AdS$_5$ metric
(with a wall).  At saturation $z_s\approx 1/Q_s$. The transverse 2-dimensional density is then defined as
$Q_s^2\equiv z_sn_\perp$. 

For curved AdS$_5$, the Pomeron intercept is $D_\perp(\lambda\approx 40)/12\approx =0.17$,  
and (\ref{nTT}) at saturation gives  $l_sQ_s\approx l_s/l_p\approx 1/x^{\frac 13}$.
This is to be compared with 
$l_sQ^{GW}_s\approx l_s/l_p\approx 1/x^{0.144}$ obtained empirically by 
Golec-Wustoff~\cite{GolecBiernat:1998js,GolecBiernat:1999qd}, and  $l_sQ^{SZ}_s\approx l_s/l_p\approx 1/x^{0.114}$ obtained by Stoffers and one of us~\cite{Stoffers:2012zw,Stoffers:2012ai,Stoffers:2013tla}. For curved $D_\perp(\lambda)$,  (\ref{nTX}) yields at saturation

\be
l_pQ_s(\lambda)\approx \left(\frac 1x\right)^{\frac{3 (D_\perp -1)^2}{2D_\perp\sqrt{\lambda}D_\perp(\lambda)}}\rightarrow \left(\frac 1x\right)^{0.155}
\label{QSADS}
\ee
using $D_\perp=3$ and $\lambda\approx 40$~\cite{Stoffers:2012zw,Stoffers:2012ai,Stoffers:2013tla}. 
(\ref{QSADS}) is overall consistent with the full AdS$_5$ curved analysis carried 
in~\cite{Stoffers:2012zw,Stoffers:2012ai,Stoffers:2013tla}, and remarkably close to the empirical result
~\cite{GolecBiernat:1998js,GolecBiernat:1999qd}.

The saturation of the Bekenstein bound  maybe viewed as the string dual to the gluon saturation description in the color glass condensate model for fixed impact parameter using the Pomeron or string slope as a scale~\cite{Rezaeian:2012ji,Kowalski:2003hm}.
The large string bit density (\ref{nT}) may upset the integrity of the string. Perhaps a more
appropriate description is in terms of a fluid of string bits. However, three generic stringy ingredients need to be retained:
1) the string provides  for a key property of the wee partons namely their transverse (Gribov) diffusion 
with a diffusion constant ${\bf D}=l_s^2/2$ set by the string length;
2) the exponential rise in the string density of states with its mass, provides for the  most efficient 
mechanism  to scramble information and reach the Bekenstein bound and thus  saturation;
3) the self-interacting string in the mean-field approximation maybe the dual of a Pomeron branching into 
multiple Pomerons or fan-diagrams in Reggeon calculus~\cite{9780511534959}.

\section{\label{epsilon} Angular Deformations}

 The fluctuating string with fixed end-points exhibit azimuthal deformations in the transverse plane that can be characterized by the azimuthal moment~\cite{Bzdak:2013rya, Kalaydzhyan:2014tfa}

\be
\epsilon_n = \frac{ \frac{1}{N}\sum_{i}^N  e^{i n \phi_i}  \left( r^\perp_i \right)^n  }{  r_\perp^n }
\ee
where
\be
r_\perp^n = \frac{1}{N}\sum_{i}^N     \left( r^\perp_i \right)^n
\ee
with $\phi$ the azimuthal angle as measured from the impact parameter line along ${\bf b}$. $r_\perp$ is the averaged size of the string on the transverse plane. For $b=0$, we have $ \left< r_\perp^2  \right> /2  = R_\perp^2 / D_\perp$, where $\left< \cdots \right>$ is the average over string ensembles. Specifically,  define $x  \equiv x_\perp^{i=1}$ 
and $y  \equiv x_\perp^{i=2}$ in the transverse plane, where $x $ is parallel to the impact parameter $\bold{b}$ and $y $ perpendicular to it,
\bea
x_\perp (k , \tau) &=&     \sum_{n=1}^{N-1}  X_{n}  (\tau) \sin \left( \frac{n k}{N } \pi\right)+ b   \frac{k}{N }   \nonumber\\
y_\perp (k , \tau) &=&      \sum_{n=1}^{N-1}  Y_{n}  (\tau) \sin \left( \frac{n k}{N } \pi\right)    
\eea
Both $X_n, Y_n$ are normally distributed with width $1/2\omega_n$~(\ref{VAR1}) or

\be
X_n \sim   \mathcal{N} \left( 0 , \frac{1}{2 \omega_n} \right) \ \ \ \ \ \ \ \ \ \ \ \  Y_n \sim \mathcal{N} \left( 0 , \frac{1}{2 \omega_n} \right)
\ee
satisfy the normal distributions. We obtain
\bea
x_\perp (k , \tau)  \sim  \mathcal{N} \left( b   \frac{k}{N } ,  \Sigma_k^2 \right) \ \ \ \ \ \ \ \ \ \ \ \  y_\perp (k , \tau)  \sim   \mathcal{N} \left(0 , \Sigma_k^2  \right)
\eea
where
\be
\Sigma_k^2  = \sum_{n=1}^{N-1} \frac{\sin^2 \left( \frac{n k}{N } \pi\right) }{2 \omega_n}
\ee

\begin{figure}[!htb]
\minipage{0.48\textwidth}
\includegraphics[width=60mm]{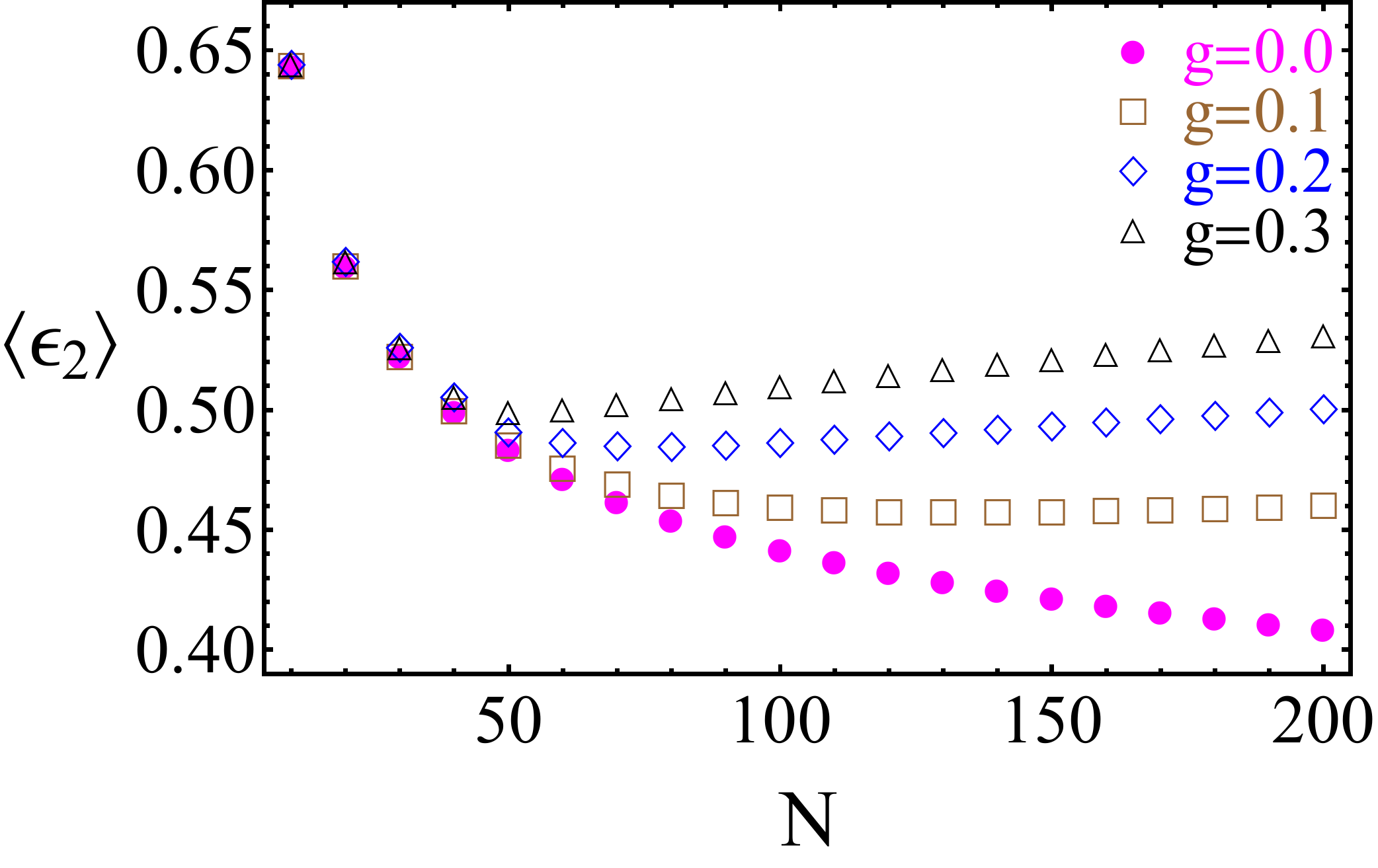}
\endminipage\hfill
\minipage{0.48\textwidth}
\includegraphics[width=60mm]{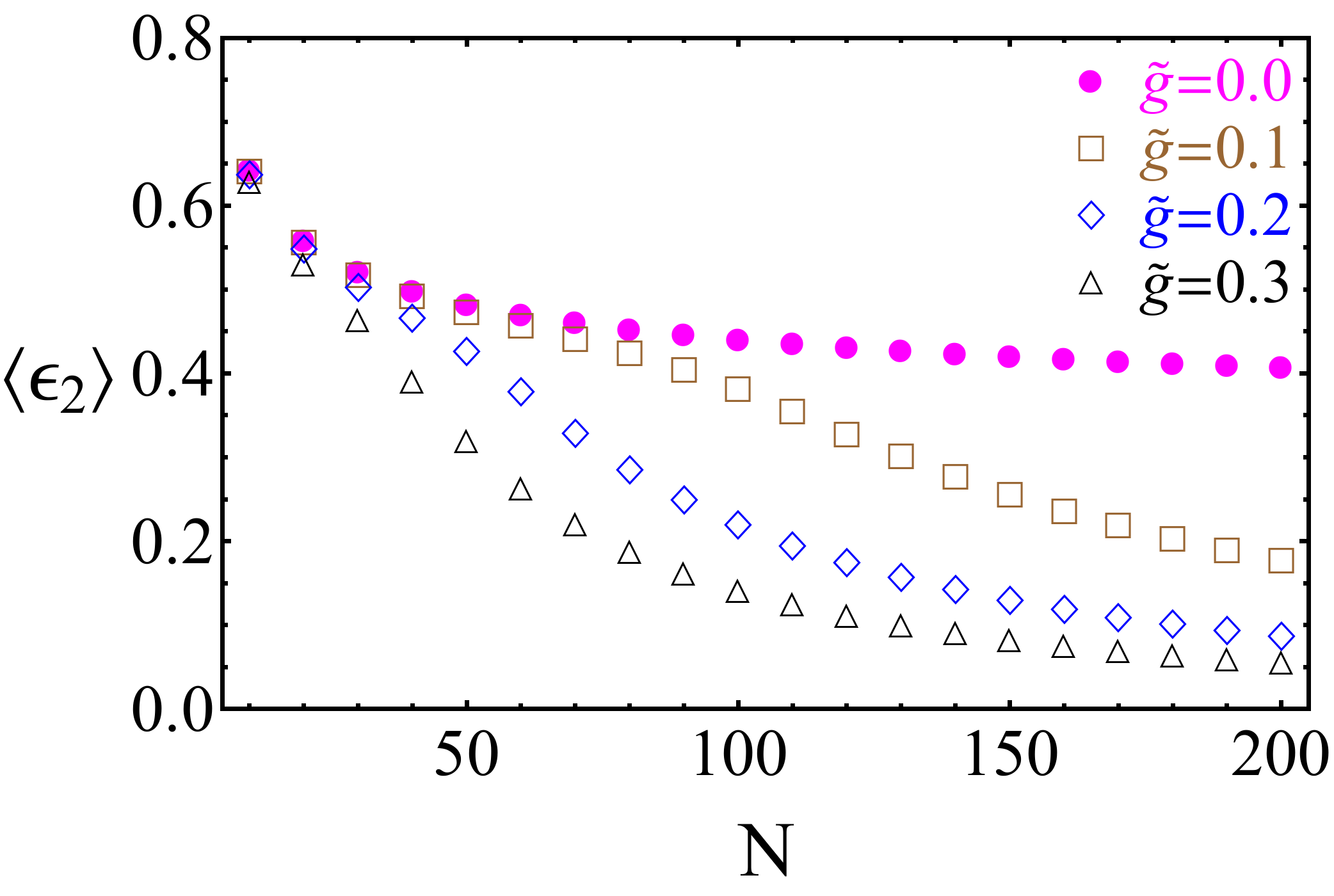}
\endminipage
 \caption{Attractive interaction: $g = 0.1, 0.2, 0.3$~($\bold{Left}$). \,\,   Repulsive interaction: $\tilde{g} = 0.1, 0.2, 0.3$~($\bold{Right}$). } \label{E2Diffg}
\end{figure}

\begin{figure}[!htb]
\minipage{0.48\textwidth}
\includegraphics[width=60mm]{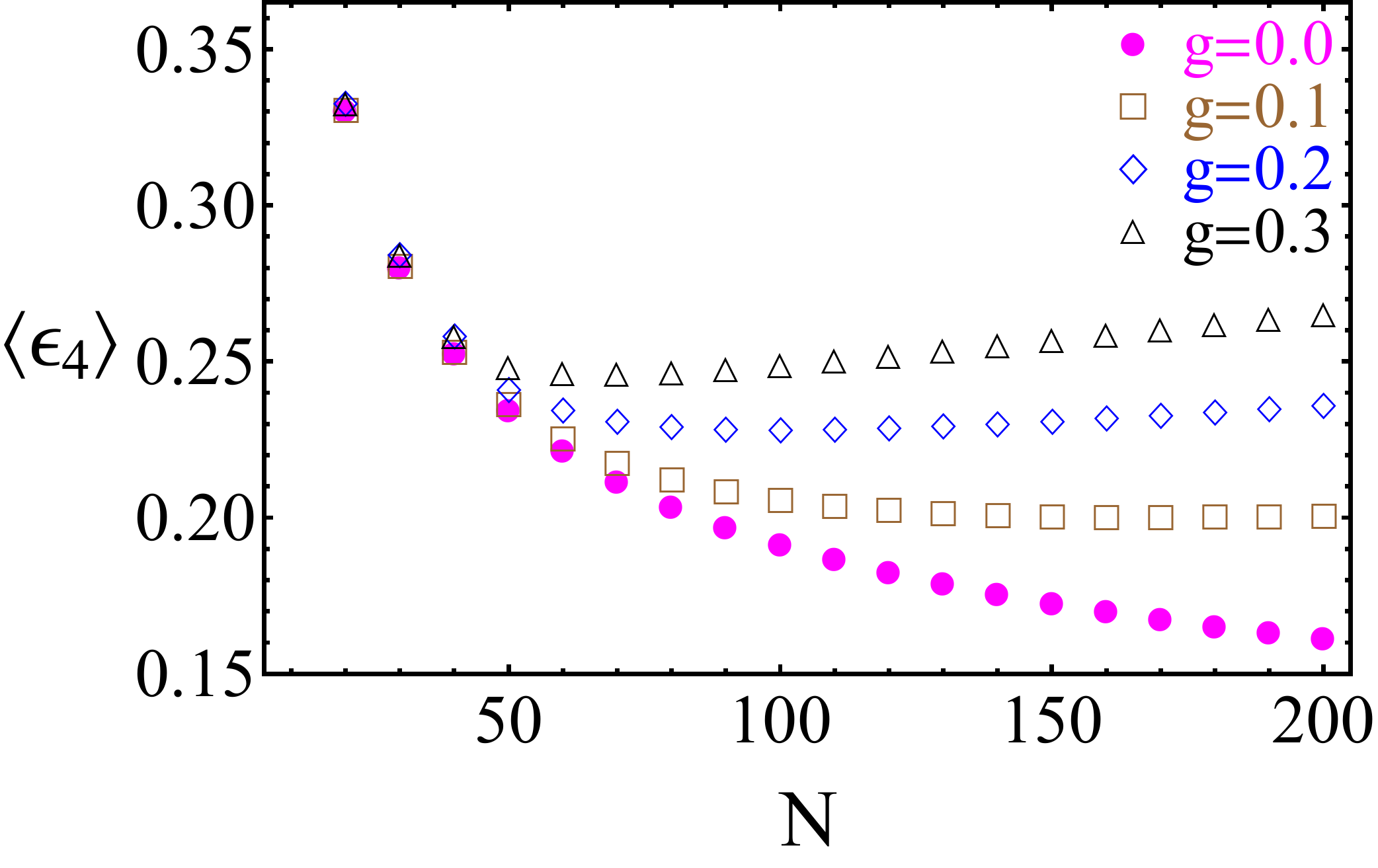}
\endminipage\hfill
\minipage{0.48\textwidth}
\includegraphics[width=60mm]{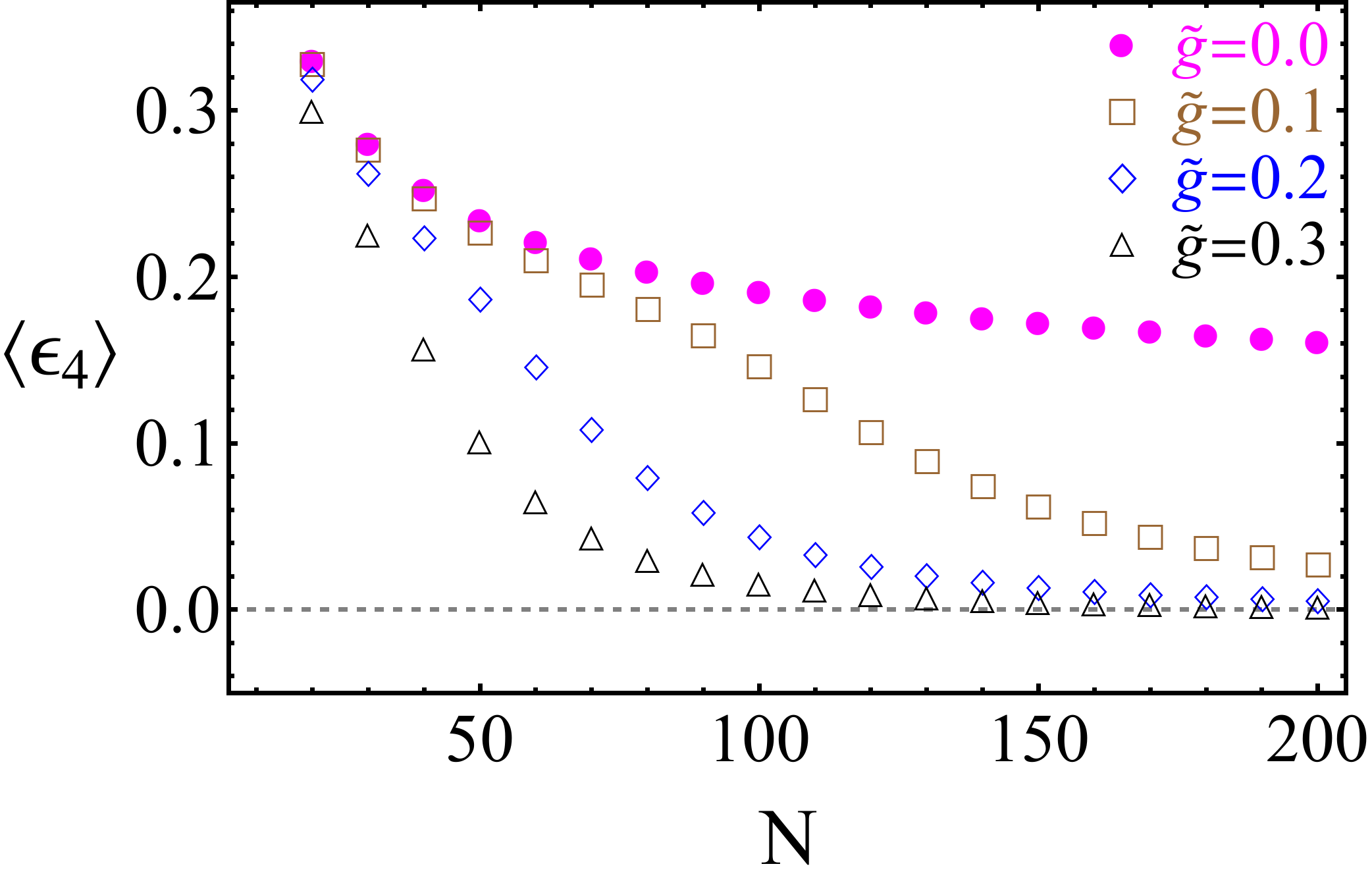}
\endminipage
 \caption{Attractive interaction: $g = 0.1, 0.2, 0.3$~($\bold{Left}$). \,\,   Repulsive interaction: $\tilde{g} = 0.1, 0.2, 0.3$~($\bold{Right}$). } \label{E4Diffg}
\end{figure}

For large $N$, each of the transverse coordinates $x_\perp (k,\tau)$ are almost independent. The azimuthal moments averaged over 
the independent transverse coordinates read

\bea
\left<\epsilon_n \right> &=&  \frac{1}{\left< r_T^n  \right>} \left[ \frac{1}{N+1}  \sum_{k=1}^{N-1} \int_0^\infty d r \int_0^{2 \pi}  d \phi ~  r^{n+1} \cos (n \phi) ~ \rho \left( r\cos \phi + \frac{b}{2} , r \sin \phi, k \right) + \frac{1 + (-1)^n}{N+1} \left( \frac{b}{2}  \right)^n    \right] \nonumber\\
&& +   \frac{i}{\left< r_T^n  \right>}  \frac{1}{N+1}  \sum_{k=1}^{N-1} \int_0^\infty d r \int_0^{2 \pi}  d \phi ~  r^{n+1} \sin (n \phi) ~ \rho \left( r\cos \phi + \frac{b}{2} , r \sin \phi, k \right) 
\eea
where
\be
\left< r_T^n  \right> = \frac{1}{N+1}  \sum_{k=1}^{N-1} \int_0^\infty d r \int_0^{2 \pi}  d \phi ~  r^{n+1} ~ \rho \left( r\cos \phi + \frac{b}{2} , r \sin \phi, k \right) + \frac{2}{N+1} \left( \frac{b}{2}  \right)^n
\ee
and
\be
\rho(x , y, k) = \frac{1}{2 \pi} \frac{1}{ \Sigma_k^2}  \exp \left[ -  \frac{\left( x - b \frac{k}{N}   \right)^2 + y^2}{2  \Sigma_k^2 } \right]
\ee
The Gaussian integrations can be done leading to

\be\la{epsilonanalytic}
\left<\epsilon_n\right> = \frac{b^n}{\left< r_T^n  \right> }  \left[  \frac{1}{N+1}  \sum_{k=1}^{N-1}   \left(  \frac{1}{2} -  \frac{k}{N} \right)^n  + \frac{1+ (-1)^n}{N+1} \left( \frac{1}{2}  \right)^n \right]
\ee
Note that the moments $\left<\epsilon_n\right>$ are real and that all  the odd moments vanish, i.e. $\left<\epsilon_n\right>=0$ for odd $n$. 
Simple algebra yields

\be
 \frac{\left< r_T^2  \right> } {b^2} = \frac{1}{N+1}  \sum_{k=1}^{N-1}   \left(  \frac{1}{2} -  \frac{k}{N} \right)^2  + \frac{2}{N+1}  \sum_{k=1}^{N-1}  \frac{ \Sigma_k^2 }{b^2}  + \frac{2}{N+1} \left( \frac{1}{2}  \right)^2 
\ee
and
\be
 \frac{\left< r_T^4  \right> } {b^4} = \frac{1}{N+1}  \sum_{k=1}^{N-1}   \left(  \frac{1}{2} -  \frac{k}{N} \right)^4  + \frac{8}{N+1}  \sum_{k=1}^{N-1}  \frac{ \Sigma_k^2 }{b^2}\left(  \frac{1}{2} -  \frac{k}{N} \right)^2 +  \frac{8}{N+1}  \sum_{k=1}^{N-1}  \frac{ \Sigma_k^4 }{b^4}   + \frac{2}{N+1} \left( \frac{1}{2}  \right)^4
\ee
In the limit $N \longrightarrow \infty$,  the moments simplify

\be
\Sigma_{\tilde{k}}^2 \approx \sum_{n=1}^N \frac{1}{4 n} = \frac{R_\perp^2}{D_\perp}
\ee
so that~(for even $n$)
\be
\left<\epsilon_n \right> \approx \frac{b^n}{\left< r_T^n  \right> }   \int_0^1 d \tilde{k}       \left(  \frac{1}{2} -  \tilde{k} \right)^n      
= \frac{b^n}{\left< r_T^n  \right> }  \frac{1}{ 2^n (1 + n)}
\ee
 
\be
 \frac{\left< r_T^2  \right> } {b^2} \approx \frac{1}{12}  +    \frac{2}{ D_\perp} \frac{ R_\perp^2 }{b^2}   
\ee

\be
 \frac{\left< r_T^4  \right> } {b^4} \approx \frac{1}{80}  + \frac{2}{3}   \frac{R_\perp^2 }{b^2 D_\perp }  +   8     \frac{ R_\perp^4 }{ D_\perp^2 b^4}  
\ee
For small $b$, we obtain 

\be
\left<\epsilon_2 \right>  \approx \frac{D_\perp}{24}  \frac{b^2}{    R_\perp^2  }
\ee

\be
\left<\epsilon_4\right>  \approx \frac{D_\perp^2}{640}  \frac{b^4}{    R_\perp^4  }
\ee
For general $b$, the numerical results of   $\left< \epsilon_2 \right>$ and   $\left< \epsilon_4 \right>$   are displayed in   Fig.~\ref{E2Diffg}  and Fig.~\ref{E4Diffg}  respectively.

To show the transverse cross correlations it is also useful to use the cross moments~\cite{Bzdak:2013rya, Kalaydzhyan:2014tfa}
\bea\la{epsilonnm}
\left(\epsilon_n \{  2 \} \right)^2 &=&  \left< \left|\epsilon_n\right|^2  \right> \nonumber\\
\left(\epsilon_n \{  4 \} \right)^4 &=&   -   \left< \left|\epsilon_n\right|^4  \right>    + 2 \left< \left|\epsilon_n\right|^2  \right>^2 \nonumber\\
\left(\epsilon_n \{  6 \} \right)^6 &=& \frac{1}{4} \left[  \left< \left|\epsilon_n\right|^6  \right>  -  9 \left< \left|\epsilon_n\right|^4 \right>  \left< \left| \epsilon_n\right|^2  \right> + 12 \left< \left|\epsilon_n\right|^2  \right>^3   \right] \nonumber\\
\left(\epsilon_n \{  8 \} \right)^8 &=&   \frac{1}{33} \left[ - \left< \left|\epsilon_n\right|^8  \right> + 16 \left< \left|\epsilon_n\right|^6  \right> \left< \left|\epsilon_n\right|^2  \right>  + 18  \left<  \left|\epsilon_n\right|^4  \right>^2 - 144 \left< \left|\epsilon_n\right|^4  \right> \left< \left|\epsilon_n\right|^2  \right>^2 + 144\left< \left|\epsilon_n\right|^2  \right>^4  \right]
\eea

\begin{figure}[!htb]
\minipage{0.33\textwidth}
\includegraphics[width=53mm]{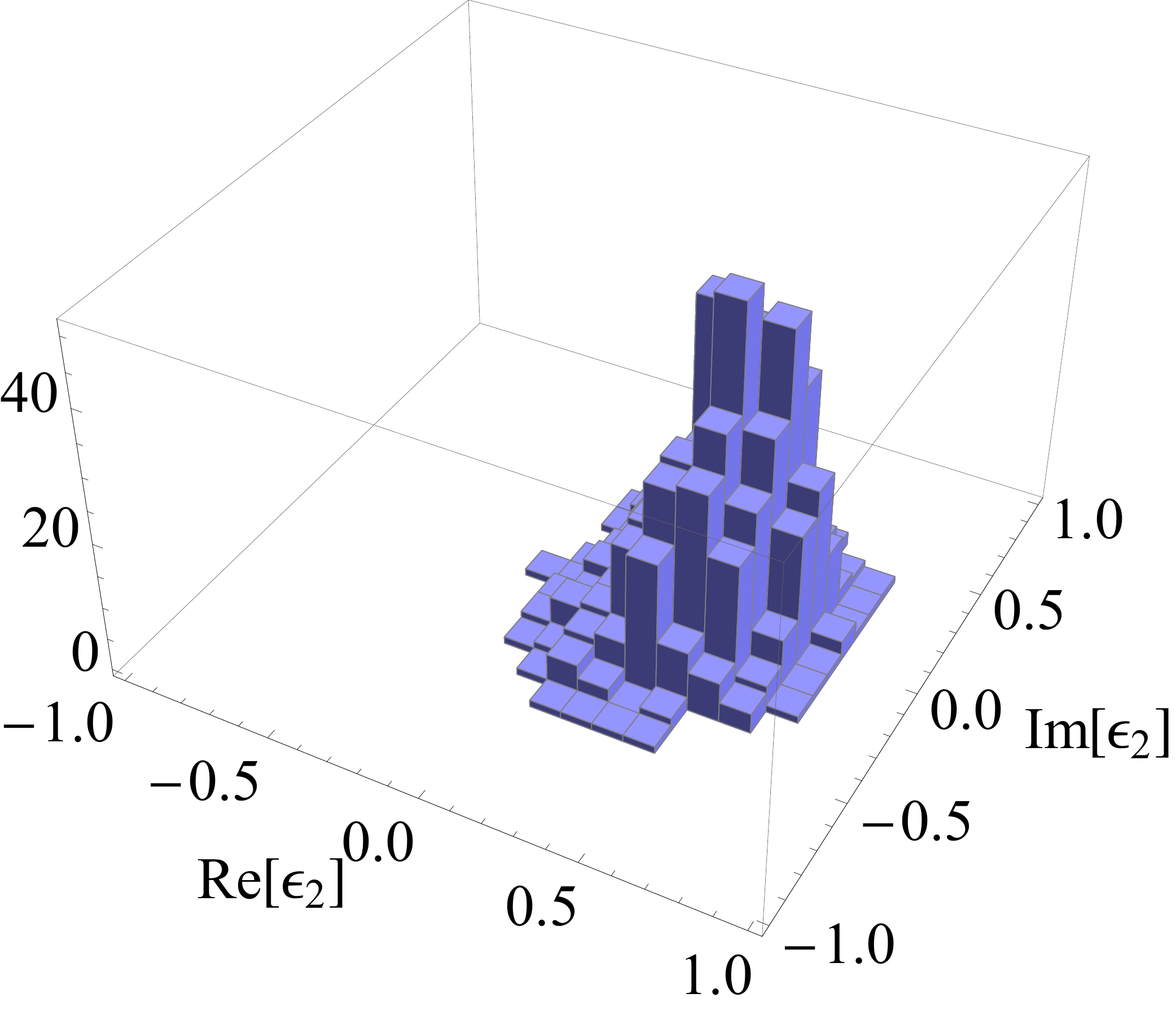}
\endminipage\hfill
\minipage{0.33\textwidth}
\includegraphics[width=53mm]{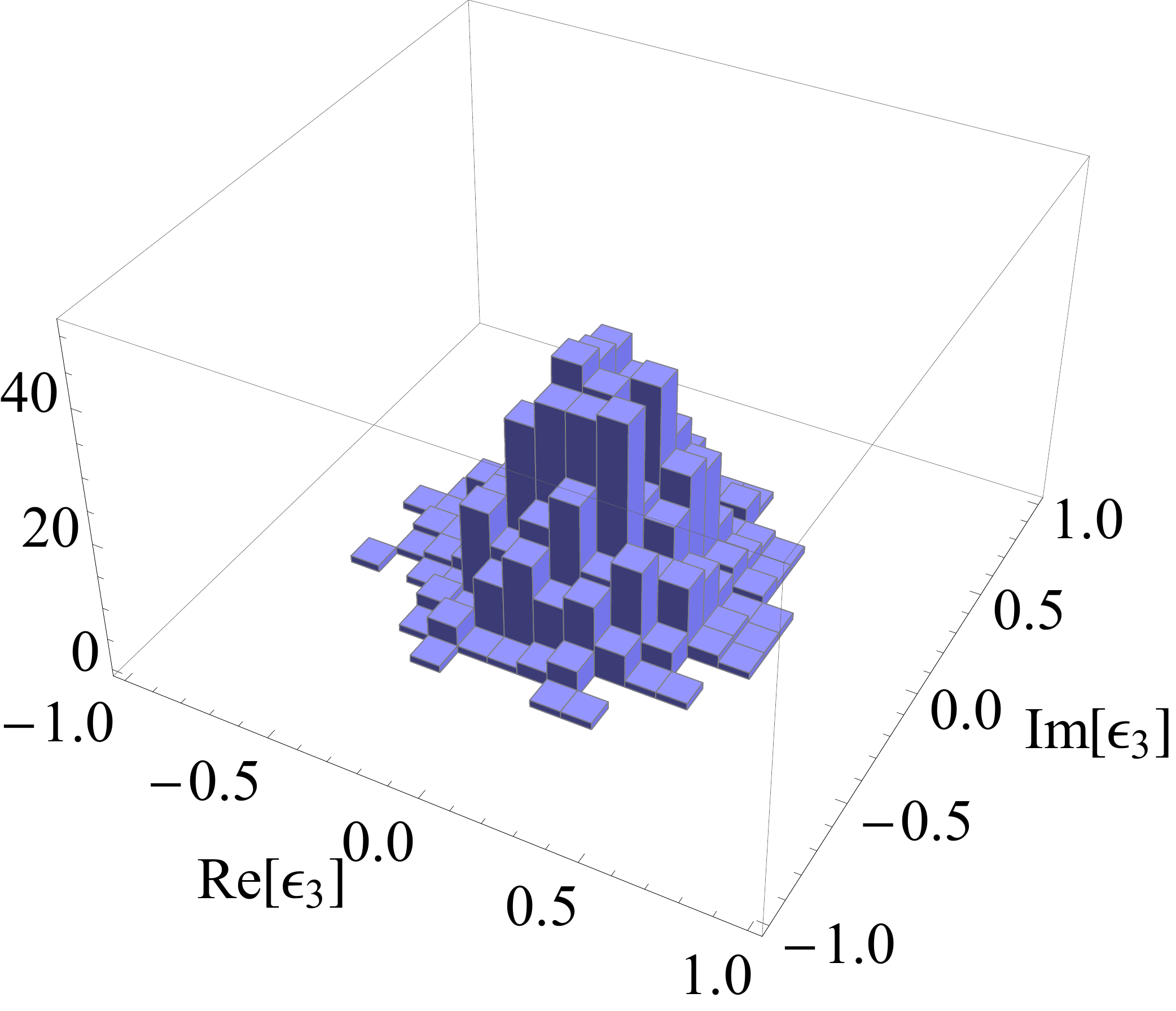}
\endminipage
\minipage{0.33\textwidth}
\includegraphics[width=53mm]{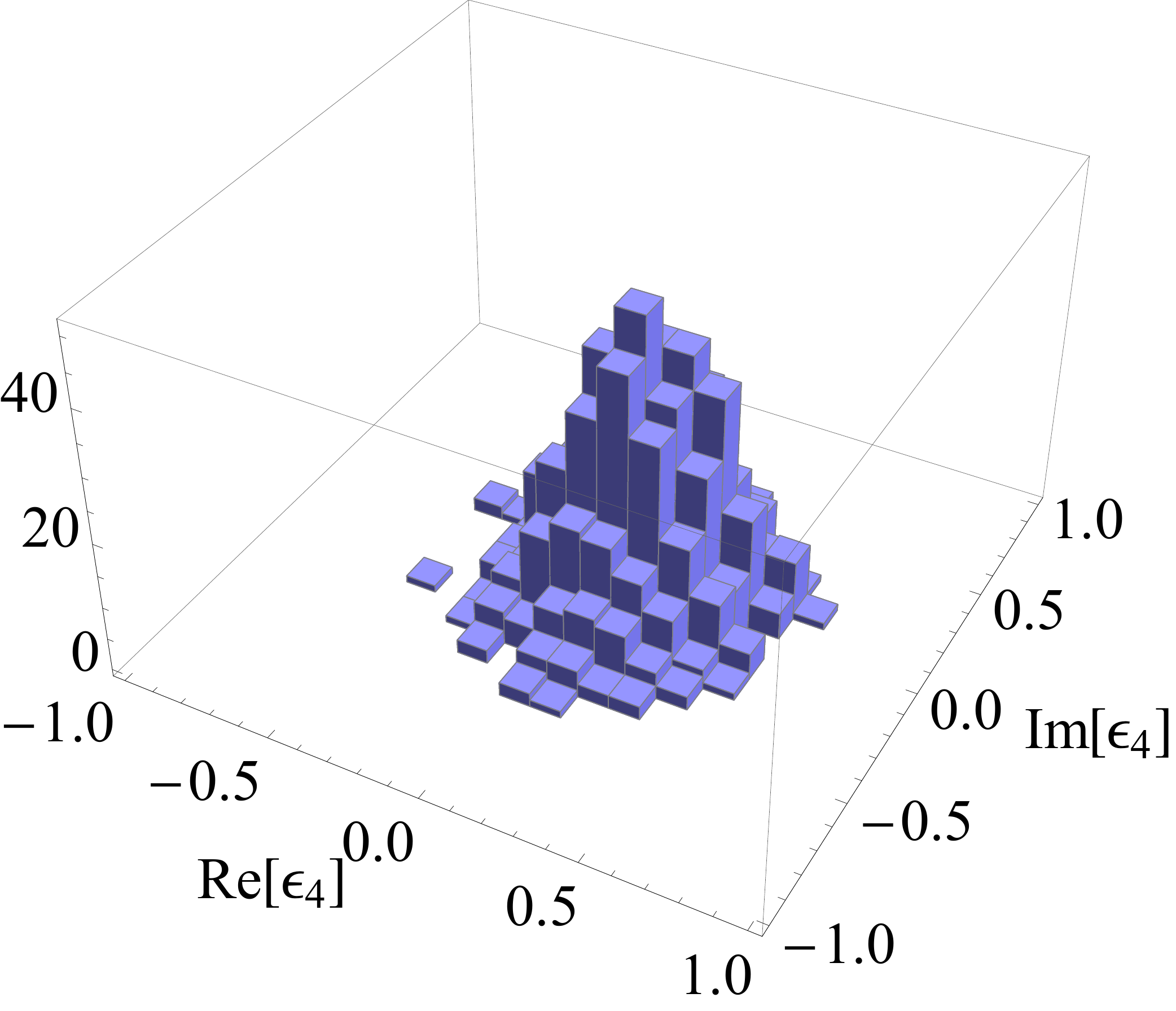}
\endminipage
  \caption{3D Histograms, 1000 random generated strings. N=200.  }\label{Histogram3D}
\end{figure}

\begin{figure}[!htb]
\minipage{0.33\textwidth}
\includegraphics[width=53mm]{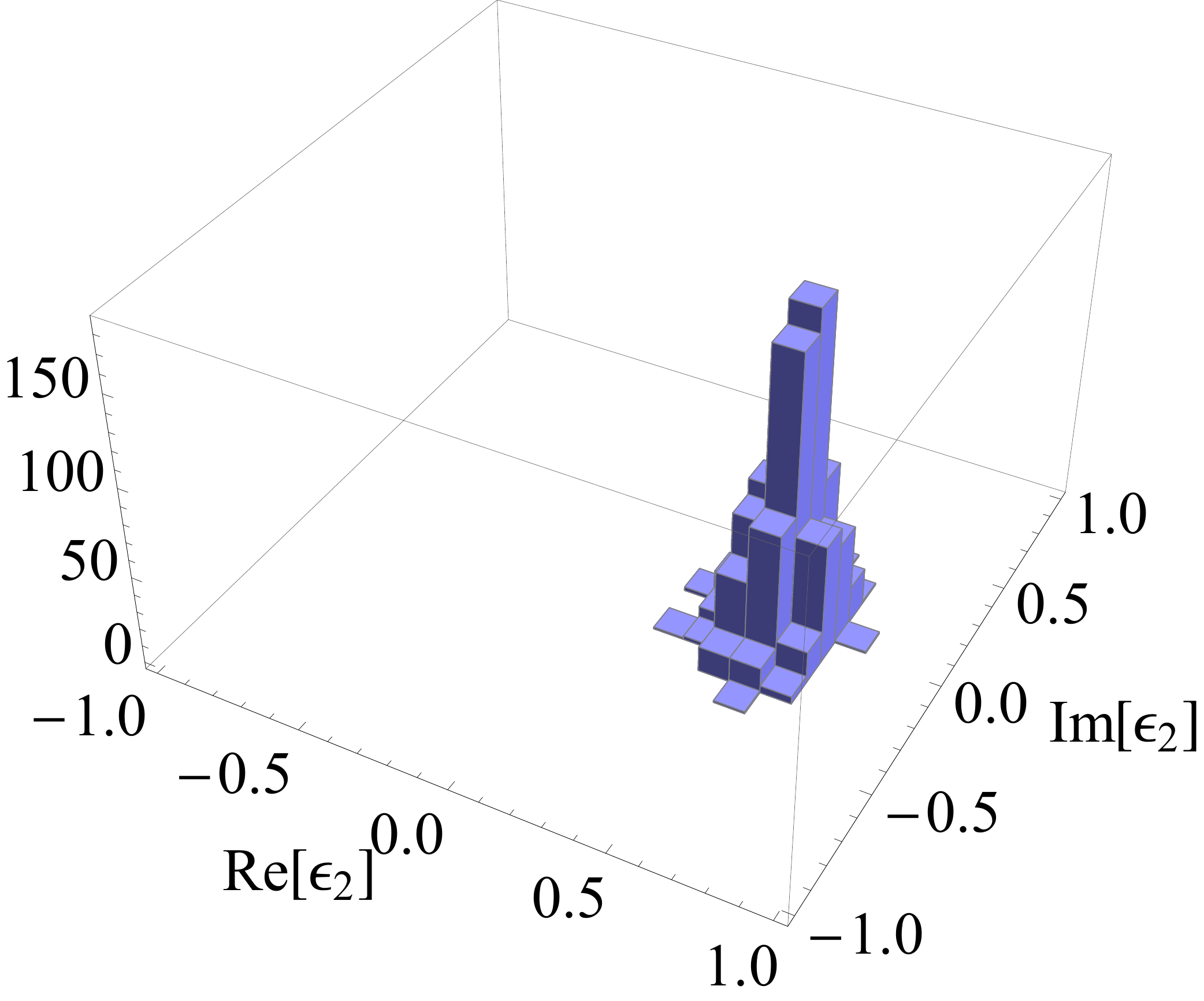}
\endminipage\hfill
\minipage{0.33\textwidth}
\includegraphics[width=53mm]{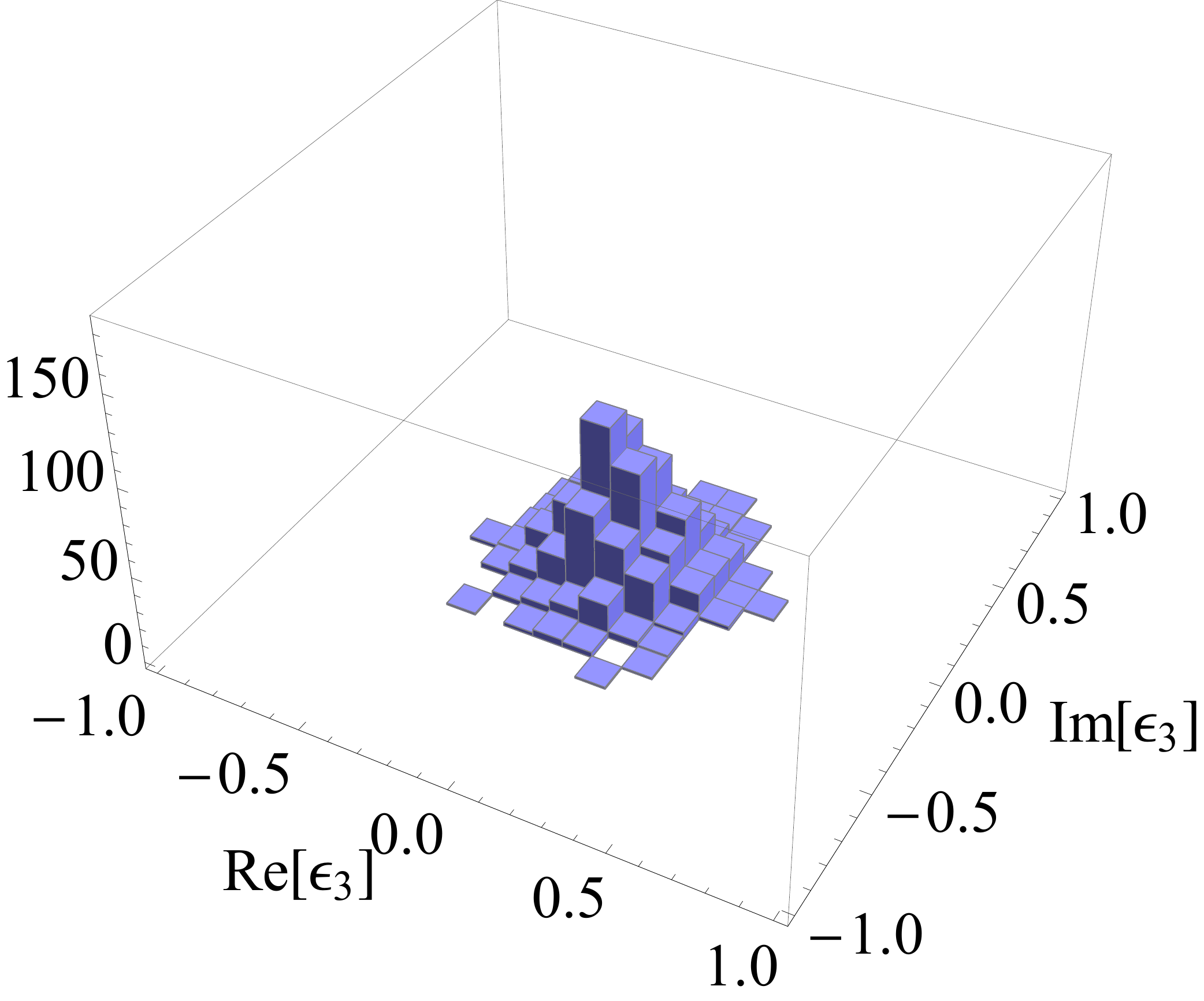}
\endminipage
\minipage{0.33\textwidth}
\includegraphics[width=53mm]{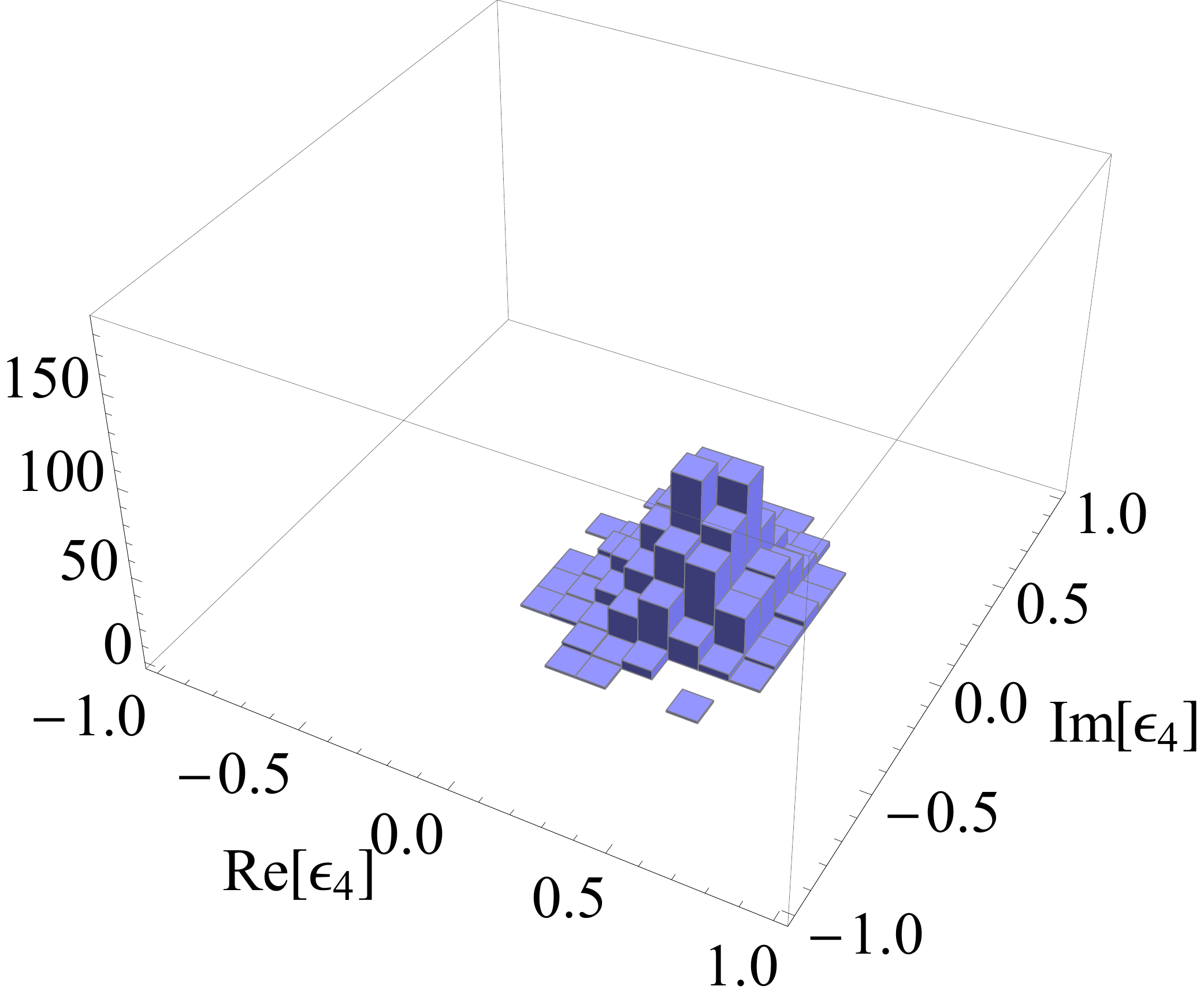}
\endminipage
  \caption{3D Histograms, 1000 random generated strings. N=200. Attractive interaction $g=0.3$.  } \label{Histogram3Dg}
\end{figure}

\begin{figure}[!htb]
\minipage{0.33\textwidth}
\includegraphics[width=53mm]{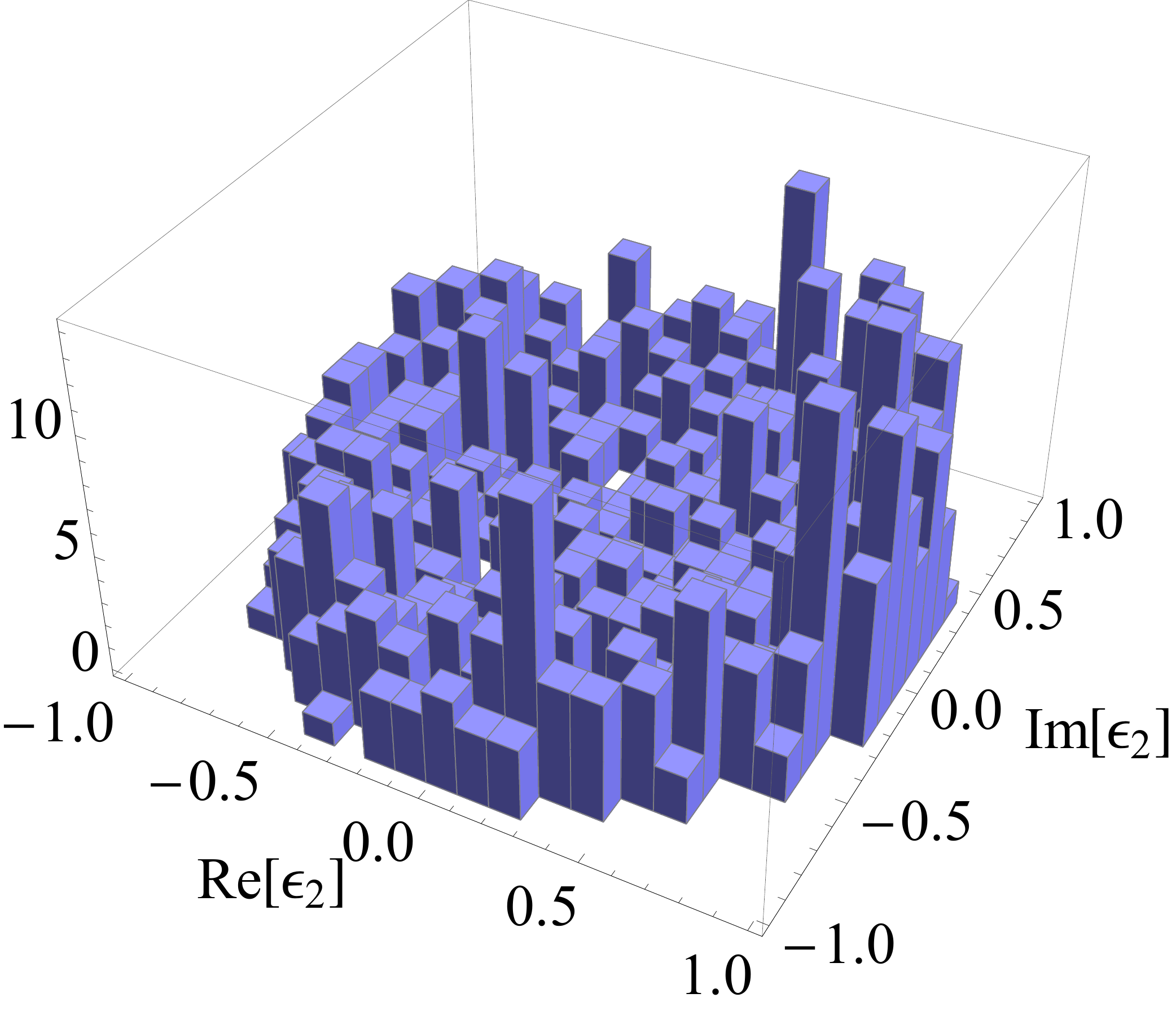}
\endminipage\hfill
\minipage{0.33\textwidth}
\includegraphics[width=53mm]{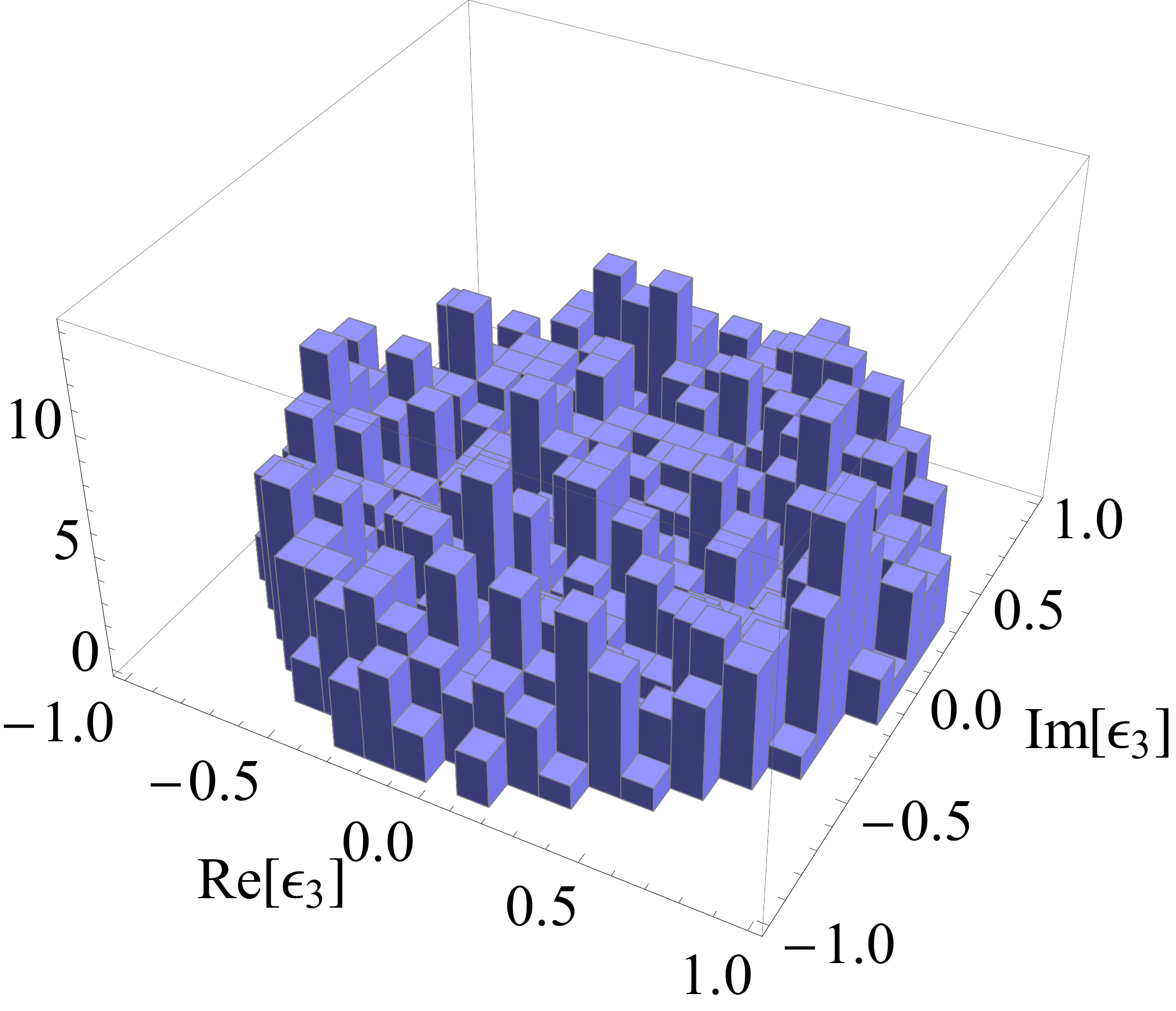}
\endminipage
\minipage{0.33\textwidth}
\includegraphics[width=53mm]{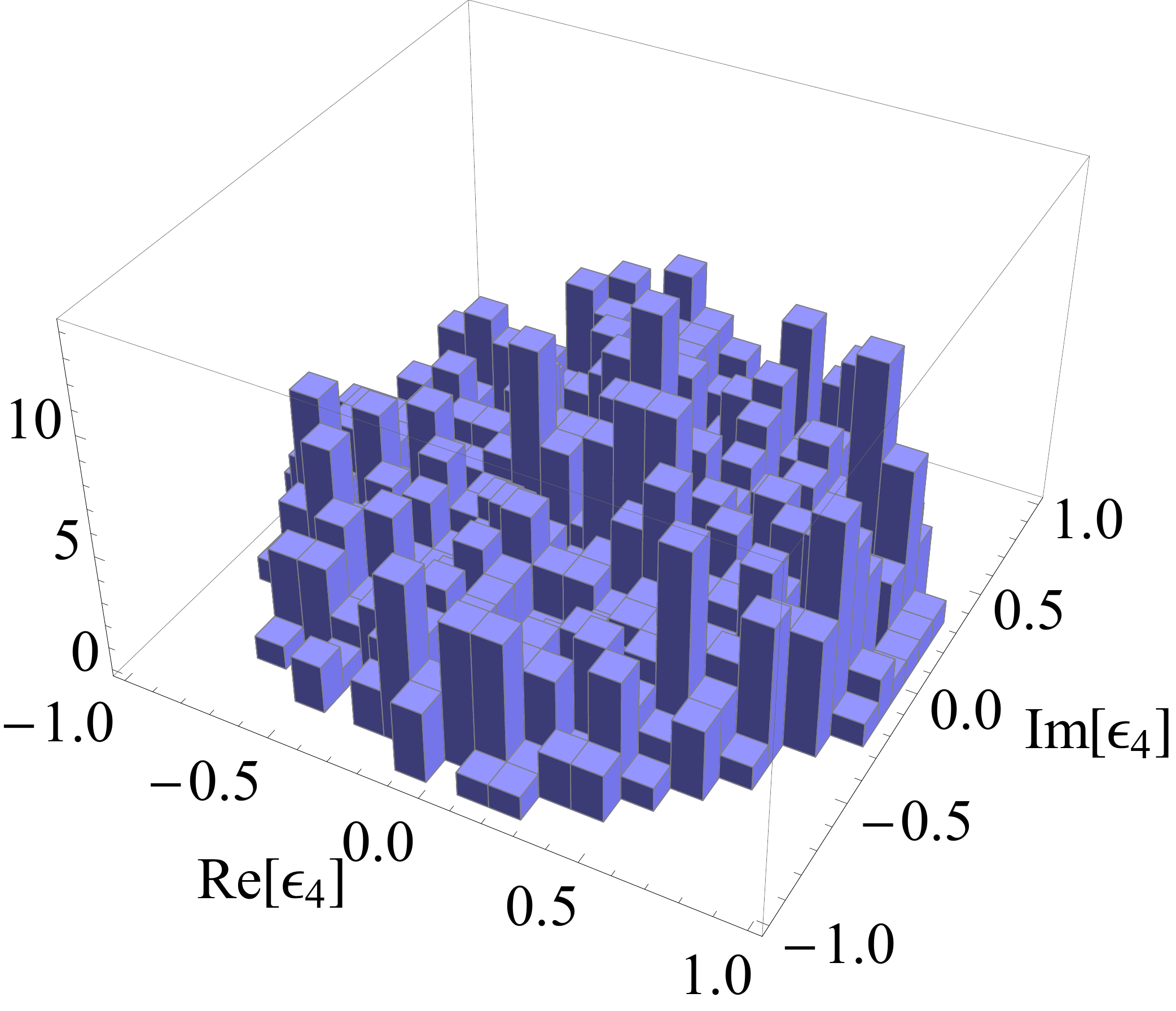}
\endminipage
  \caption{3D Histograms, 1000 random generated strings. N=200.   Repulsive interaction $\tilde{g} = 0.3$.  } \label{Histogram3Dig}
\end{figure}

    To characterize the initial azimuthal deformation of the string
bits in the transverse collision plane, we show in Fig.~\ref{Histogram3D} the pdf distributions of 1000 randomly generated strings 
at a resolution of $N=200$  with no self-interactions $g/\tilde{g}=0$. The pdf shown are for
the distributions in $\epsilon_{2,3,4}$ respectively. 
We also show in Fig.~\ref{Histogram3Dg} the pdf distributions of 1000 randomly generated strings 
at a resolution of $N=200$ undergoing string bit attractions with $g=0.3$ in the mean-field approximation. Note the strong dipole deformation in the leftmost figure. The same pdf for the repulsive case
with $\tilde{g}=0.3$ are shown in Fig.~\ref{Histogram3Dig}. The linear spreading of the string bits with the resolution $N$ causes the
azimuthal deformations to be relatively uniform.

\begin{figure}[!htb]
\minipage{0.33\textwidth}
\includegraphics[width=41mm]{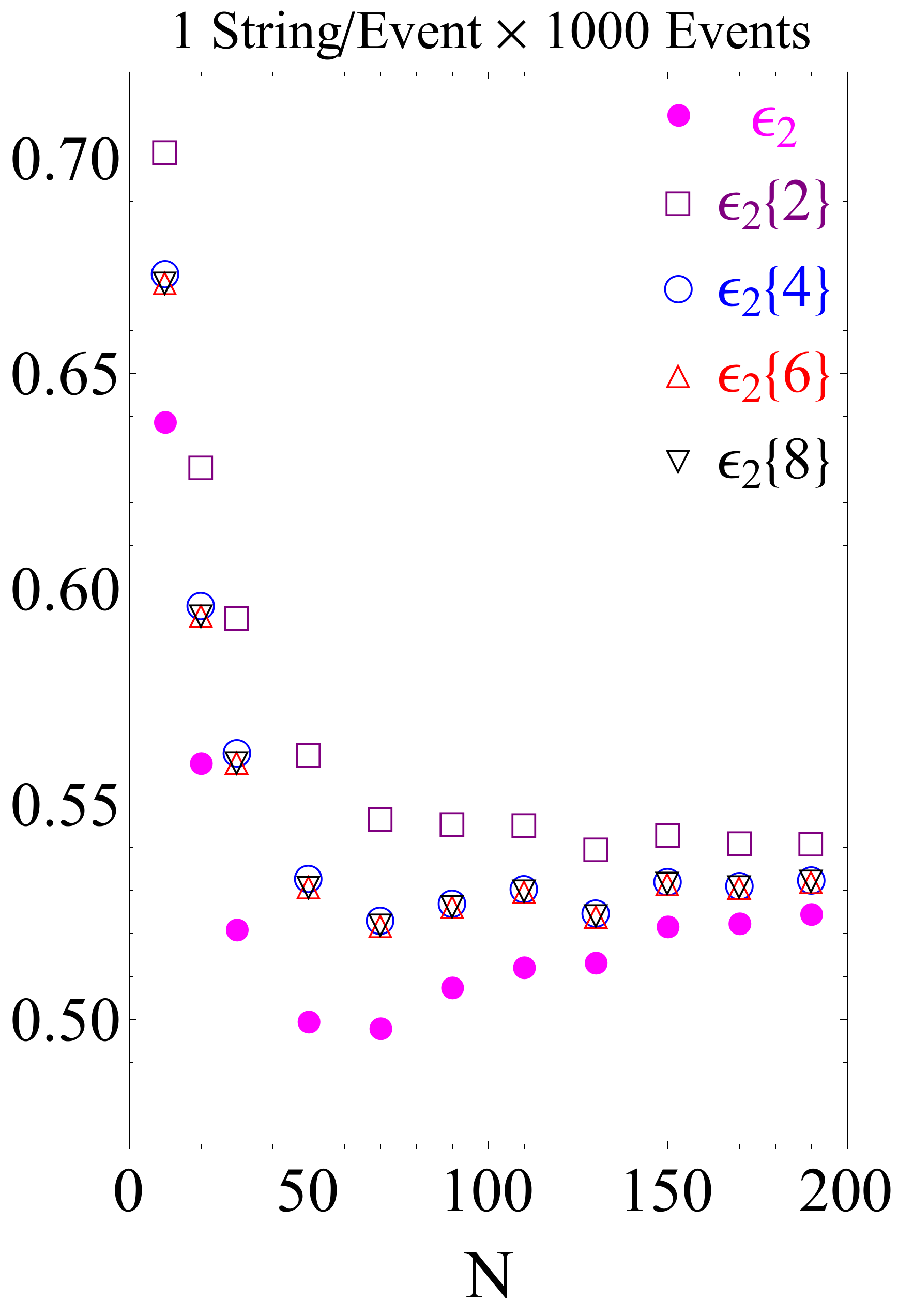}
\endminipage\hfill
\minipage{0.33\textwidth}
\includegraphics[width=41mm]{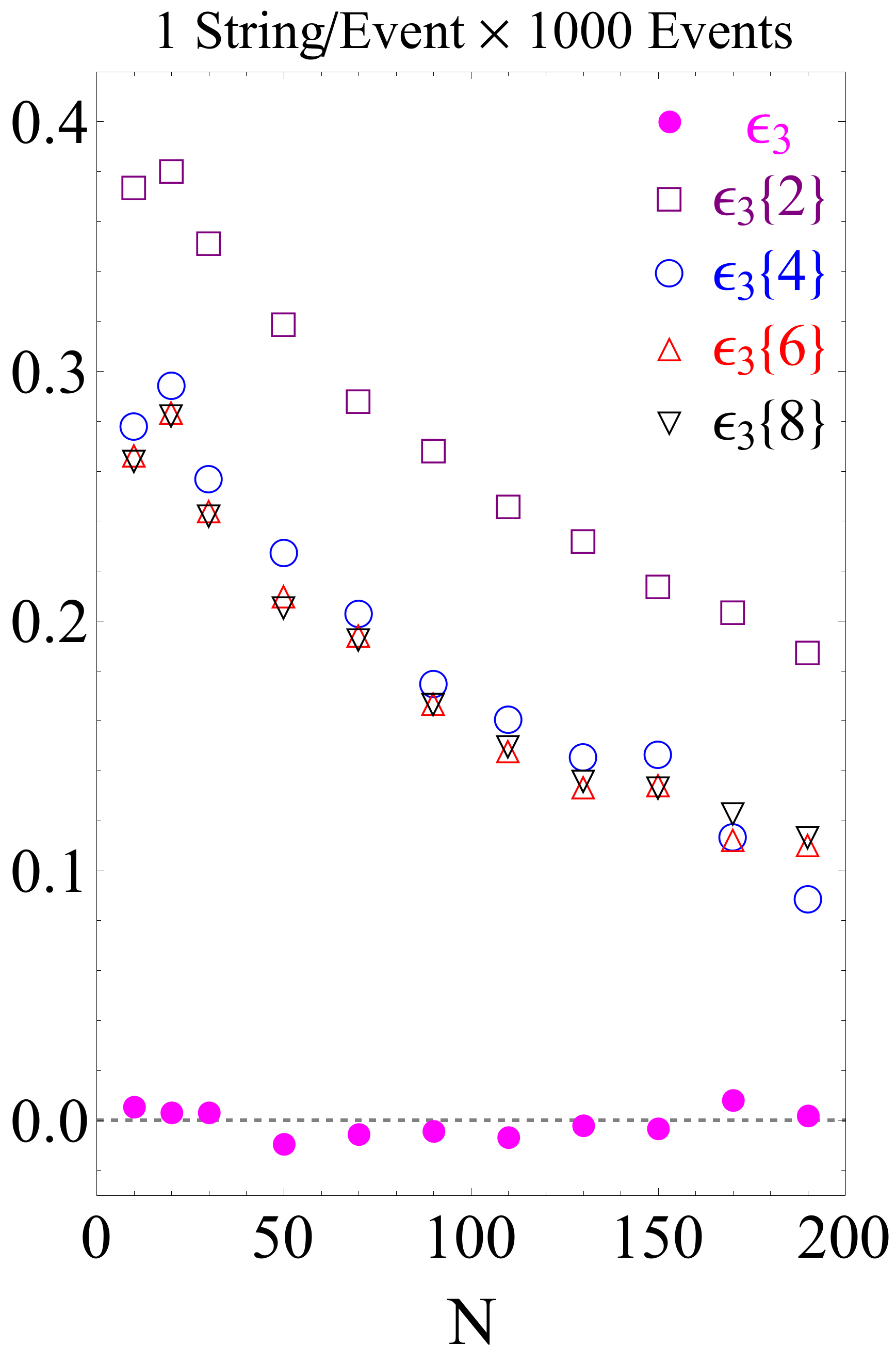}
\endminipage
\minipage{0.33\textwidth}
\includegraphics[width=41mm]{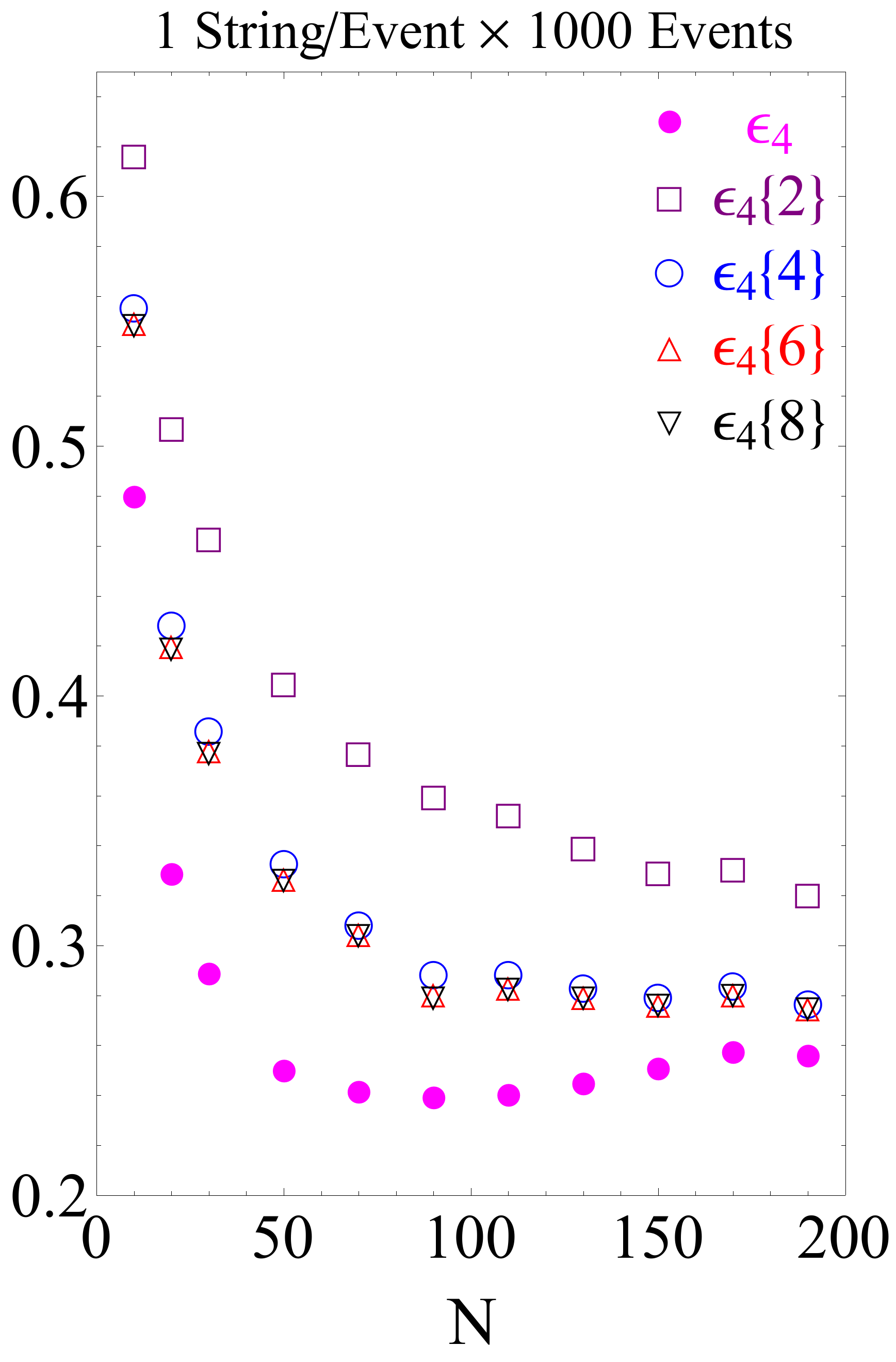}
\endminipage
  \caption{   Attractive interaction $g=0.3$. }  \label{MOMENT6a}
\end{figure}

\begin{figure}[!htb]
\minipage{0.33\textwidth}
\includegraphics[width=41mm]{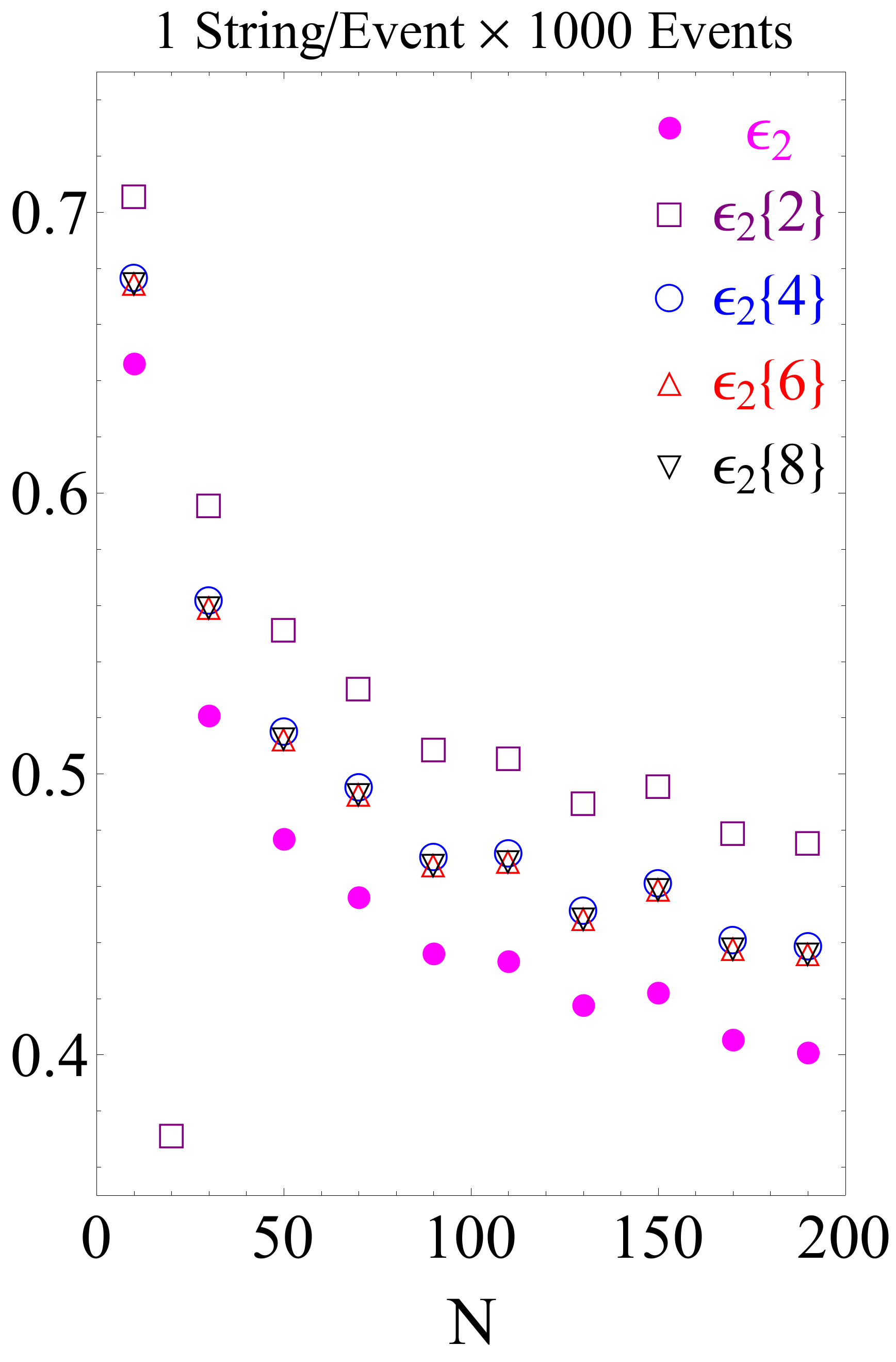}
\endminipage\hfill
\minipage{0.33\textwidth}
\includegraphics[width=41mm]{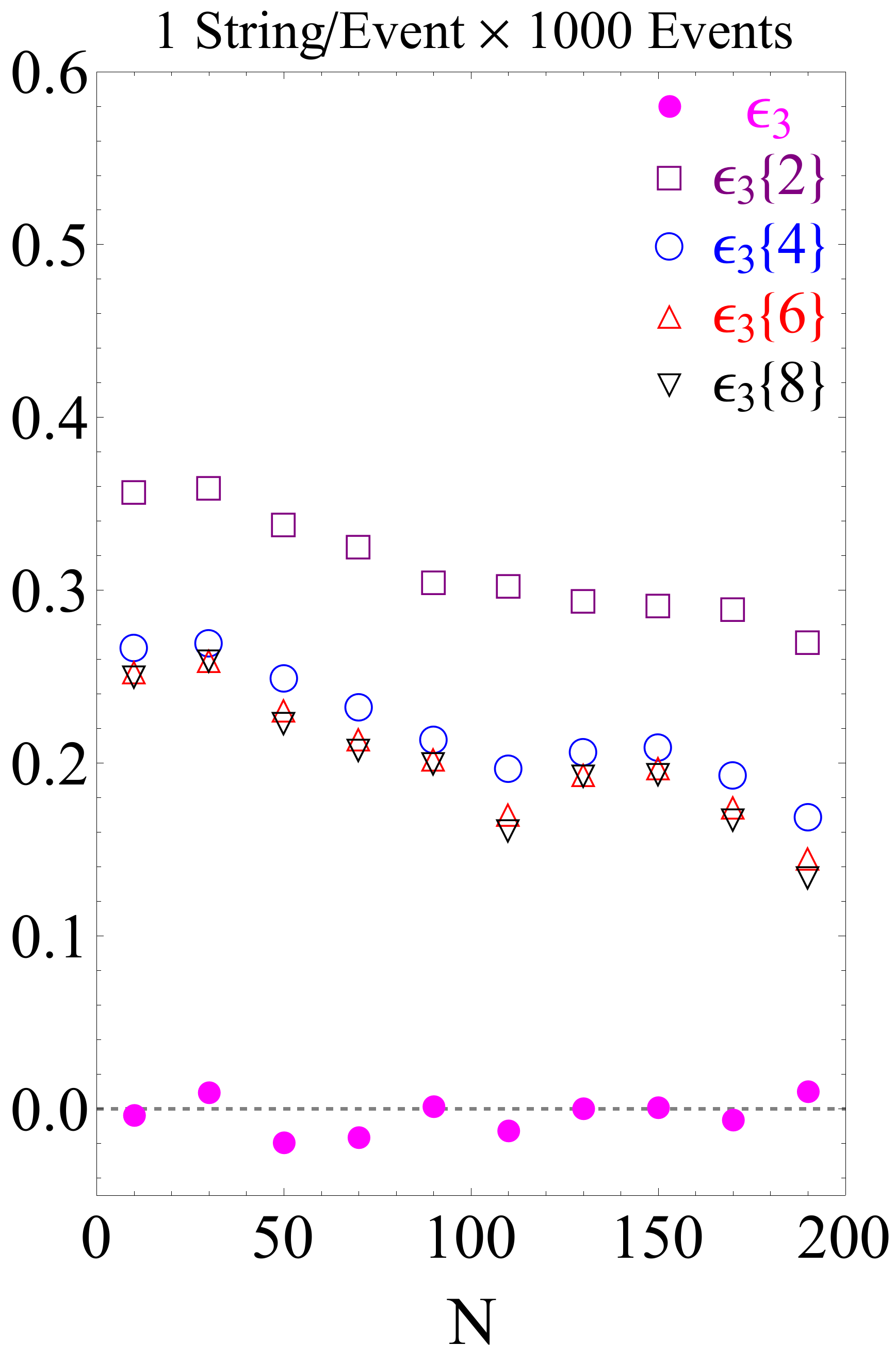}
\endminipage
\minipage{0.33\textwidth}
\includegraphics[width=41mm]{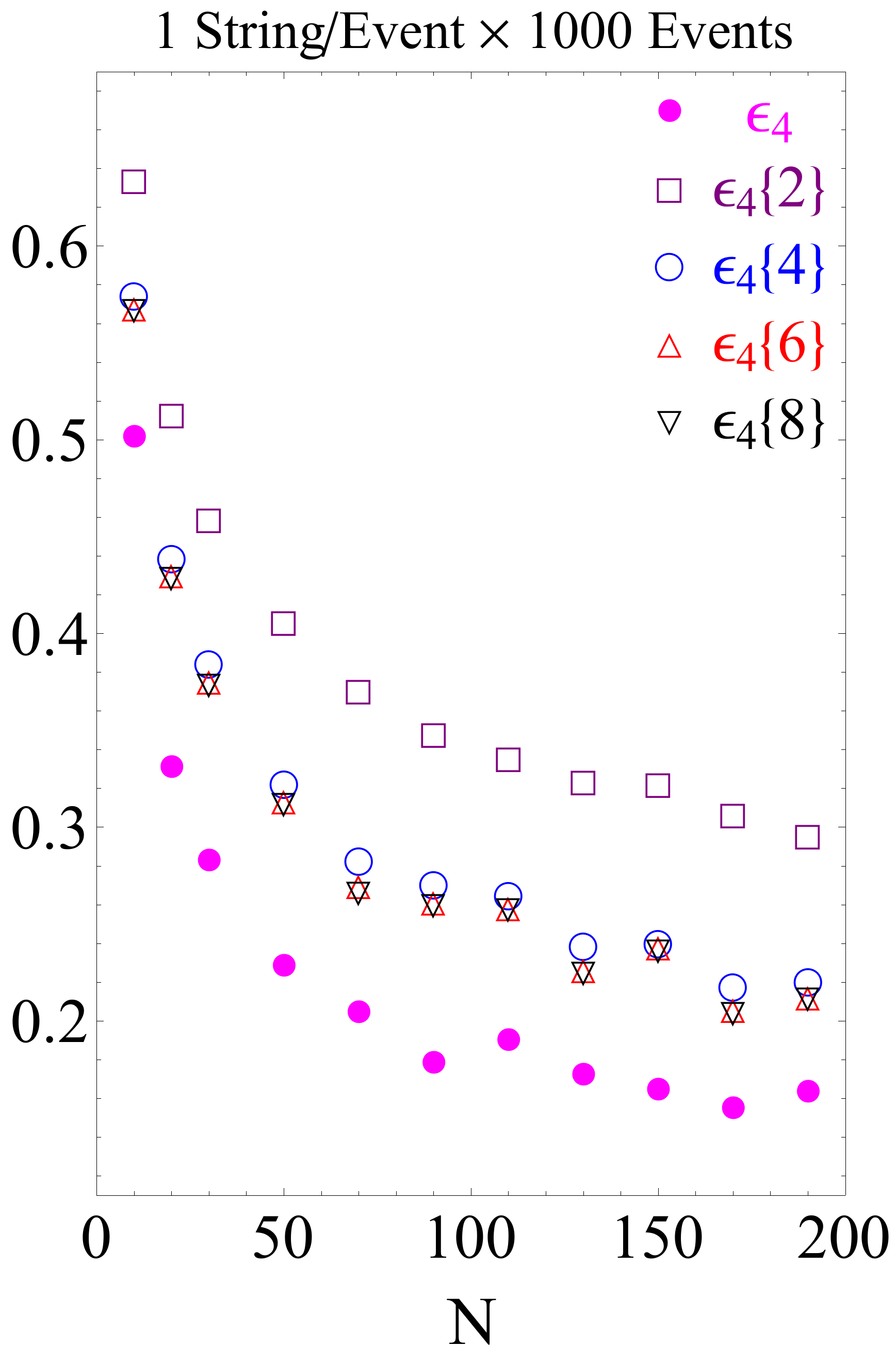}
\endminipage
  \caption{Non-interacting $g/\tilde{g} = 0$.}\label{EnRandoma}
\end{figure}

\begin{figure}[!htb]
\minipage{0.33\textwidth}
\includegraphics[width=41mm]{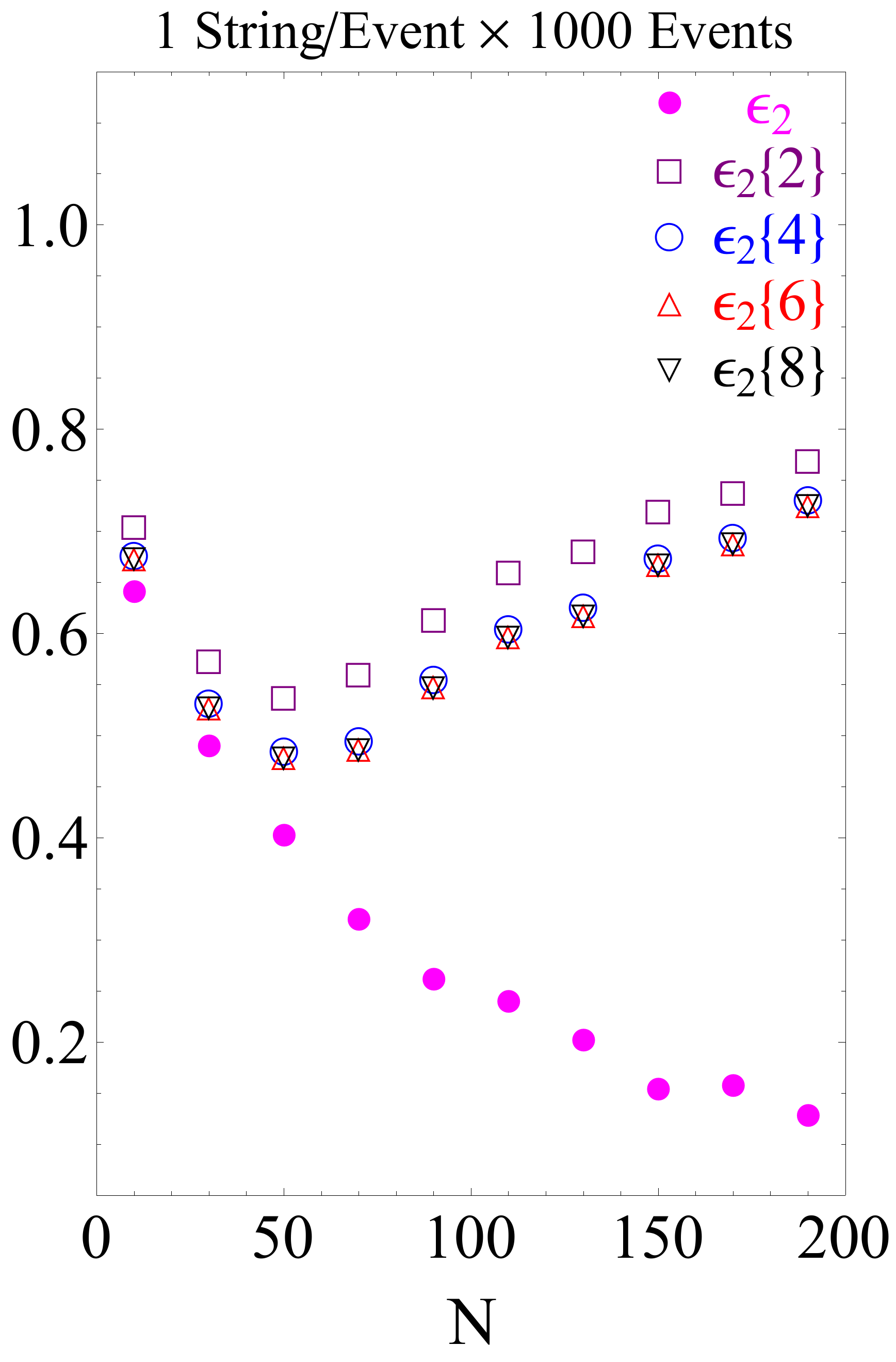}
\endminipage\hfill
\minipage{0.33\textwidth}
\includegraphics[width=41mm]{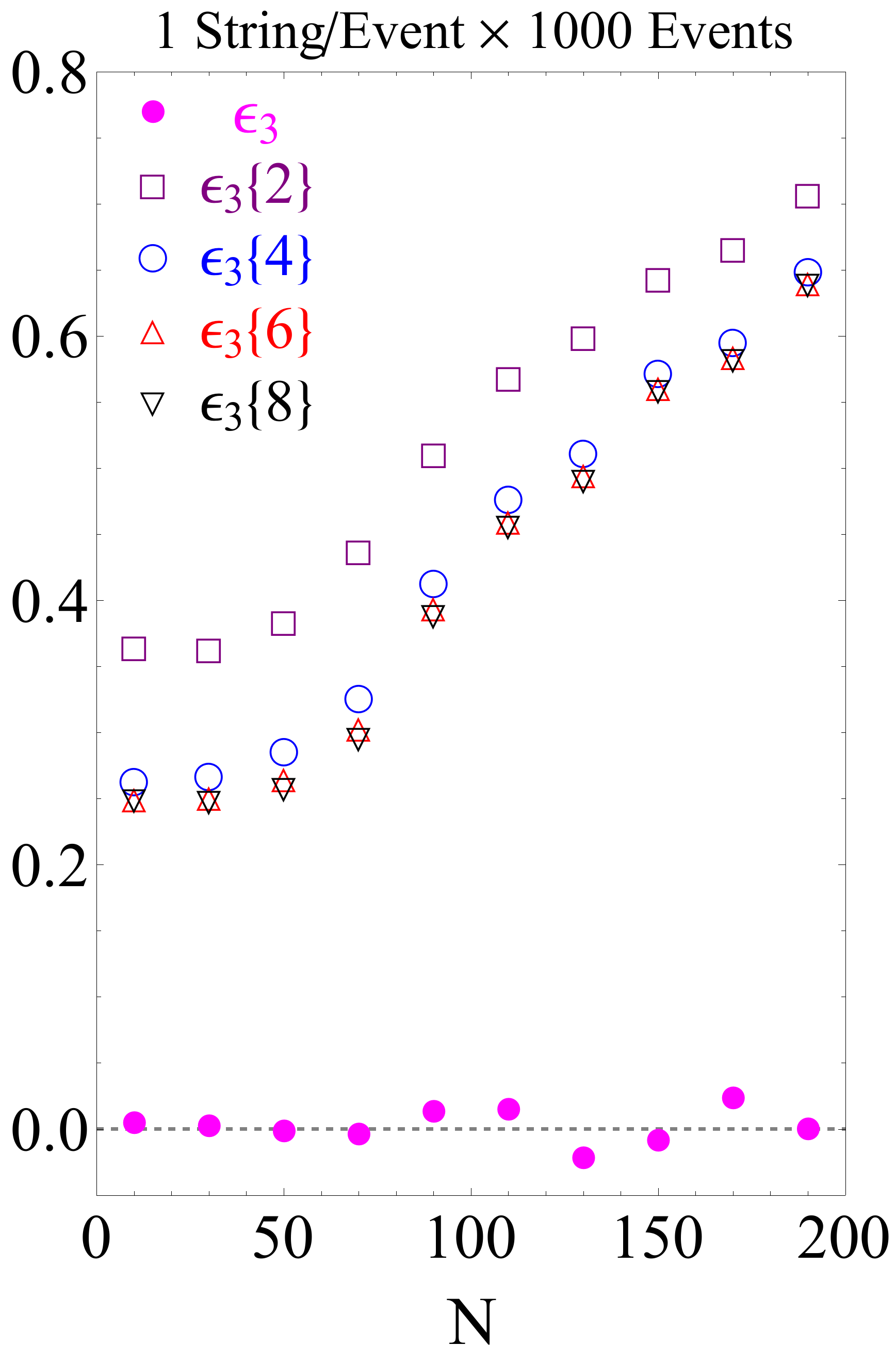}
\endminipage
\minipage{0.33\textwidth}
\includegraphics[width=41mm]{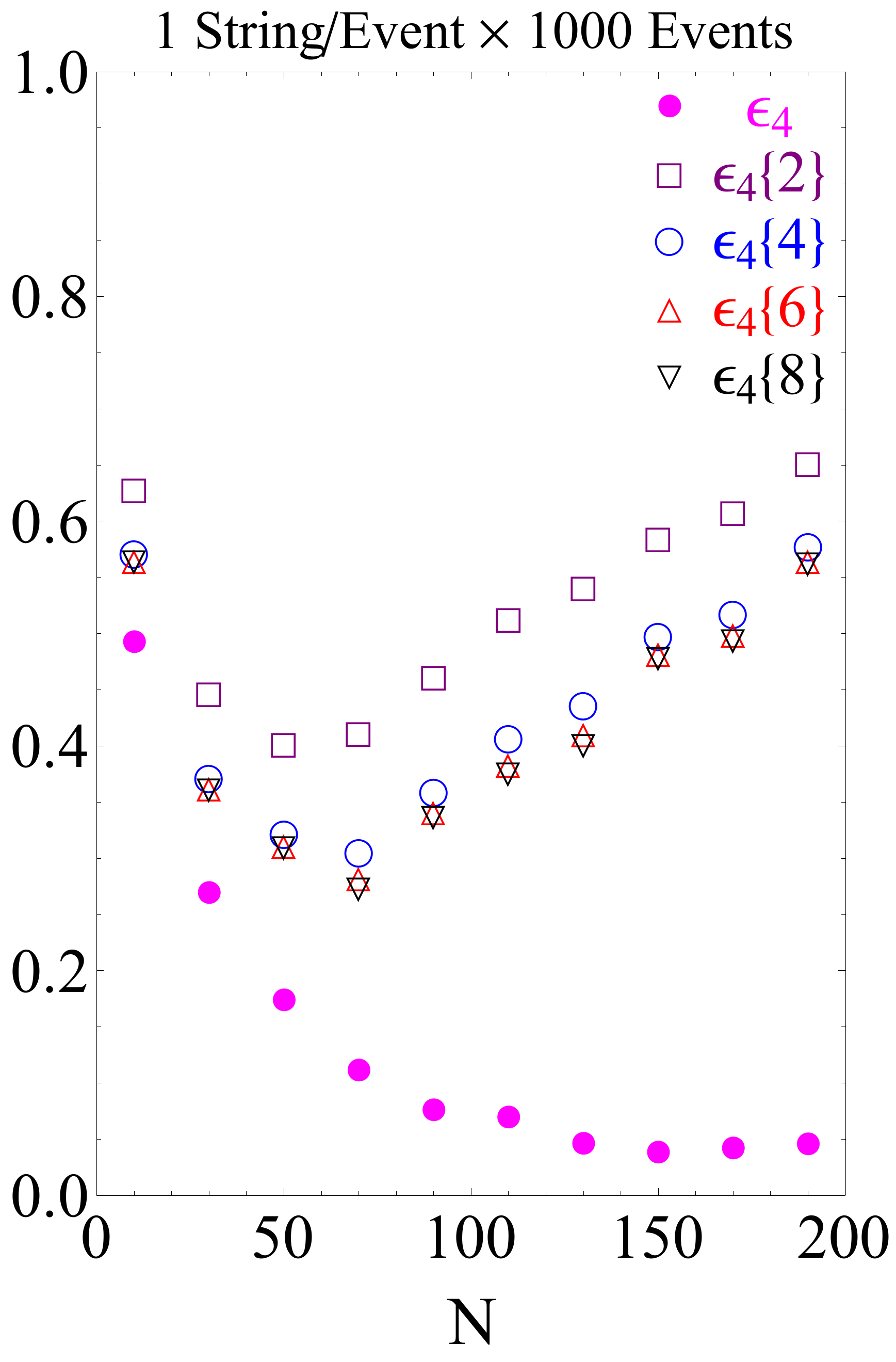}
\endminipage
  \caption{   Repulsive interaction $\tilde{g} = 0.3$.  } \label{MOMENT3a}
\end{figure}

\begin{figure}[!htb]
\minipage{0.33\textwidth}
\includegraphics[width=41mm]{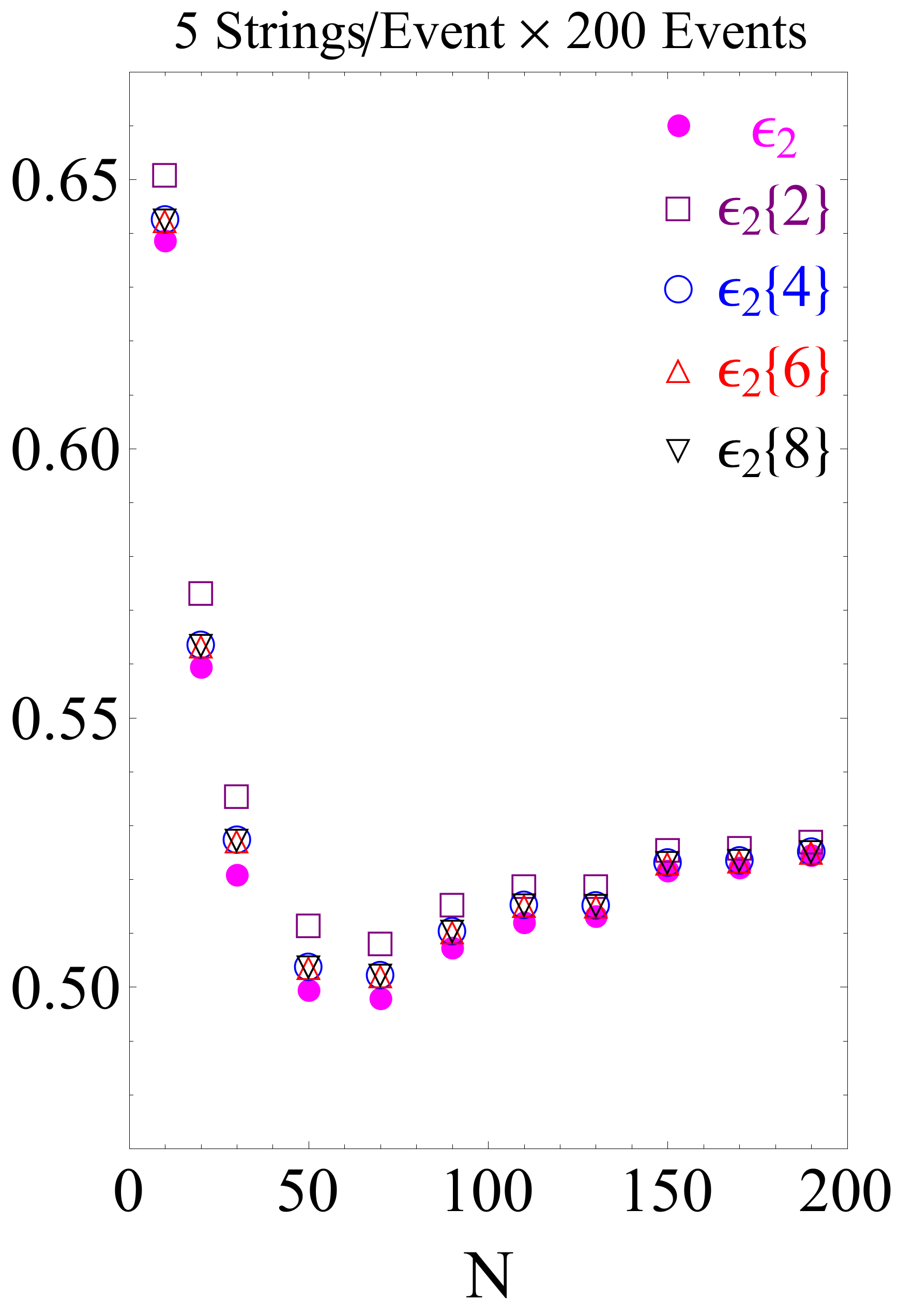}
\endminipage\hfill
\minipage{0.33\textwidth}
\includegraphics[width=41mm]{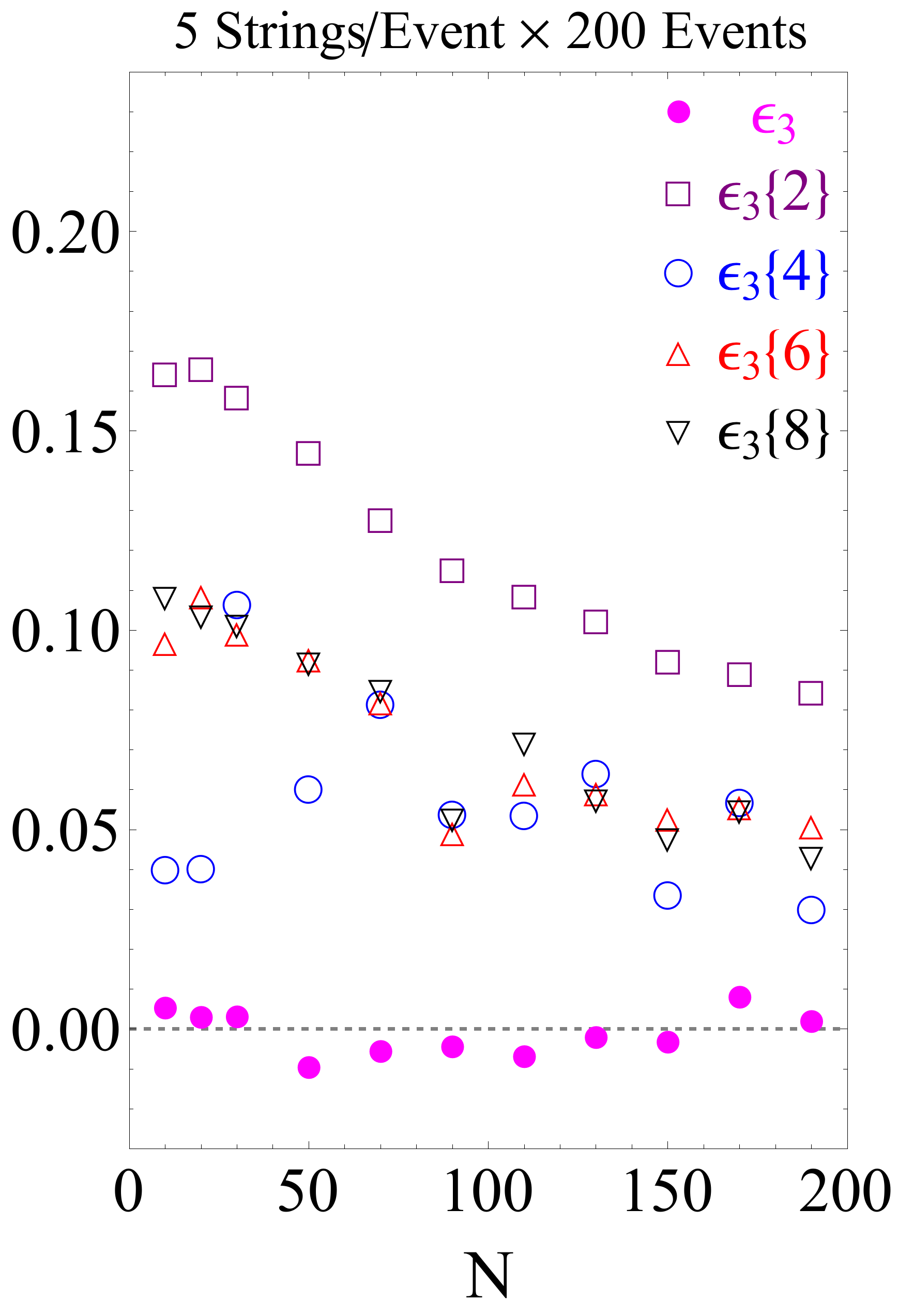}
\endminipage
\minipage{0.33\textwidth}
\includegraphics[width=41mm]{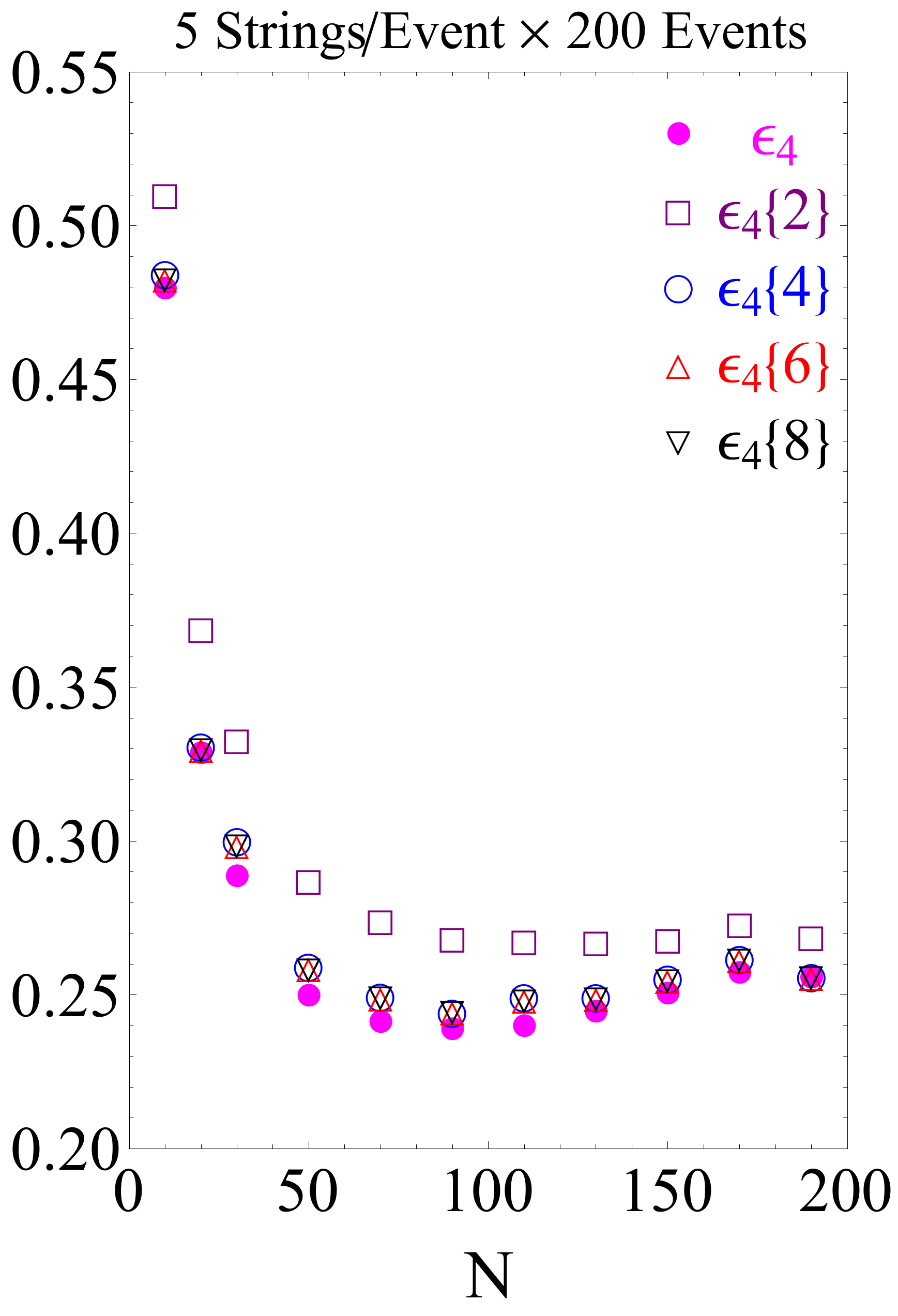}
\endminipage
  \caption{ Attractive interaction $g=0.3$.  }  \label{MOMENT6b}
\end{figure}

\begin{figure}[!htb]
\minipage{0.33\textwidth}
\includegraphics[width=41mm]{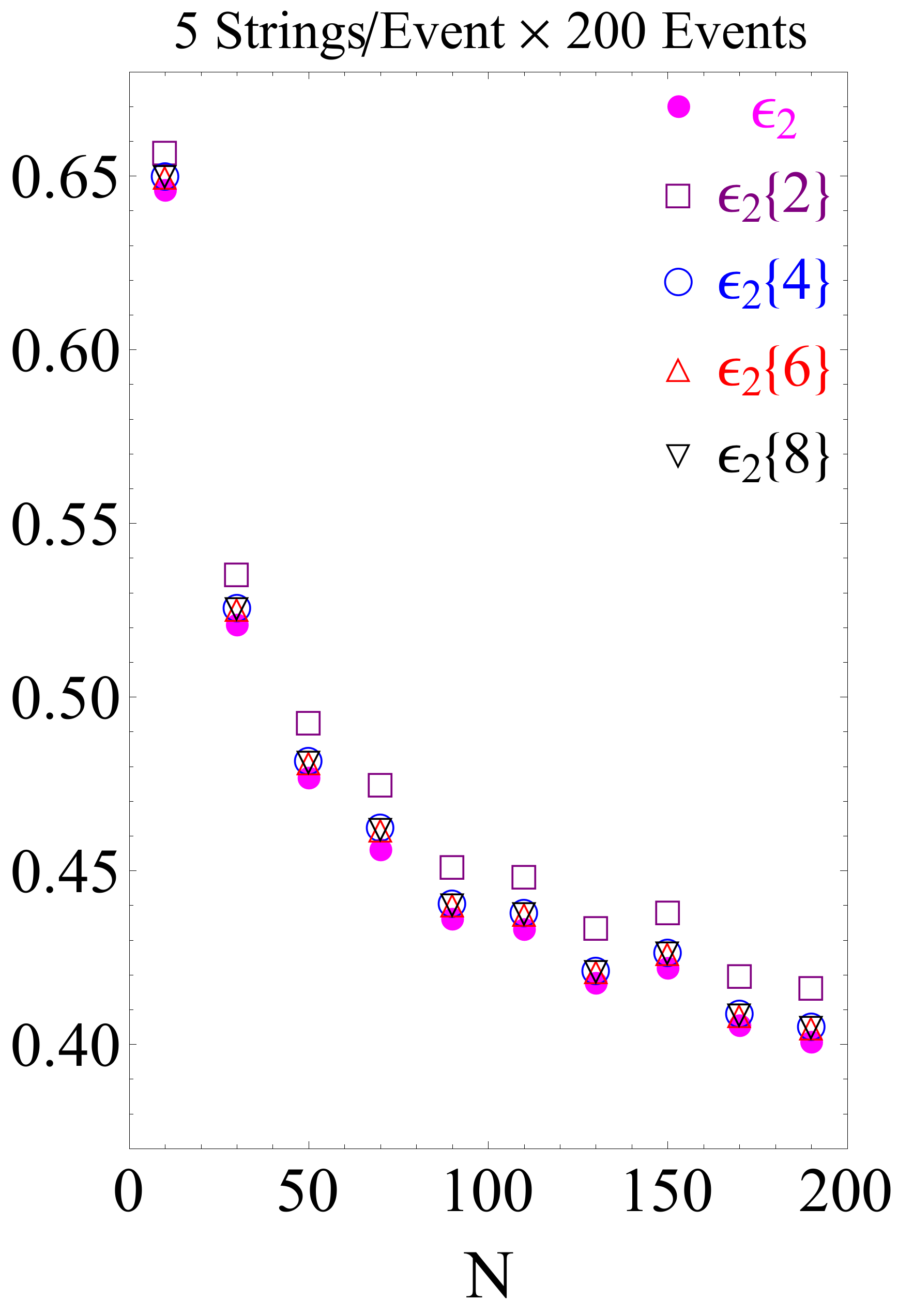}
\endminipage\hfill
\minipage{0.33\textwidth}
\includegraphics[width=41mm]{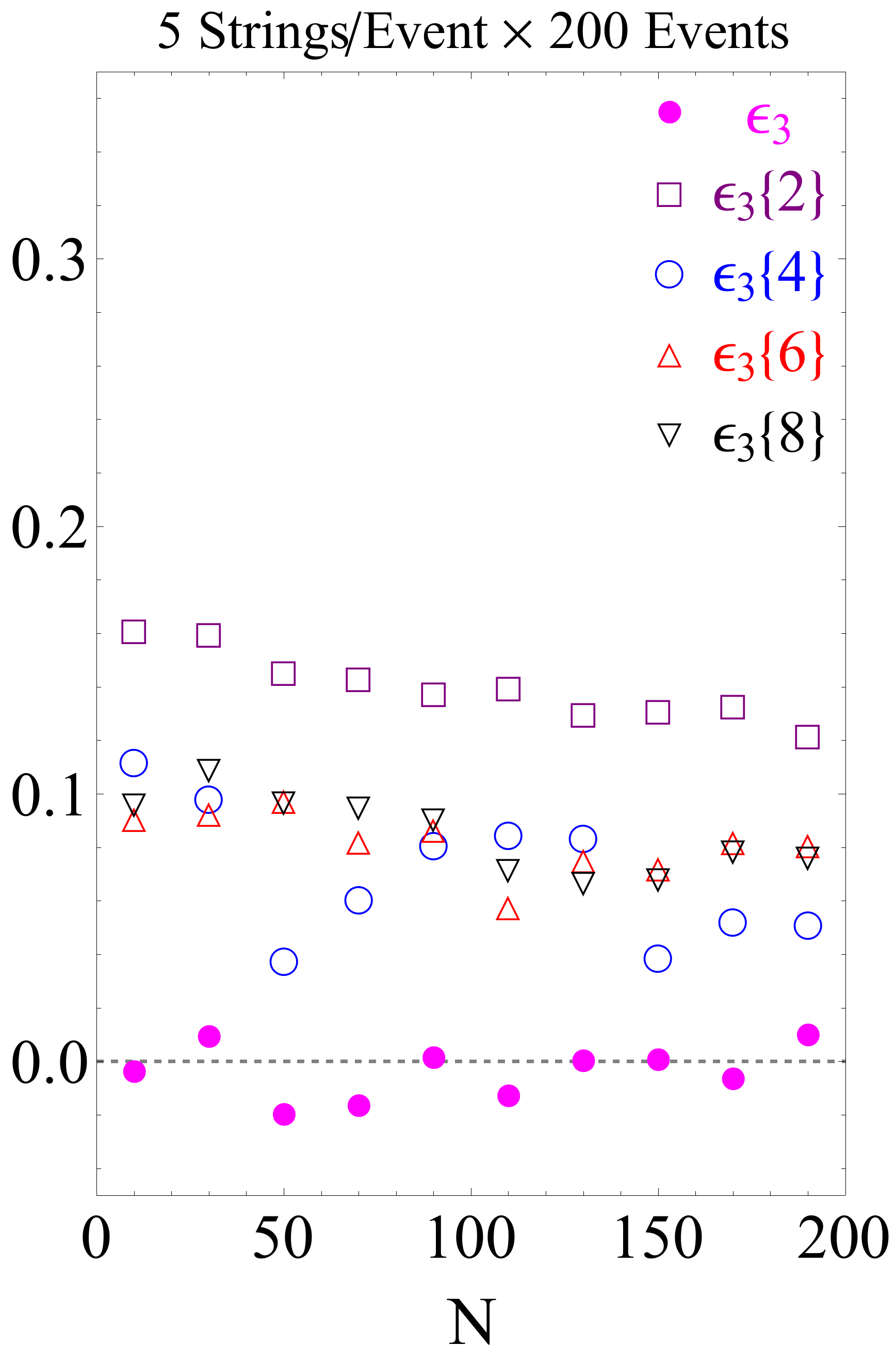}
\endminipage
\minipage{0.33\textwidth}
\includegraphics[width=41mm]{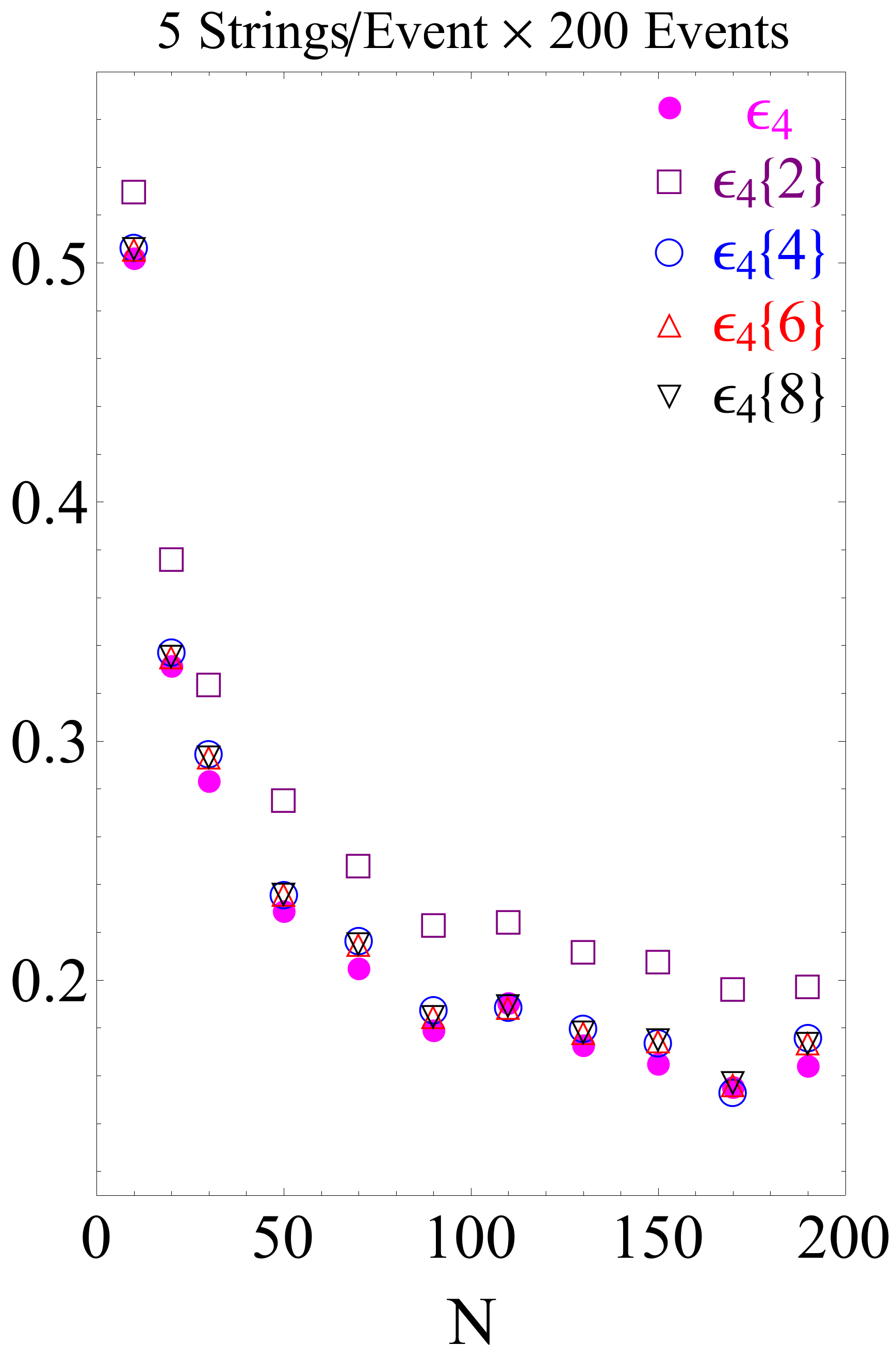}
\endminipage
  \caption{Non-interacting $g/\tilde{g} = 0$. }\label{EnRandomb}
\end{figure}

\begin{figure}[!htb]
\minipage{0.33\textwidth}
\includegraphics[width=41mm]{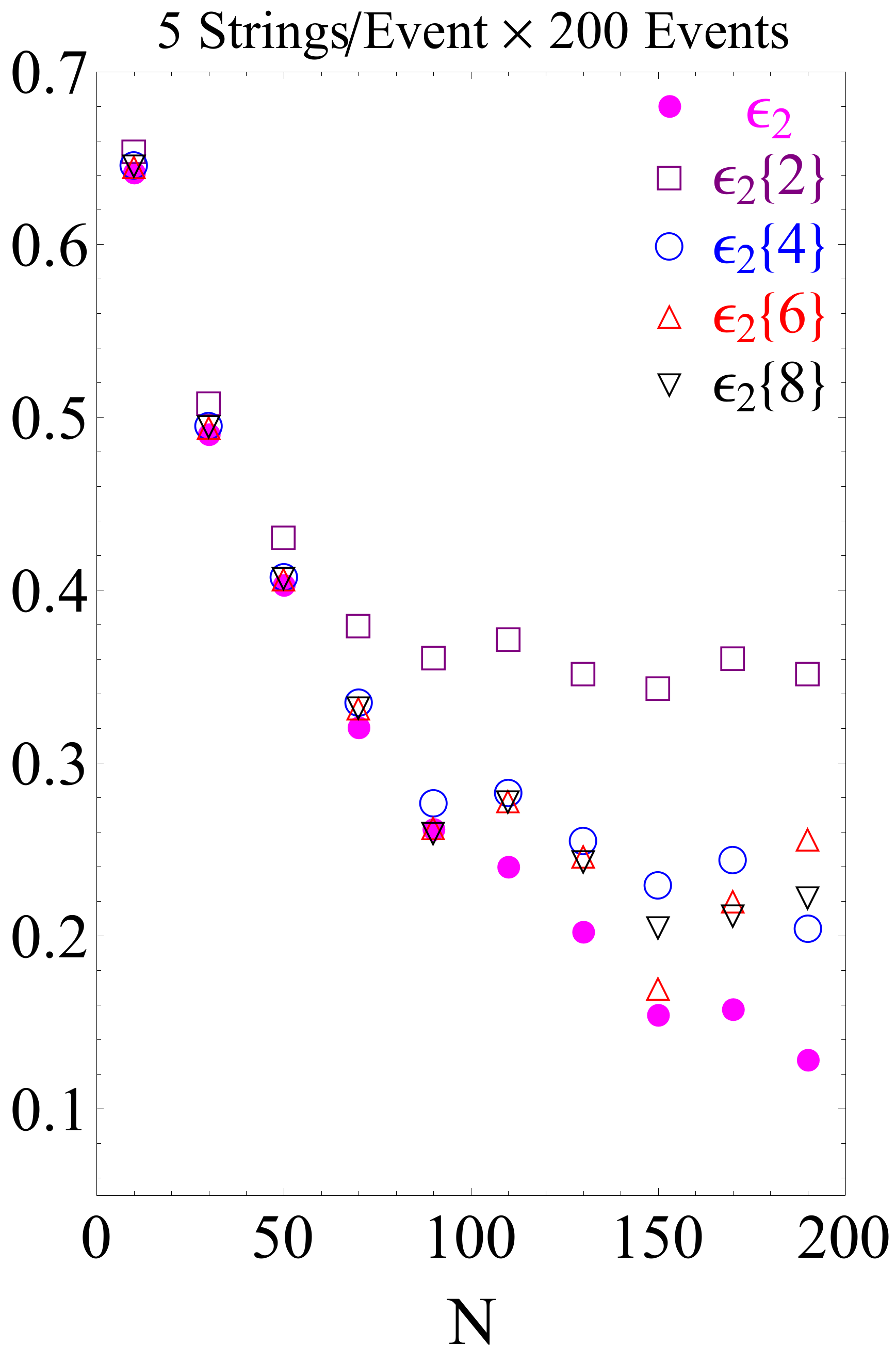}
\endminipage\hfill
\minipage{0.33\textwidth}
\includegraphics[width=41mm]{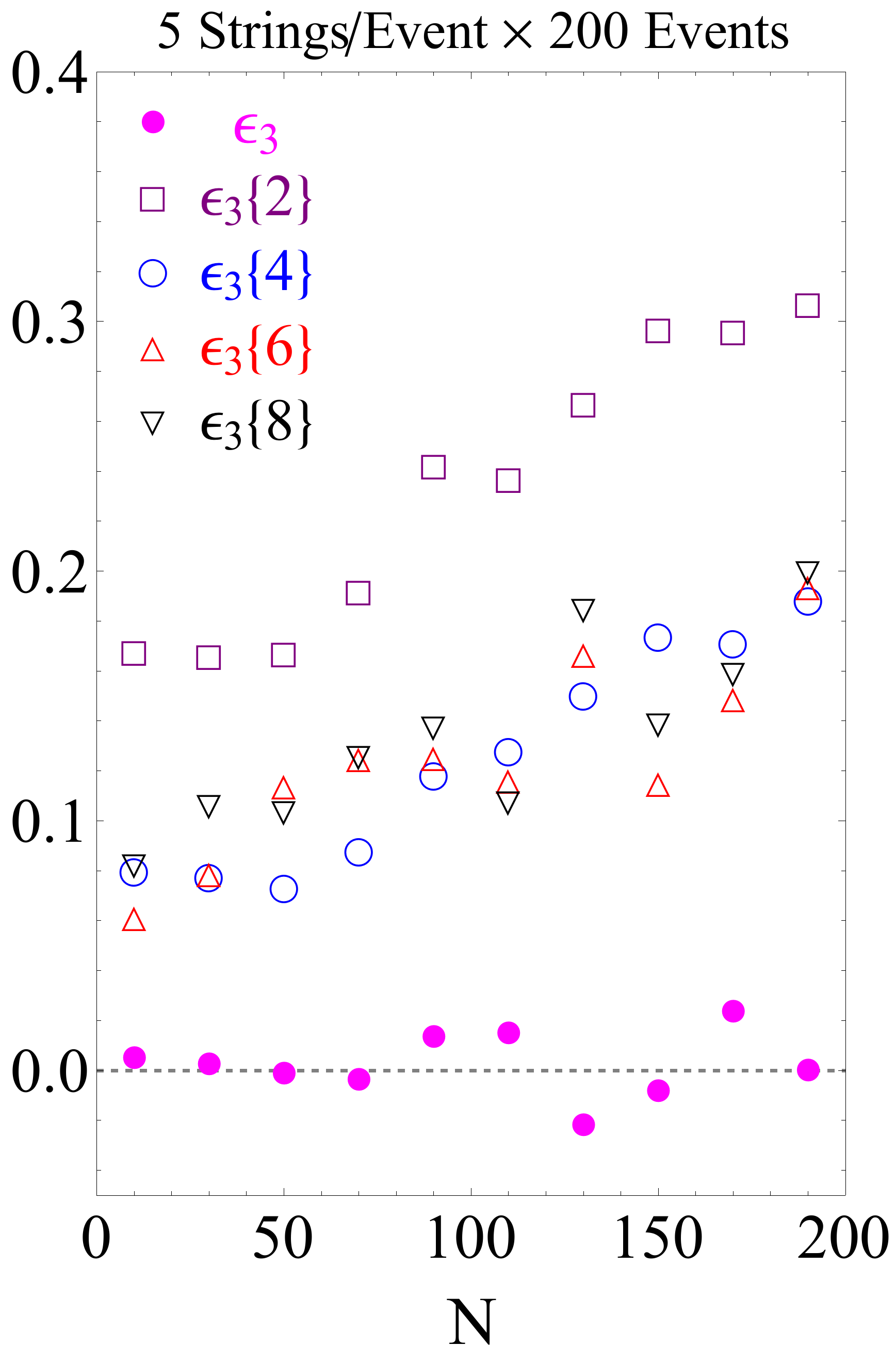}
\endminipage
\minipage{0.33\textwidth}
\includegraphics[width=41mm]{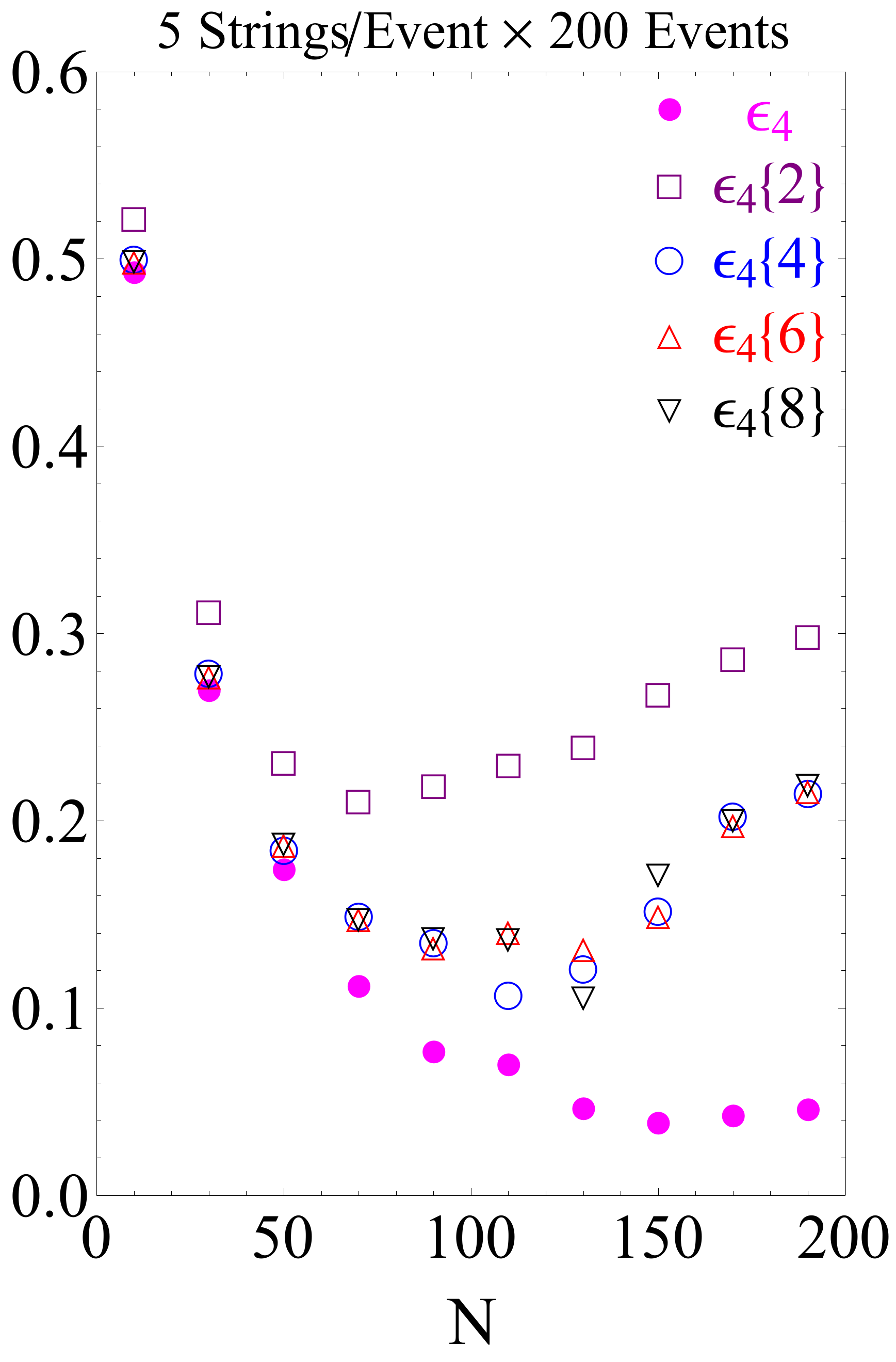}
\endminipage
  \caption{     Repulsive interaction $\tilde{g} = 0.3$.  }  \label{MOMENT3b}
\end{figure}

\begin{figure}[!htb]
\minipage{0.33\textwidth}
\includegraphics[width=41mm]{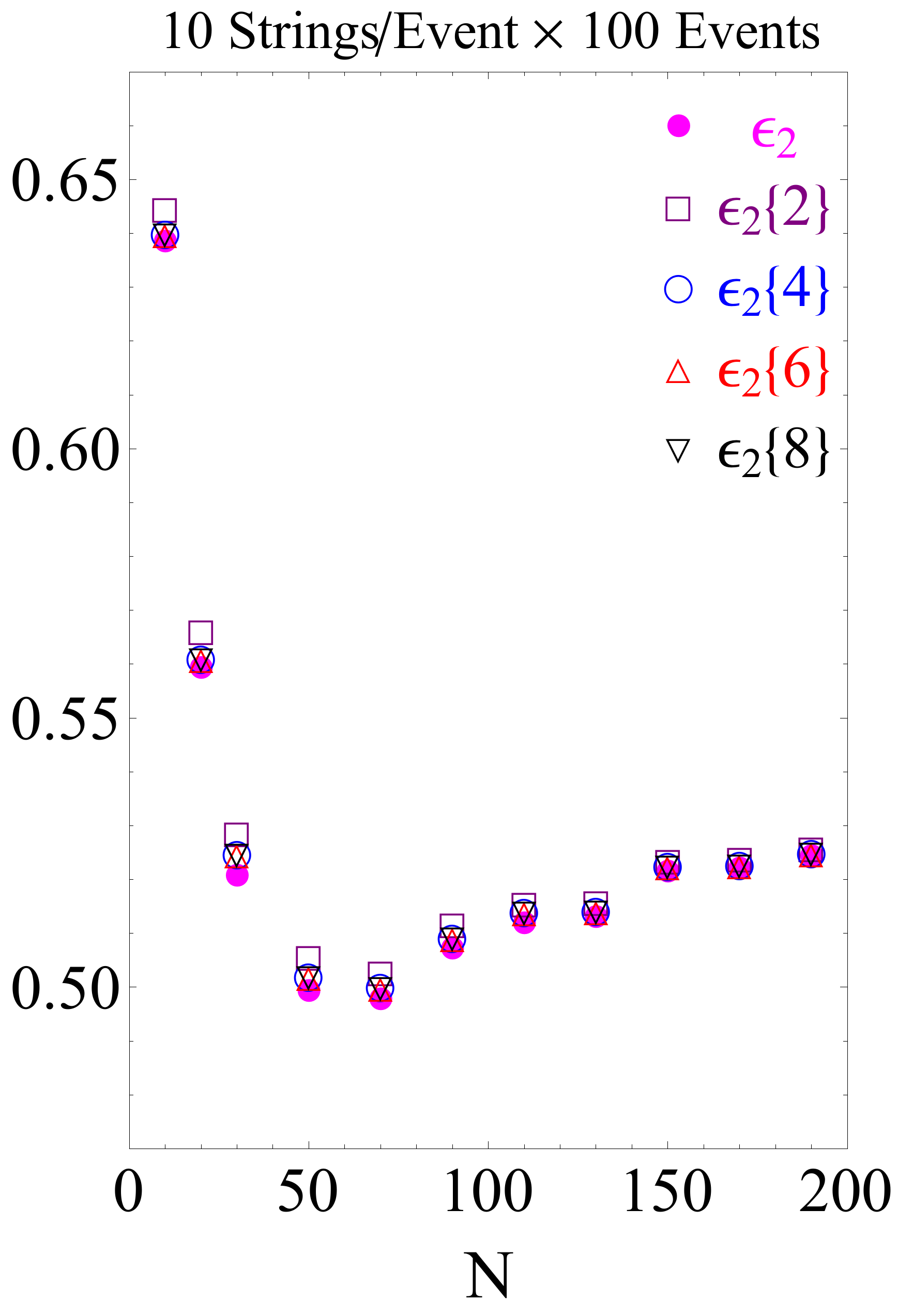}
\endminipage\hfill
\minipage{0.33\textwidth}
\includegraphics[width=41mm]{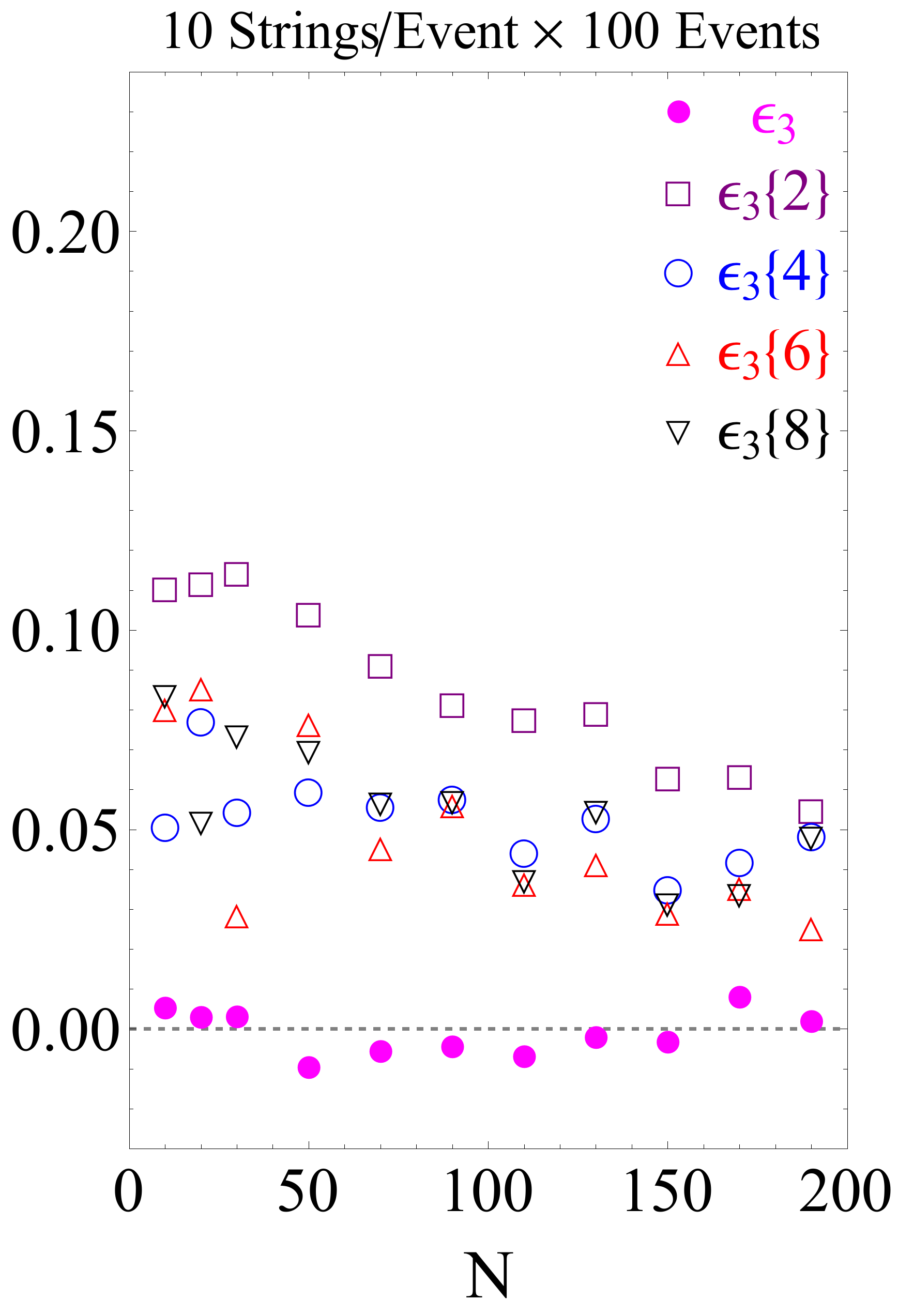}
\endminipage
\minipage{0.33\textwidth}
\includegraphics[width=41mm]{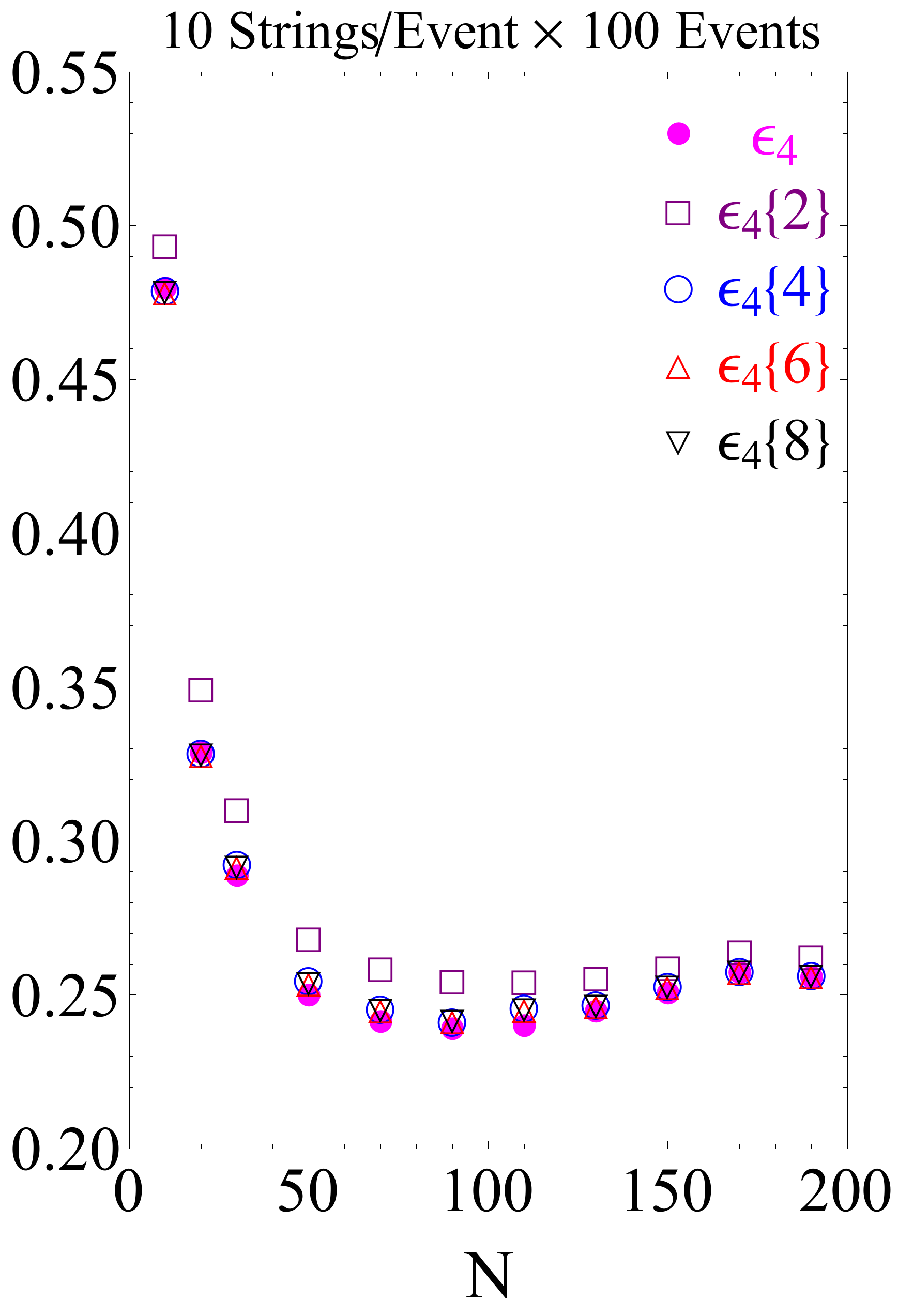}
\endminipage
  \caption{Attractive interaction $g=0.3$.  }  \label{MOMENT6c}
\end{figure}

\begin{figure}[!htb]
\minipage{0.33\textwidth}
\includegraphics[width=41mm]{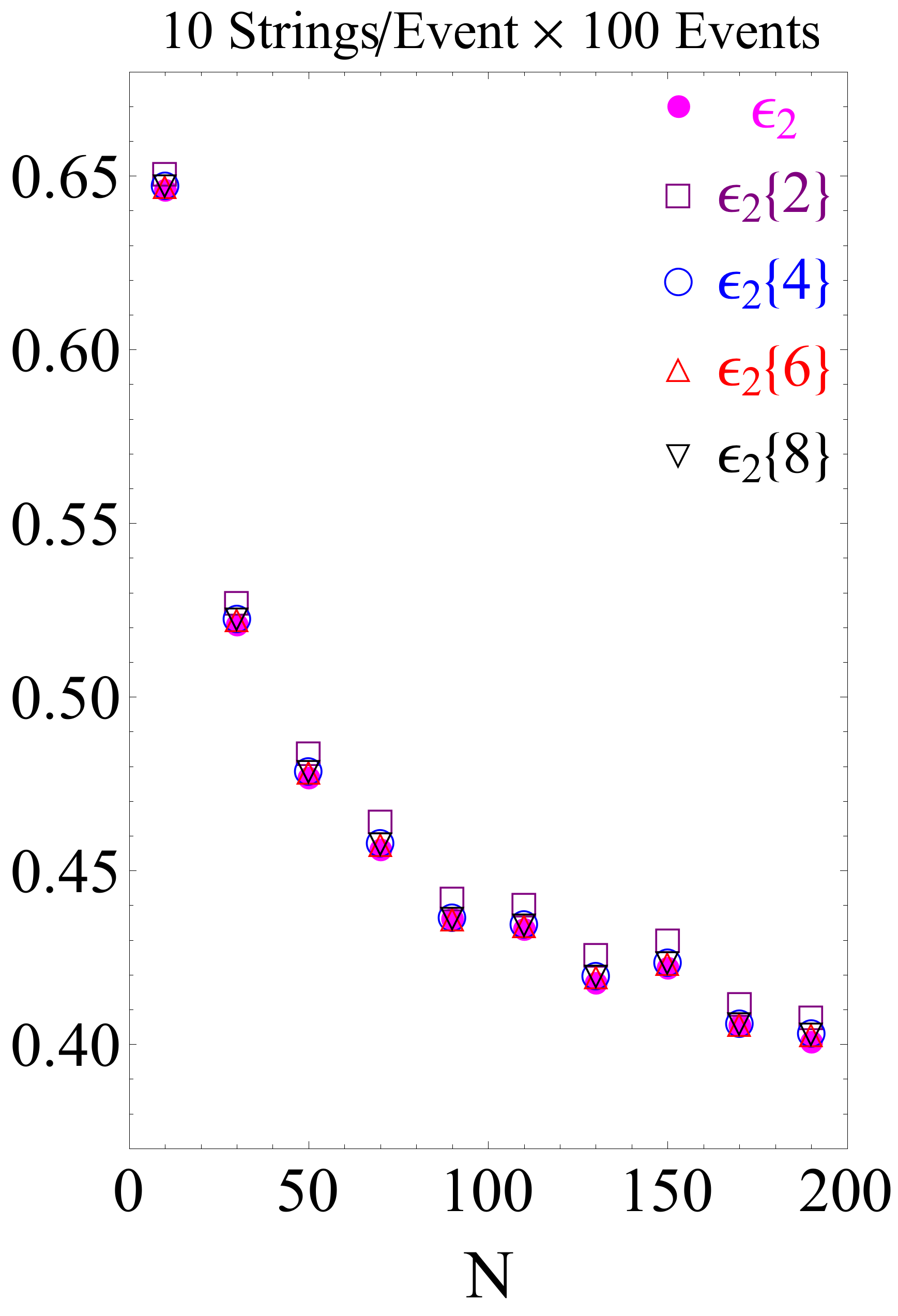}
\endminipage\hfill
\minipage{0.33\textwidth}
\includegraphics[width=41mm]{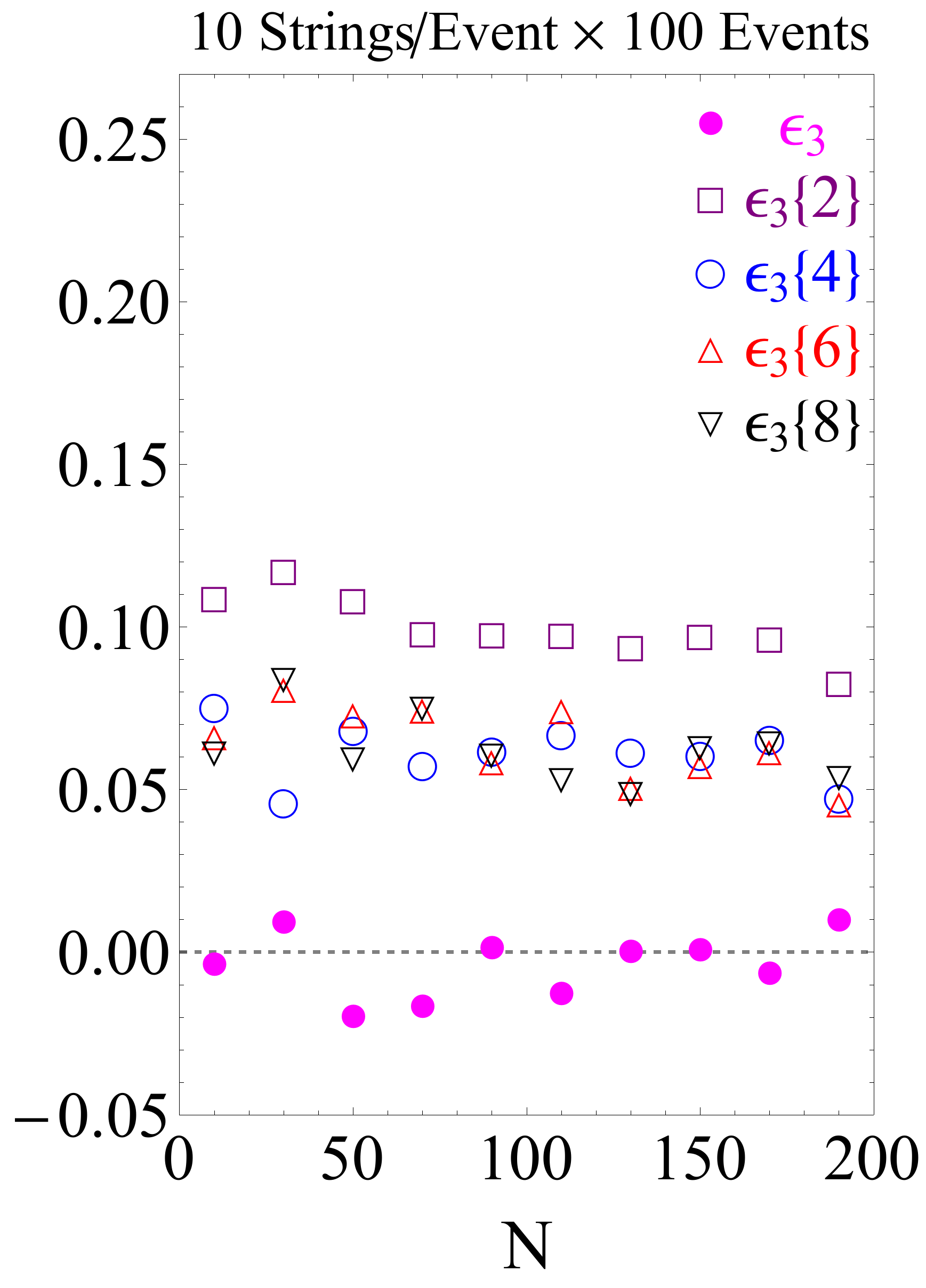}
\endminipage
\minipage{0.33\textwidth}
\includegraphics[width=41mm]{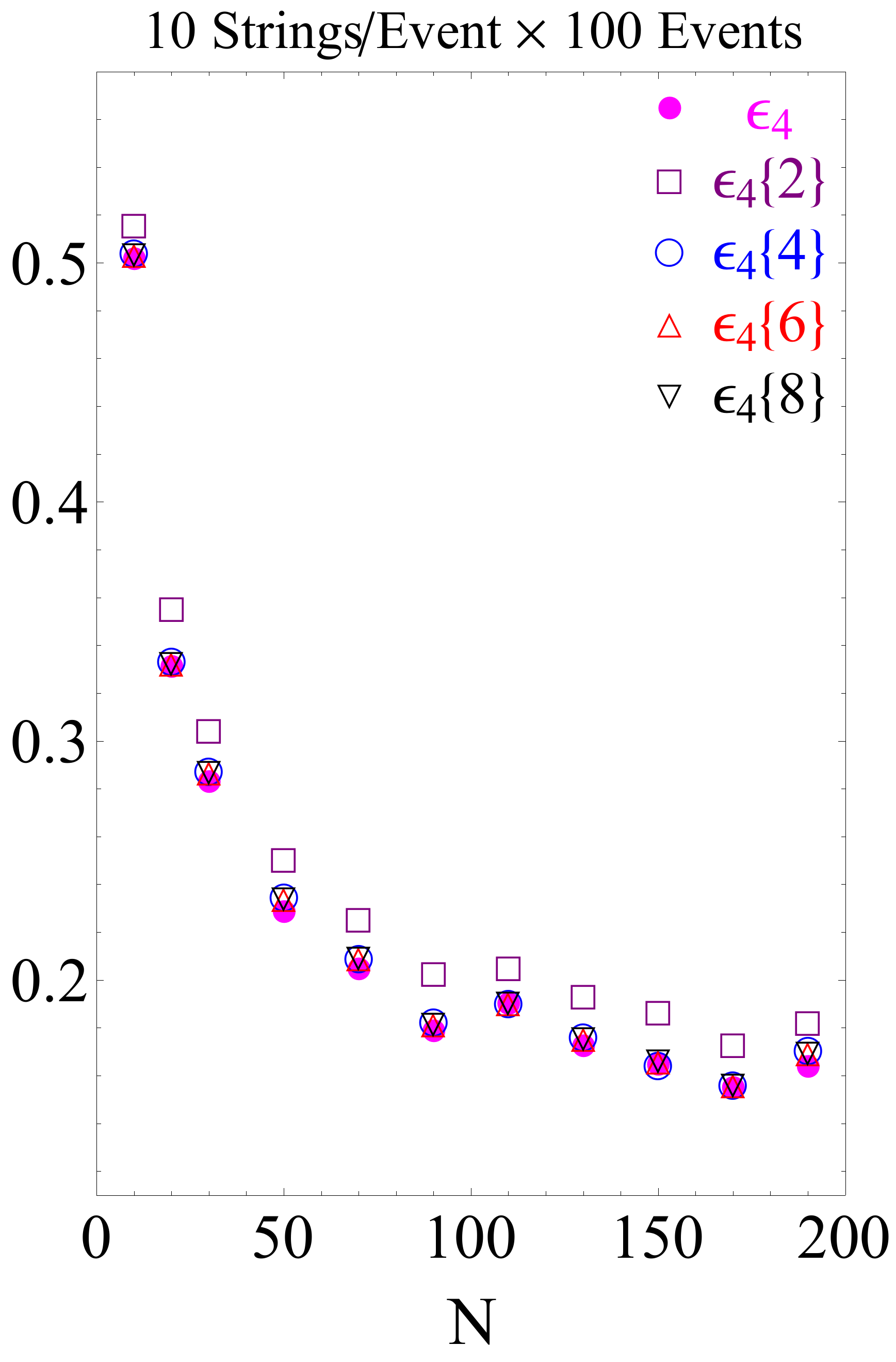}
\endminipage
  \caption{Non-interacting $g/\tilde{g} = 0$.}\label{EnRandomc}
\end{figure}

\begin{figure}[!htb]
\minipage{0.33\textwidth}
\includegraphics[width=41mm]{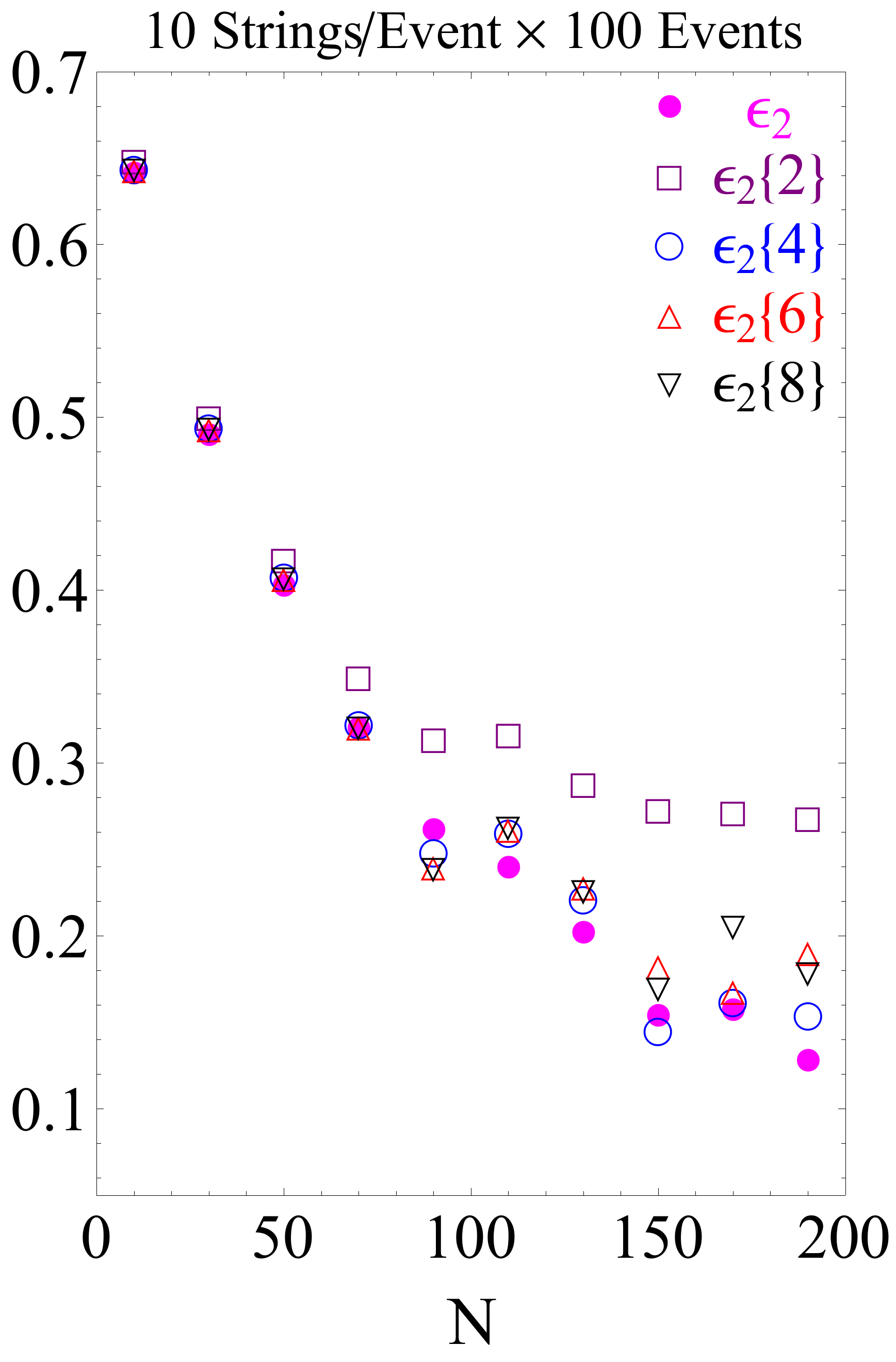}
\endminipage\hfill
\minipage{0.33\textwidth}
\includegraphics[width=41mm]{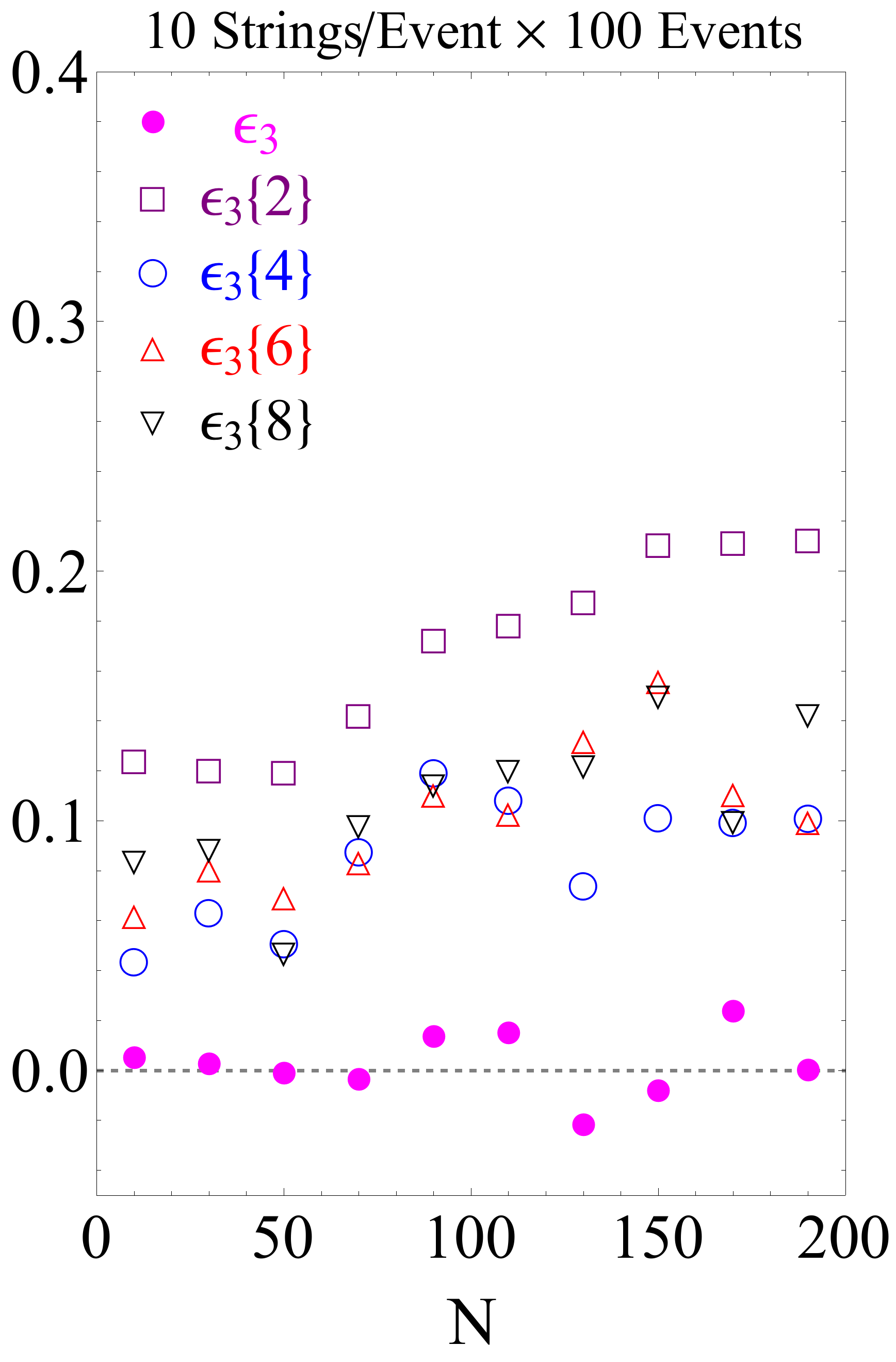}
\endminipage
\minipage{0.33\textwidth}
\includegraphics[width=41mm]{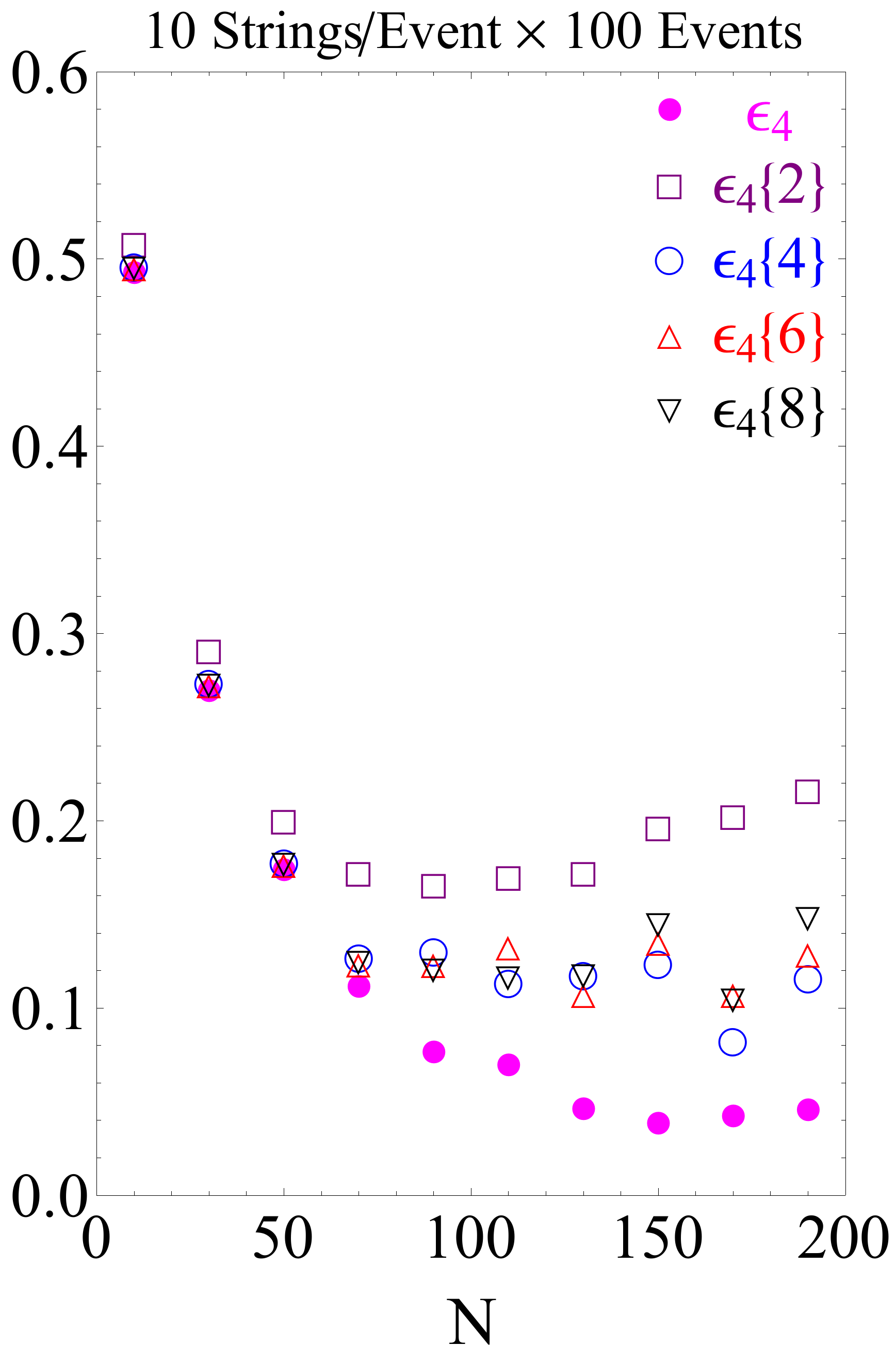}
\endminipage
  \caption{    Repulsive interaction $\tilde{g} = 0.3$.  }  \label{MOMENT3c}
\end{figure}

For completeness we show the behavior of the cross moments with the resolution for attractive, non-interacting and repulsive strings in
Fig.~\ref{MOMENT6a}, Fig.~\ref{EnRandoma} and Fig.~\ref{MOMENT3a} respectively by sampling 1000 times a single string streched 
with $b=5$. The attraction is set at $g=0.3$ while the repulsion at $\tilde {g}=0.3$ for the infinite range case with $m=0$. 
Recall that the realistic case of a massive glueball or scalar mass $m$ is amenable to $m=0$ by appropriately decreasing $g$
or $\tilde g$.   In a typical $pp$ collision at collider energies, we expect to exchange about 10 such long strings~\cite{Stoffers:2012zw,Stoffers:2012ai,Stoffers:2013tla}.  In Fig.~\ref{MOMENT6b}, Fig.~\ref{EnRandomb} and Fig.~\ref{MOMENT3b} we show the same cross
moments following from the exchange of 5 typical strings streched at $b=5$ sampled 200 times for the attractive, non-interacting and repulsive case
respectively. The case where 10 string are exchanged is shown in Fig.~\ref{MOMENT6c},  Fig.~\ref{EnRandomc}  and Fig.~\ref{MOMENT3c} for the
same arrangements of parameters with each 10 string event sampled 100 times.   The critical moments for the pre-saturation coupling
$g\approx 1/\sqrt{N}\approx 0.01$ are not much different from the $g=0$ presented here. 
We note that $\epsilon_n \{ 4 \} \approx \epsilon_n \{ 6 \} \approx \epsilon_n \{ 8 \} $ 
in agreement with suggestion made in~\cite{Bzdak:2013rya}. 
The more string exchanges, the denser and more symmetric the transverse string bit distribution for a fixed resolution $N$, the smaller the cross moments. Fig.~\ref{EnRandomc} should represent typical cross moments in $pp$ collisions at collider energies such as RHIC and LHC 
for minimum bias events. For the high multiplicity $pp$ and $pA$ events reported at LHC hot string configurations near the Hagedorn temperature are needed. They will be discussed in a sequel.

\section{\label{conclusion} Conclusions}

Long holographic strings in walled AdS$_5$  and $D_\perp=3$ provides for a dual description of diffractive scattering and production as well as low-x DIS~\cite{Stoffers:2012zw}. Although a key aspect of  AdS$_5$ is its conformality  which translates to the conformal character of QCD in the UV, the essentials of the walled AdS$_5$ construction for the holographic string with a large rapidity interval can be captured by a relatively cold transverse string with an effective transverse dimension $2<D_\perp<3$. The Pomeron intercept follows from the zero point
motion or Luscher term of the free transverse string with $D_\perp/12$, and the Pomeron slope is fixed by the string tension.

At low-x, DIS scattering of a small dipole of size $1/Q$ scattering off a fixed target dipole can be regarded as the
exchange of a streched string with fixed parameter $b$ with a large rapidity interval ${\rm ln}(1/x)$. Although the DIS structure function
averages over all impact parameters, the dominant contribution to this averaging stems from relatively large $b/l_s\gg 1$
in units of the string scale $l_s\approx 0.1$ fm. Therefore low-x studies could be turned to the studies of a transverse holographic string 
at higher and higher resolution, a dual description to the wee parton description in perturbative QCD.  An essential aspect of the partonic
description is Gribov transverse diffusion which arises naturally in the quantum string description as emphasized by Susskind and others~\cite{Susskind:1994hb,Susskind:1994vu,Damour:1999aw,Horowitz:1996nw}.

The wee parton description at low-x can be mapped on a discretized transverse string in $N=1/x$ string bits~\cite{Susskind:1994hb,Karliner:1988hd}. 
We have shown that the quantum description of a free transverse and discretized string in $D_\perp=3$ results in  highly deformed 
geometries for the string sizes and shapes when the string is streched at fixed impact parameter. We have suggested that at high 
resolution or string bits, string self-interactions can be captured by a mean-field pair interaction between the string bits. The pair interaction
is characterized by a coupling $g$ and a range $1/m$ both of which are inter-changeble by re-scaling.  The holographic origin of the transverse string allows for the identification of $g$ with the bulk Newtonian-like  constant for $m=0$. As a result a Planck scale
emerges for holographic strings at high resolution in $D_\perp=3$. In terms of the gauge coupling $\alpha_s$, the holographic Planck
scale is identified as $l_P\approx \alpha_s^{2/3}l_s$.

In flat dimensions and for relatively weak self-interactions, 
 we have found that the string initial diffusive growth undergoes a first order change into a smaller
and fixed size transverse string of the order of few string lengths at a resolution of $x_c\approx 0.001$ and for a small string coupling
$g_s\approx 0.1$. We have identified this change with a pre-saturation stage whereby the string geometry is fixed and small, but the transverse string bit density is still dilute on the Planck scale $l_p^3n_\perp\approx 0.01$.
At a much higher resolution or $x_s\approx 10^{-5}$ we have found that the transverse string bit density saturates the Bekenstein bound
of one bit per Planck scale. We have identified this point with the saturation scale. 

In curved dimensions, 
a simple estimate can be made by noting that the curvature causes the
string interaction to take place effectively  in lower dimension with $D_\perp\rightarrow D_\perp(\lambda)$. A similar observation
was  made in~\cite{Stoffers:2012zw,Stoffers:2012ai,Stoffers:2013tla} for the Pomeron intercept. The result is a smoothening of the transition to the Bekenstein bound observed in flat $D_\perp=3$. Saturation was found to take place at a higher value of small-x or $x_s\approx 10^{-3}$.

The geometry of the string bit distributions emerging from streched strings for a typical impact parameter of $b=5\equiv 10\,l_s$
is rich in structure and transverse deformation.  We have presented a detailed study of its transverse moments and moment distributions
for single and multiple string exchanges. These prompt and deformed distributions can be used to initialize the prompt parton distributions
in current $pp$ and $pA$ collisions in colliders at high resolution or low-x.  The large deformations observed in this analysis show that
they can yield large transverse asymmetries in prompt multi-particle production in the Pomeron kinematics. Also they may translate to
large transverse momentum asymmetries in the flow analyses of multiplicity at current collider energies. We plan to address some of these
issues next.

\section{\label{acknowledgements} Acknowledgements}

We thank Dima Kharzeev and Edward Shuryak for discussions.
This work was supported by the U.S. Department of Energy under Contract No.
DE-FG-88ER40388.

\section{\label{appendix} Appendix}

In this Appendix we discuss a simple phenomenological way of introducing the effects of AdS$_5$
warping on the transverse oscillators in (\ref{XS1}) that reproduces the key property of Gribov
diffusion derived in~\cite{Stoffers:2012zw,Stoffers:2012ai,Stoffers:2013tla}. For that 
we introduce the rescalings $\tau \rightarrow \lambda_\tau \tau$ and $b \rightarrow \tilde{b}$, 
so that (\ref{XS1}) now reads

\be
S_\perp = \frac{\sigma_T}{2}   \int d \tau \int_0^\pi d \sigma \ \ \left[ \frac{1}{\lambda_\tau^2} \left( \dot{x}_\perp  \right)^2  +  \left( {x'}_\perp \right)^2    \right]
\ee
with the end-point condition 
\be
x_\perp^i (\sigma=0, \tau) = 0 \ \ \ \ \ \ x_\perp^i (\sigma=\pi, \tau) = \bold{\tilde{b}}^i
\ee
The Lagrangian for the discretized string is now
\be 
\mathcal{L}_\perp  =   \frac{1}{\lambda_\tau^2}  \frac{1}{N}    \sum_{k=0}^{N}  \left( \dot{x}_\perp^i (k) \right)^2 -   \frac{1}{N}     \sum_{k=1}^N \left(  \frac{x_\perp^i(k) - x_\perp^i (k-1)}{\frac{\pi}{N}}    \right)^2 
\ee
The mode decompostion for the amplitudes  $x_\perp^i$ reads
\be
x_\perp^i (k , \tau) =  \lambda_\tau   \bold{\tilde{b}}^i \frac{k}{N }  +  \lambda_\tau \sum_{n=1}^{N-1}  X_{n}^i (\tau) \sin \left( \frac{n k}{N } \pi\right)   \ \ \ \ \ \ \ \ \    (k=0,1, \cdots, N)
\label{N1}
\ee
 and their conjugate momenta are

\be
p_\perp^i (k, \tau) = \frac{\partial\mathcal{L}}{\partial  \dot{x}_\perp^i} = \frac{1}{\lambda_\tau^2}  \frac{2}{N}   \dot{x}_\perp^i =  \frac{1}{\lambda_\tau} \frac{2}{N}   \sum_{n=1}^{N-1} \dot{X}_{n}^i (\tau) \sin \left( \frac{n k}{N } \pi\right) \equiv  \frac{1}{\lambda_\tau} \frac{2}{N}  \sum_{n=1}^{N-1}  P_{n}^i (\tau) \sin \left( \frac{n k}{N } \pi\right)
\ee
Thus, the Hamiltonian

\bea
\mathcal{H}_\perp  =  \frac{1}{2 }  \sum_{n=1}^{N-1}   \left(      P _{n}^i (\tau) P_{n }^i (\tau)   + 
\lambda_\tau^2 \Omega_n^2  {X}_{n}^i (\tau)  {X}_{n }^i (\tau) \right)  + \lambda_\tau^2  \frac{  \tilde{b}^2}{  \pi^2 } 
\eea
 
 The ground state of this dangling  N-string   is a product of warped Gaussians

\be
\Psi[ \lambda_\tau; X]=\prod_{n, i} \Psi (\lambda_\tau; X_n^i) = \prod_{n, i} \left( \frac{\lambda_\tau \Omega_n}{\pi}  \right)^{\frac{1}{4}} \exp\left[  - \frac{\lambda_\tau \Omega_n}{2} (X_n^i)^2   \right]
\ee
leading to the ground state energy
\be
\left< \mathcal{H}_\perp \right>   =  \frac{D_\perp \lambda_\tau}{2 } \sum_{n=1}^{N-1}    \Omega_n     +    \frac{\lambda_\tau^2  \tilde{b}^2}{  \pi^2 }
\la{rescaleH}
\ee
(\ref{hga}) is recovered for $\lambda_\tau=1$ as it should. If we set $\lambda_\tau = {D_\perp(\lambda)}/{D_\perp}$ and 
$\tilde{b} =  {b}/{\lambda_\tau}$, (\ref{rescaleH}) reads as

\be
\left< \mathcal{H}_\perp \right>   =  \frac{D_\perp(\lambda)}{2 } \sum_{n=1}^{N-1}    \Omega_n     +    \frac{b^2}{  \pi^2 }
\ee
 The string transverse squared size (\ref{RT0}) is now
\be 
R_\perp^2  =   \frac{1}{N} \sum_{k= 0}^N  \left< \left(  x_k^i  -   \bold{b}^i \frac{k}{N} \right)^2 \right> =  
 \lambda_\tau^2  \frac{D_\perp}{4}\sum_{n=1}^{N-1} \frac{1}{ \lambda_\tau \Omega_n}
=  \frac{D_\perp(\lambda)}{4}\sum_{n=1}^{N-1} \frac{1}{ \Omega_n}\approx \frac {D_\perp(\lambda)}{4}\,{\rm ln}(N)
\ee 
with a Pomeron intercept $D_\perp(\lambda)/12$.


\bibliography{Lowtemref}

\begin{thebibliography}{71}
\expandafter\ifx\csname natexlab\endcsname\relax\def\natexlab#1{#1}\fi
\expandafter\ifx\csname bibnamefont\endcsname\relax
  \def\bibnamefont#1{#1}\fi
\expandafter\ifx\csname bibfnamefont\endcsname\relax
  \def\bibfnamefont#1{#1}\fi
\expandafter\ifx\csname citenamefont\endcsname\relax
  \def\citenamefont#1{#1}\fi
\expandafter\ifx\csname url\endcsname\relax
  \def\url#1{\texttt{#1}}\fi
\expandafter\ifx\csname urlprefix\endcsname\relax\def\urlprefix{URL }\fi
\providecommand{\bibinfo}[2]{#2}
\providecommand{\eprint}[2][]{\url{#2}}

\bibitem[{\citenamefont{Donnachie and Landshoff}(1992)}]{Donnachie:1992ny}
\bibinfo{author}{\bibfnamefont{A.}~\bibnamefont{Donnachie}} \bibnamefont{and}
  \bibinfo{author}{\bibfnamefont{P.}~\bibnamefont{Landshoff}},
  \bibinfo{journal}{Phys.Lett.} \textbf{\bibinfo{volume}{B296}},
  \bibinfo{pages}{227} (\bibinfo{year}{1992}), \eprint{hep-ph/9209205}.

\bibitem[{\citenamefont{Gribov and Lipatov}(1972)}]{Gribov:1972ri}
\bibinfo{author}{\bibfnamefont{V.}~\bibnamefont{Gribov}} \bibnamefont{and}
  \bibinfo{author}{\bibfnamefont{L.}~\bibnamefont{Lipatov}},
  \bibinfo{journal}{Sov.J.Nucl.Phys.} \textbf{\bibinfo{volume}{15}},
  \bibinfo{pages}{438} (\bibinfo{year}{1972}).

\bibitem[{\citenamefont{Gribov}(2003)}]{9780511534959}
\bibinfo{author}{\bibfnamefont{V.~N.} \bibnamefont{Gribov}},
  \emph{\bibinfo{title}{The Theory of Complex Angular Momenta}}
  (\bibinfo{publisher}{Cambridge University Press}, \bibinfo{year}{2003}), ISBN
  \bibinfo{isbn}{9780511534959}, \bibinfo{note}{cambridge Books Online},
  \urlprefix\url{http://dx.doi.org/10.1017/CBO9780511534959}.

\bibitem[{\citenamefont{Kuraev et~al.}(1976)\citenamefont{Kuraev, Lipatov, and
  Fadin}}]{Kuraev:1976ge}
\bibinfo{author}{\bibfnamefont{E.~A.} \bibnamefont{Kuraev}},
  \bibinfo{author}{\bibfnamefont{L.~N.} \bibnamefont{Lipatov}},
  \bibnamefont{and} \bibinfo{author}{\bibfnamefont{V.~S.} \bibnamefont{Fadin}},
  \bibinfo{journal}{Sov.Phys.JETP} \textbf{\bibinfo{volume}{44}},
  \bibinfo{pages}{443} (\bibinfo{year}{1976}).

\bibitem[{\citenamefont{Lipatov}(1976)}]{Lipatov:1976zz}
\bibinfo{author}{\bibfnamefont{L.}~\bibnamefont{Lipatov}},
  \bibinfo{journal}{Sov.J.Nucl.Phys.} \textbf{\bibinfo{volume}{23}},
  \bibinfo{pages}{338} (\bibinfo{year}{1976}).

\bibitem[{\citenamefont{Sterman}(1999)}]{Sterman:1999yc}
\bibinfo{author}{\bibfnamefont{G.~F.} \bibnamefont{Sterman}}
  (\bibinfo{year}{1999}), \eprint{hep-ph/9905548}.

\bibitem[{\citenamefont{Fadin et~al.}(1975)\citenamefont{Fadin, Kuraev, and
  Lipatov}}]{Fadin:1975cb}
\bibinfo{author}{\bibfnamefont{V.~S.} \bibnamefont{Fadin}},
  \bibinfo{author}{\bibfnamefont{E.}~\bibnamefont{Kuraev}}, \bibnamefont{and}
  \bibinfo{author}{\bibfnamefont{L.}~\bibnamefont{Lipatov}},
  \bibinfo{journal}{Phys.Lett.} \textbf{\bibinfo{volume}{B60}},
  \bibinfo{pages}{50} (\bibinfo{year}{1975}).

\bibitem[{\citenamefont{Balitsky and Lipatov}(1978)}]{Balitsky:1978ic}
\bibinfo{author}{\bibfnamefont{I.}~\bibnamefont{Balitsky}} \bibnamefont{and}
  \bibinfo{author}{\bibfnamefont{L.}~\bibnamefont{Lipatov}},
  \bibinfo{journal}{Sov.J.Nucl.Phys.} \textbf{\bibinfo{volume}{28}},
  \bibinfo{pages}{822} (\bibinfo{year}{1978}).

\bibitem[{\citenamefont{Veneziano}(1968)}]{Veneziano:1968yb}
\bibinfo{author}{\bibfnamefont{G.}~\bibnamefont{Veneziano}},
  \bibinfo{journal}{Nuovo Cim.} \textbf{\bibinfo{volume}{A57}},
  \bibinfo{pages}{190} (\bibinfo{year}{1968}).

\bibitem[{\citenamefont{Greensite}(1985)}]{Greensite:1984sb}
\bibinfo{author}{\bibfnamefont{J.}~\bibnamefont{Greensite}},
  \bibinfo{journal}{Nucl.Phys.} \textbf{\bibinfo{volume}{B249}},
  \bibinfo{pages}{263} (\bibinfo{year}{1985}).

\bibitem[{\citenamefont{Maldacena}(1998)}]{Maldacena:1998im}
\bibinfo{author}{\bibfnamefont{J.~M.} \bibnamefont{Maldacena}},
  \bibinfo{journal}{Phys.Rev.Lett.} \textbf{\bibinfo{volume}{80}},
  \bibinfo{pages}{4859} (\bibinfo{year}{1998}), \eprint{hep-th/9803002}.

\bibitem[{\citenamefont{Rho et~al.}(1999)\citenamefont{Rho, Sin, and
  Zahed}}]{Rho:1999jm}
\bibinfo{author}{\bibfnamefont{M.}~\bibnamefont{Rho}},
  \bibinfo{author}{\bibfnamefont{S.-J.} \bibnamefont{Sin}}, \bibnamefont{and}
  \bibinfo{author}{\bibfnamefont{I.}~\bibnamefont{Zahed}},
  \bibinfo{journal}{Phys.Lett.} \textbf{\bibinfo{volume}{B466}},
  \bibinfo{pages}{199} (\bibinfo{year}{1999}), \eprint{hep-th/9907126}.

\bibitem[{\citenamefont{Janik and Peschanski}(2000)}]{Janik:2000aj}
\bibinfo{author}{\bibfnamefont{R.}~\bibnamefont{Janik}} \bibnamefont{and}
  \bibinfo{author}{\bibfnamefont{R.~B.} \bibnamefont{Peschanski}},
  \bibinfo{journal}{Nucl.Phys.} \textbf{\bibinfo{volume}{B586}},
  \bibinfo{pages}{163} (\bibinfo{year}{2000}), \eprint{hep-th/0003059}.

\bibitem[{\citenamefont{Janik}(2001)}]{Janik:2000pp}
\bibinfo{author}{\bibfnamefont{R.~A.} \bibnamefont{Janik}},
  \bibinfo{journal}{Phys.Lett.} \textbf{\bibinfo{volume}{B500}},
  \bibinfo{pages}{118} (\bibinfo{year}{2001}), \eprint{hep-th/0010069}.

\bibitem[{\citenamefont{Polchinski and Strassler}(2002)}]{Polchinski:2001tt}
\bibinfo{author}{\bibfnamefont{J.}~\bibnamefont{Polchinski}} \bibnamefont{and}
  \bibinfo{author}{\bibfnamefont{M.~J.} \bibnamefont{Strassler}},
  \bibinfo{journal}{Phys.Rev.Lett.} \textbf{\bibinfo{volume}{88}},
  \bibinfo{pages}{031601} (\bibinfo{year}{2002}), \eprint{hep-th/0109174}.

\bibitem[{\citenamefont{Polchinski and Susskind}(2001)}]{Polchinski:2001ju}
\bibinfo{author}{\bibfnamefont{J.}~\bibnamefont{Polchinski}} \bibnamefont{and}
  \bibinfo{author}{\bibfnamefont{L.}~\bibnamefont{Susskind}}, pp.
  \bibinfo{pages}{105--114} (\bibinfo{year}{2001}), \eprint{hep-th/0112204}.

\bibitem[{\citenamefont{Polchinski and Strassler}(2003)}]{Polchinski:2002jw}
\bibinfo{author}{\bibfnamefont{J.}~\bibnamefont{Polchinski}} \bibnamefont{and}
  \bibinfo{author}{\bibfnamefont{M.~J.} \bibnamefont{Strassler}},
  \bibinfo{journal}{JHEP} \textbf{\bibinfo{volume}{0305}}, \bibinfo{pages}{012}
  (\bibinfo{year}{2003}), \eprint{hep-th/0209211}.

\bibitem[{\citenamefont{Brower et~al.}(2007)\citenamefont{Brower, Polchinski,
  Strassler, and Tan}}]{Brower:2006ea}
\bibinfo{author}{\bibfnamefont{R.~C.} \bibnamefont{Brower}},
  \bibinfo{author}{\bibfnamefont{J.}~\bibnamefont{Polchinski}},
  \bibinfo{author}{\bibfnamefont{M.~J.} \bibnamefont{Strassler}},
  \bibnamefont{and} \bibinfo{author}{\bibfnamefont{C.-I.} \bibnamefont{Tan}},
  \bibinfo{journal}{JHEP} \textbf{\bibinfo{volume}{0712}}, \bibinfo{pages}{005}
  (\bibinfo{year}{2007}), \eprint{hep-th/0603115}.

\bibitem[{\citenamefont{Brower et~al.}(2009)\citenamefont{Brower, Strassler,
  and Tan}}]{Brower:2007xg}
\bibinfo{author}{\bibfnamefont{R.~C.} \bibnamefont{Brower}},
  \bibinfo{author}{\bibfnamefont{M.~J.} \bibnamefont{Strassler}},
  \bibnamefont{and} \bibinfo{author}{\bibfnamefont{C.-I.} \bibnamefont{Tan}},
  \bibinfo{journal}{JHEP} \textbf{\bibinfo{volume}{0903}}, \bibinfo{pages}{092}
  (\bibinfo{year}{2009}), \eprint{0710.4378}.

\bibitem[{\citenamefont{Brower et~al.}(2010)\citenamefont{Brower, Djuric,
  Sarcevic, and Tan}}]{Brower:2010wf}
\bibinfo{author}{\bibfnamefont{R.~C.} \bibnamefont{Brower}},
  \bibinfo{author}{\bibfnamefont{M.}~\bibnamefont{Djuric}},
  \bibinfo{author}{\bibfnamefont{I.}~\bibnamefont{Sarcevic}}, \bibnamefont{and}
  \bibinfo{author}{\bibfnamefont{C.-I.} \bibnamefont{Tan}},
  \bibinfo{journal}{JHEP} \textbf{\bibinfo{volume}{1011}}, \bibinfo{pages}{051}
  (\bibinfo{year}{2010}), \eprint{1007.2259}.

\bibitem[{\citenamefont{Brower et~al.}(2011)\citenamefont{Brower, Djuric,
  Sarcevic, and Tan}}]{Brower:2011dx}
\bibinfo{author}{\bibfnamefont{R.~C.} \bibnamefont{Brower}},
  \bibinfo{author}{\bibfnamefont{M.}~\bibnamefont{Djuric}},
  \bibinfo{author}{\bibfnamefont{I.}~\bibnamefont{Sarcevic}}, \bibnamefont{and}
  \bibinfo{author}{\bibfnamefont{C.-I.} \bibnamefont{Tan}}
  (\bibinfo{year}{2011}), \eprint{1106.5681}.

\bibitem[{\citenamefont{Hatta et~al.}(2008{\natexlab{a}})\citenamefont{Hatta,
  Iancu, and Mueller}}]{Hatta:2007cs}
\bibinfo{author}{\bibfnamefont{Y.}~\bibnamefont{Hatta}},
  \bibinfo{author}{\bibfnamefont{E.}~\bibnamefont{Iancu}}, \bibnamefont{and}
  \bibinfo{author}{\bibfnamefont{A.}~\bibnamefont{Mueller}},
  \bibinfo{journal}{JHEP} \textbf{\bibinfo{volume}{0801}}, \bibinfo{pages}{063}
  (\bibinfo{year}{2008}{\natexlab{a}}), \eprint{0710.5297}.

\bibitem[{\citenamefont{Hatta et~al.}(2008{\natexlab{b}})\citenamefont{Hatta,
  Iancu, and Mueller}}]{Hatta:2007he}
\bibinfo{author}{\bibfnamefont{Y.}~\bibnamefont{Hatta}},
  \bibinfo{author}{\bibfnamefont{E.}~\bibnamefont{Iancu}}, \bibnamefont{and}
  \bibinfo{author}{\bibfnamefont{A.}~\bibnamefont{Mueller}},
  \bibinfo{journal}{JHEP} \textbf{\bibinfo{volume}{0801}}, \bibinfo{pages}{026}
  (\bibinfo{year}{2008}{\natexlab{b}}), \eprint{0710.2148}.

\bibitem[{\citenamefont{Albacete et~al.}(2008)\citenamefont{Albacete,
  Kovchegov, and Taliotis}}]{Albacete:2008ze}
\bibinfo{author}{\bibfnamefont{J.~L.} \bibnamefont{Albacete}},
  \bibinfo{author}{\bibfnamefont{Y.~V.} \bibnamefont{Kovchegov}},
  \bibnamefont{and} \bibinfo{author}{\bibfnamefont{A.}~\bibnamefont{Taliotis}},
  \bibinfo{journal}{JHEP} \textbf{\bibinfo{volume}{0807}}, \bibinfo{pages}{074}
  (\bibinfo{year}{2008}), \eprint{0806.1484}.

\bibitem[{\citenamefont{Albacete et~al.}(2009)\citenamefont{Albacete,
  Kovchegov, and Taliotis}}]{Albacete:2008vv}
\bibinfo{author}{\bibfnamefont{J.~L.} \bibnamefont{Albacete}},
  \bibinfo{author}{\bibfnamefont{Y.~V.} \bibnamefont{Kovchegov}},
  \bibnamefont{and} \bibinfo{author}{\bibfnamefont{A.}~\bibnamefont{Taliotis}},
  \bibinfo{journal}{AIP Conf.Proc.} \textbf{\bibinfo{volume}{1105}},
  \bibinfo{pages}{356} (\bibinfo{year}{2009}), \eprint{0811.0818}.

\bibitem[{\citenamefont{Basar et~al.}(2012)\citenamefont{Basar, Kharzeev, Yee,
  and Zahed}}]{Basar:2012jb}
\bibinfo{author}{\bibfnamefont{G.}~\bibnamefont{Basar}},
  \bibinfo{author}{\bibfnamefont{D.~E.} \bibnamefont{Kharzeev}},
  \bibinfo{author}{\bibfnamefont{H.-U.} \bibnamefont{Yee}}, \bibnamefont{and}
  \bibinfo{author}{\bibfnamefont{I.}~\bibnamefont{Zahed}},
  \bibinfo{journal}{Phys.Rev.} \textbf{\bibinfo{volume}{D85}},
  \bibinfo{pages}{105005} (\bibinfo{year}{2012}), \eprint{1202.0831}.

\bibitem[{\citenamefont{Stoffers and
  Zahed}(2013{\natexlab{a}})}]{Stoffers:2012zw}
\bibinfo{author}{\bibfnamefont{A.}~\bibnamefont{Stoffers}} \bibnamefont{and}
  \bibinfo{author}{\bibfnamefont{I.}~\bibnamefont{Zahed}},
  \bibinfo{journal}{Phys.Rev.} \textbf{\bibinfo{volume}{D87}},
  \bibinfo{pages}{075023} (\bibinfo{year}{2013}{\natexlab{a}}),
  \eprint{1205.3223}.

\bibitem[{\citenamefont{Stoffers and Zahed}(2012)}]{Stoffers:2012ai}
\bibinfo{author}{\bibfnamefont{A.}~\bibnamefont{Stoffers}} \bibnamefont{and}
  \bibinfo{author}{\bibfnamefont{I.}~\bibnamefont{Zahed}}
  (\bibinfo{year}{2012}), \eprint{1210.3724}.

\bibitem[{\citenamefont{Stoffers and
  Zahed}(2013{\natexlab{b}})}]{Stoffers:2013tla}
\bibinfo{author}{\bibfnamefont{A.}~\bibnamefont{Stoffers}} \bibnamefont{and}
  \bibinfo{author}{\bibfnamefont{I.}~\bibnamefont{Zahed}},
  \bibinfo{journal}{Acta Phys.Polon.Supp.} \textbf{\bibinfo{volume}{6}},
  \bibinfo{pages}{7} (\bibinfo{year}{2013}{\natexlab{b}}).

\bibitem[{\citenamefont{Qian and Zahed}(2012)}]{Zahed:2012sg}
\bibinfo{author}{\bibfnamefont{Y.}~\bibnamefont{Qian}} \bibnamefont{and}
  \bibinfo{author}{\bibfnamefont{I.}~\bibnamefont{Zahed}}
  (\bibinfo{year}{2012}), \eprint{1211.6421}.

\bibitem[{\citenamefont{Shuryak and Zahed}(2014)}]{Shuryak:2013sra}
\bibinfo{author}{\bibfnamefont{E.}~\bibnamefont{Shuryak}} \bibnamefont{and}
  \bibinfo{author}{\bibfnamefont{I.}~\bibnamefont{Zahed}},
  \bibinfo{journal}{Phys.Rev.} \textbf{\bibinfo{volume}{D89}},
  \bibinfo{pages}{094001} (\bibinfo{year}{2014}), \eprint{1311.0836}.

\bibitem[{\citenamefont{Bergman and Thorn}(1997)}]{Bergman:1997ki}
\bibinfo{author}{\bibfnamefont{O.}~\bibnamefont{Bergman}} \bibnamefont{and}
  \bibinfo{author}{\bibfnamefont{C.~B.} \bibnamefont{Thorn}},
  \bibinfo{journal}{Nucl.Phys.} \textbf{\bibinfo{volume}{B502}},
  \bibinfo{pages}{309} (\bibinfo{year}{1997}), \eprint{hep-th/9702068}.

\bibitem[{\citenamefont{Karliner et~al.}(1988)\citenamefont{Karliner, Klebanov,
  and Susskind}}]{Karliner:1988hd}
\bibinfo{author}{\bibfnamefont{M.}~\bibnamefont{Karliner}},
  \bibinfo{author}{\bibfnamefont{I.~R.} \bibnamefont{Klebanov}},
  \bibnamefont{and} \bibinfo{author}{\bibfnamefont{L.}~\bibnamefont{Susskind}},
  \bibinfo{journal}{Int.J.Mod.Phys.} \textbf{\bibinfo{volume}{A3}},
  \bibinfo{pages}{1981} (\bibinfo{year}{1988}).

\bibitem[{\citenamefont{Susskind and Griffin}(1994)}]{Susskind:1994hb}
\bibinfo{author}{\bibfnamefont{L.}~\bibnamefont{Susskind}} \bibnamefont{and}
  \bibinfo{author}{\bibfnamefont{P.}~\bibnamefont{Griffin}}
  (\bibinfo{year}{1994}), \eprint{hep-ph/9410306}.

\bibitem[{\citenamefont{Susskind}(1995)}]{Susskind:1994vu}
\bibinfo{author}{\bibfnamefont{L.}~\bibnamefont{Susskind}},
  \bibinfo{journal}{J.Math.Phys.} \textbf{\bibinfo{volume}{36}},
  \bibinfo{pages}{6377} (\bibinfo{year}{1995}), \eprint{hep-th/9409089}.

\bibitem[{\citenamefont{Froissart}(1961)}]{Froissart:1961ux}
\bibinfo{author}{\bibfnamefont{M.}~\bibnamefont{Froissart}},
  \bibinfo{journal}{Phys.Rev.} \textbf{\bibinfo{volume}{123}},
  \bibinfo{pages}{1053} (\bibinfo{year}{1961}).

\bibitem[{\citenamefont{Bekenstein}(1973)}]{Bekenstein:1973ur}
\bibinfo{author}{\bibfnamefont{J.~D.} \bibnamefont{Bekenstein}},
  \bibinfo{journal}{Phys.Rev.} \textbf{\bibinfo{volume}{D7}},
  \bibinfo{pages}{2333} (\bibinfo{year}{1973}).

\bibitem[{\citenamefont{Bekenstein}(1972)}]{Bekenstein:1972tm}
\bibinfo{author}{\bibfnamefont{J.}~\bibnamefont{Bekenstein}},
  \bibinfo{journal}{Lett.Nuovo Cim.} \textbf{\bibinfo{volume}{4}},
  \bibinfo{pages}{737} (\bibinfo{year}{1972}).

\bibitem[{\citenamefont{Bekenstein}(1974)}]{Bekenstein:1974ax}
\bibinfo{author}{\bibfnamefont{J.~D.} \bibnamefont{Bekenstein}},
  \bibinfo{journal}{Phys.Rev.} \textbf{\bibinfo{volume}{D9}},
  \bibinfo{pages}{3292} (\bibinfo{year}{1974}).

\bibitem[{\citenamefont{Hawking}(1974)}]{Hawking:1974rv}
\bibinfo{author}{\bibfnamefont{S.}~\bibnamefont{Hawking}},
  \bibinfo{journal}{Nature} \textbf{\bibinfo{volume}{248}}, \bibinfo{pages}{30}
  (\bibinfo{year}{1974}).

\bibitem[{\citenamefont{Hawking}(1975)}]{Hawking:1974sw}
\bibinfo{author}{\bibfnamefont{S.}~\bibnamefont{Hawking}},
  \bibinfo{journal}{Commun.Math.Phys.} \textbf{\bibinfo{volume}{43}},
  \bibinfo{pages}{199} (\bibinfo{year}{1975}).

\bibitem[{\citenamefont{Cornalba
  et~al.}(2010{\natexlab{a}})\citenamefont{Cornalba, Costa, and
  Penedones}}]{Cornalba:2010vk}
\bibinfo{author}{\bibfnamefont{L.}~\bibnamefont{Cornalba}},
  \bibinfo{author}{\bibfnamefont{M.~S.} \bibnamefont{Costa}}, \bibnamefont{and}
  \bibinfo{author}{\bibfnamefont{J.}~\bibnamefont{Penedones}},
  \bibinfo{journal}{Phys.Rev.Lett.} \textbf{\bibinfo{volume}{105}},
  \bibinfo{pages}{072003} (\bibinfo{year}{2010}{\natexlab{a}}),
  \eprint{1001.1157}.

\bibitem[{\citenamefont{Cornalba
  et~al.}(2010{\natexlab{b}})\citenamefont{Cornalba, Costa, and
  Penedones}}]{Cornalba:2009ax}
\bibinfo{author}{\bibfnamefont{L.}~\bibnamefont{Cornalba}},
  \bibinfo{author}{\bibfnamefont{M.~S.} \bibnamefont{Costa}}, \bibnamefont{and}
  \bibinfo{author}{\bibfnamefont{J.}~\bibnamefont{Penedones}},
  \bibinfo{journal}{JHEP} \textbf{\bibinfo{volume}{1003}}, \bibinfo{pages}{133}
  (\bibinfo{year}{2010}{\natexlab{b}}), \eprint{0911.0043}.

\bibitem[{\citenamefont{Cornalba and Costa}(2008)}]{Cornalba:2008sp}
\bibinfo{author}{\bibfnamefont{L.}~\bibnamefont{Cornalba}} \bibnamefont{and}
  \bibinfo{author}{\bibfnamefont{M.~S.} \bibnamefont{Costa}},
  \bibinfo{journal}{Phys.Rev.} \textbf{\bibinfo{volume}{D78}},
  \bibinfo{pages}{096010} (\bibinfo{year}{2008}), \eprint{0804.1562}.

\bibitem[{\citenamefont{McLerran}(2002)}]{McLerran:2001sr}
\bibinfo{author}{\bibfnamefont{L.~D.} \bibnamefont{McLerran}},
  \bibinfo{journal}{Lect.Notes Phys.} \textbf{\bibinfo{volume}{583}},
  \bibinfo{pages}{291} (\bibinfo{year}{2002}), \eprint{hep-ph/0104285}.

\bibitem[{\citenamefont{Tapia~Takaki}(2010)}]{TapiaTakaki:2010zz}
\bibinfo{author}{\bibfnamefont{J.}~\bibnamefont{Tapia~Takaki}}
  (\bibinfo{collaboration}{ALICE}), \bibinfo{journal}{J.Phys.}
  \textbf{\bibinfo{volume}{G37}}, \bibinfo{pages}{094050}
  (\bibinfo{year}{2010}).

\bibitem[{\citenamefont{Iancu and Venugopalan}(2003)}]{Iancu:2003xm}
\bibinfo{author}{\bibfnamefont{E.}~\bibnamefont{Iancu}} \bibnamefont{and}
  \bibinfo{author}{\bibfnamefont{R.}~\bibnamefont{Venugopalan}}
  (\bibinfo{year}{2003}), \eprint{hep-ph/0303204}.

\bibitem[{\citenamefont{Iancu et~al.}(2002)\citenamefont{Iancu, Leonidov, and
  McLerran}}]{Iancu:2002xk}
\bibinfo{author}{\bibfnamefont{E.}~\bibnamefont{Iancu}},
  \bibinfo{author}{\bibfnamefont{A.}~\bibnamefont{Leonidov}}, \bibnamefont{and}
  \bibinfo{author}{\bibfnamefont{L.}~\bibnamefont{McLerran}}, pp.
  \bibinfo{pages}{73--145} (\bibinfo{year}{2002}), \eprint{hep-ph/0202270}.

\bibitem[{\citenamefont{Navelet and Peschanski}(2002)}]{Navelet:2002zz}
\bibinfo{author}{\bibfnamefont{H.}~\bibnamefont{Navelet}} \bibnamefont{and}
  \bibinfo{author}{\bibfnamefont{R.~B.} \bibnamefont{Peschanski}},
  \bibinfo{journal}{Nucl.Phys.} \textbf{\bibinfo{volume}{B634}},
  \bibinfo{pages}{291} (\bibinfo{year}{2002}), \eprint{hep-ph/0201285}.

\bibitem[{\citenamefont{Iancu}(2001)}]{Iancu:2001yq}
\bibinfo{author}{\bibfnamefont{E.}~\bibnamefont{Iancu}}, pp.
  \bibinfo{pages}{184--191} (\bibinfo{year}{2001}), \eprint{hep-ph/0111400}.

\bibitem[{\citenamefont{Levin}(2001)}]{Levin:2001eq}
\bibinfo{author}{\bibfnamefont{E.}~\bibnamefont{Levin}} (\bibinfo{year}{2001}),
  \eprint{hep-ph/0105205}.

\bibitem[{\citenamefont{McLerran and Venugopalan}(1994)}]{McLerran:1993ka}
\bibinfo{author}{\bibfnamefont{L.~D.} \bibnamefont{McLerran}} \bibnamefont{and}
  \bibinfo{author}{\bibfnamefont{R.}~\bibnamefont{Venugopalan}},
  \bibinfo{journal}{Phys.Rev.} \textbf{\bibinfo{volume}{D49}},
  \bibinfo{pages}{3352} (\bibinfo{year}{1994}), \eprint{hep-ph/9311205}.

\bibitem[{\citenamefont{Gelis et~al.}(2010)\citenamefont{Gelis, Iancu,
  Jalilian-Marian, and Venugopalan}}]{Gelis:2010nm}
\bibinfo{author}{\bibfnamefont{F.}~\bibnamefont{Gelis}},
  \bibinfo{author}{\bibfnamefont{E.}~\bibnamefont{Iancu}},
  \bibinfo{author}{\bibfnamefont{J.}~\bibnamefont{Jalilian-Marian}},
  \bibnamefont{and}
  \bibinfo{author}{\bibfnamefont{R.}~\bibnamefont{Venugopalan}},
  \bibinfo{journal}{Ann.Rev.Nucl.Part.Sci.} \textbf{\bibinfo{volume}{60}},
  \bibinfo{pages}{463} (\bibinfo{year}{2010}), \eprint{1002.0333}.

\bibitem[{\citenamefont{Marquet et~al.}(2005)\citenamefont{Marquet, Mueller,
  Shoshi, and Wong}}]{Marquet:2005hu}
\bibinfo{author}{\bibfnamefont{C.}~\bibnamefont{Marquet}},
  \bibinfo{author}{\bibfnamefont{A.}~\bibnamefont{Mueller}},
  \bibinfo{author}{\bibfnamefont{A.}~\bibnamefont{Shoshi}}, \bibnamefont{and}
  \bibinfo{author}{\bibfnamefont{S.}~\bibnamefont{Wong}},
  \bibinfo{journal}{Nucl.Phys.} \textbf{\bibinfo{volume}{A762}},
  \bibinfo{pages}{252} (\bibinfo{year}{2005}), \eprint{hep-ph/0505229}.

\bibitem[{\citenamefont{Iancu and Mueller}(2004)}]{Iancu:2003uh}
\bibinfo{author}{\bibfnamefont{E.}~\bibnamefont{Iancu}} \bibnamefont{and}
  \bibinfo{author}{\bibfnamefont{A.}~\bibnamefont{Mueller}},
  \bibinfo{journal}{Nucl.Phys.} \textbf{\bibinfo{volume}{A730}},
  \bibinfo{pages}{460} (\bibinfo{year}{2004}), \eprint{hep-ph/0308315}.

\bibitem[{\citenamefont{Iancu}(2009)}]{Iancu:2009nd}
\bibinfo{author}{\bibfnamefont{E.}~\bibnamefont{Iancu}},
  \bibinfo{journal}{Nucl.Phys.Proc.Suppl.} \textbf{\bibinfo{volume}{191}},
  \bibinfo{pages}{281} (\bibinfo{year}{2009}), \eprint{0901.0986}.

\bibitem[{\citenamefont{Gelis}(2013)}]{Gelis:2012ri}
\bibinfo{author}{\bibfnamefont{F.}~\bibnamefont{Gelis}},
  \bibinfo{journal}{Int.J.Mod.Phys.} \textbf{\bibinfo{volume}{A28}},
  \bibinfo{pages}{1330001} (\bibinfo{year}{2013}), \eprint{1211.3327}.

\bibitem[{\citenamefont{Tribedy and Venugopalan}(2012)}]{Tribedy:2011aa}
\bibinfo{author}{\bibfnamefont{P.}~\bibnamefont{Tribedy}} \bibnamefont{and}
  \bibinfo{author}{\bibfnamefont{R.}~\bibnamefont{Venugopalan}},
  \bibinfo{journal}{Phys.Lett.} \textbf{\bibinfo{volume}{B710}},
  \bibinfo{pages}{125} (\bibinfo{year}{2012}), \eprint{1112.2445}.

\bibitem[{\citenamefont{Tribedy and
  Venugopalan}(2011{\natexlab{a}})}]{Tribedy:2011yn}
\bibinfo{author}{\bibfnamefont{P.}~\bibnamefont{Tribedy}} \bibnamefont{and}
  \bibinfo{author}{\bibfnamefont{R.}~\bibnamefont{Venugopalan}}
  (\bibinfo{year}{2011}{\natexlab{a}}), \eprint{1101.5922}.

\bibitem[{\citenamefont{Tribedy and
  Venugopalan}(2011{\natexlab{b}})}]{Tribedy:2010ab}
\bibinfo{author}{\bibfnamefont{P.}~\bibnamefont{Tribedy}} \bibnamefont{and}
  \bibinfo{author}{\bibfnamefont{R.}~\bibnamefont{Venugopalan}},
  \bibinfo{journal}{Nucl.Phys.} \textbf{\bibinfo{volume}{A850}},
  \bibinfo{pages}{136} (\bibinfo{year}{2011}{\natexlab{b}}),
  \eprint{1011.1895}.

\bibitem[{\citenamefont{Qian and Zahed}(2014)}]{Qian:2014jna}
\bibinfo{author}{\bibfnamefont{Y.}~\bibnamefont{Qian}} \bibnamefont{and}
  \bibinfo{author}{\bibfnamefont{I.}~\bibnamefont{Zahed}}
  (\bibinfo{year}{2014}), \eprint{1410.1092}.

\bibitem[{\citenamefont{Kalaydzhyan and Shuryak}(2014)}]{Kalaydzhyan:2014tfa}
\bibinfo{author}{\bibfnamefont{T.}~\bibnamefont{Kalaydzhyan}} \bibnamefont{and}
  \bibinfo{author}{\bibfnamefont{E.}~\bibnamefont{Shuryak}},
  \bibinfo{journal}{Phys.Rev.} \textbf{\bibinfo{volume}{D90}},
  \bibinfo{pages}{025031} (\bibinfo{year}{2014}), \eprint{1402.7363}.

\bibitem[{\citenamefont{Liu and Zahed}(2014{\natexlab{a}})}]{Liu:2014fda}
\bibinfo{author}{\bibfnamefont{Y.}~\bibnamefont{Liu}} \bibnamefont{and}
  \bibinfo{author}{\bibfnamefont{I.}~\bibnamefont{Zahed}}
  (\bibinfo{year}{2014}{\natexlab{a}}), \eprint{1408.3331}.

\bibitem[{\citenamefont{Liu and Zahed}(2014{\natexlab{b}})}]{Liu:2014qrt}
\bibinfo{author}{\bibfnamefont{Y.}~\bibnamefont{Liu}} \bibnamefont{and}
  \bibinfo{author}{\bibfnamefont{I.}~\bibnamefont{Zahed}}
  (\bibinfo{year}{2014}{\natexlab{b}}), \eprint{1407.0384}.

\bibitem[{\citenamefont{Damour and Veneziano}(2000)}]{Damour:1999aw}
\bibinfo{author}{\bibfnamefont{T.}~\bibnamefont{Damour}} \bibnamefont{and}
  \bibinfo{author}{\bibfnamefont{G.}~\bibnamefont{Veneziano}},
  \bibinfo{journal}{Nucl.Phys.} \textbf{\bibinfo{volume}{B568}},
  \bibinfo{pages}{93} (\bibinfo{year}{2000}), \eprint{hep-th/9907030}.

\bibitem[{\citenamefont{Golec-Biernat and
  Wusthoff}(1998)}]{GolecBiernat:1998js}
\bibinfo{author}{\bibfnamefont{K.~J.} \bibnamefont{Golec-Biernat}}
  \bibnamefont{and} \bibinfo{author}{\bibfnamefont{M.}~\bibnamefont{Wusthoff}},
  \bibinfo{journal}{Phys.Rev.} \textbf{\bibinfo{volume}{D59}},
  \bibinfo{pages}{014017} (\bibinfo{year}{1998}), \eprint{hep-ph/9807513}.

\bibitem[{\citenamefont{Golec-Biernat and
  Wusthoff}(1999)}]{GolecBiernat:1999qd}
\bibinfo{author}{\bibfnamefont{K.~J.} \bibnamefont{Golec-Biernat}}
  \bibnamefont{and} \bibinfo{author}{\bibfnamefont{M.}~\bibnamefont{Wusthoff}},
  \bibinfo{journal}{Phys.Rev.} \textbf{\bibinfo{volume}{D60}},
  \bibinfo{pages}{114023} (\bibinfo{year}{1999}), \eprint{hep-ph/9903358}.

\bibitem[{\citenamefont{Rezaeian et~al.}(2013)\citenamefont{Rezaeian, Siddikov,
  Van~de Klundert, and Venugopalan}}]{Rezaeian:2012ji}
\bibinfo{author}{\bibfnamefont{A.~H.} \bibnamefont{Rezaeian}},
  \bibinfo{author}{\bibfnamefont{M.}~\bibnamefont{Siddikov}},
  \bibinfo{author}{\bibfnamefont{M.}~\bibnamefont{Van~de Klundert}},
  \bibnamefont{and}
  \bibinfo{author}{\bibfnamefont{R.}~\bibnamefont{Venugopalan}},
  \bibinfo{journal}{Phys.Rev.} \textbf{\bibinfo{volume}{D87}},
  \bibinfo{pages}{034002} (\bibinfo{year}{2013}), \eprint{1212.2974}.

\bibitem[{\citenamefont{Kowalski and Teaney}(2003)}]{Kowalski:2003hm}
\bibinfo{author}{\bibfnamefont{H.}~\bibnamefont{Kowalski}} \bibnamefont{and}
  \bibinfo{author}{\bibfnamefont{D.}~\bibnamefont{Teaney}},
  \bibinfo{journal}{Phys.Rev.} \textbf{\bibinfo{volume}{D68}},
  \bibinfo{pages}{114005} (\bibinfo{year}{2003}), \eprint{hep-ph/0304189}.

\bibitem[{\citenamefont{Bzdak et~al.}(2014)\citenamefont{Bzdak, Bozek, and
  McLerran}}]{Bzdak:2013rya}
\bibinfo{author}{\bibfnamefont{A.}~\bibnamefont{Bzdak}},
  \bibinfo{author}{\bibfnamefont{P.}~\bibnamefont{Bozek}}, \bibnamefont{and}
  \bibinfo{author}{\bibfnamefont{L.}~\bibnamefont{McLerran}},
  \bibinfo{journal}{Nucl.Phys.} \textbf{\bibinfo{volume}{A927}},
  \bibinfo{pages}{15} (\bibinfo{year}{2014}), \eprint{1311.7325}.

\bibitem[{\citenamefont{Horowitz and Polchinski}(1997)}]{Horowitz:1996nw}
\bibinfo{author}{\bibfnamefont{G.~T.} \bibnamefont{Horowitz}} \bibnamefont{and}
  \bibinfo{author}{\bibfnamefont{J.}~\bibnamefont{Polchinski}},
  \bibinfo{journal}{Phys.Rev.} \textbf{\bibinfo{volume}{D55}},
  \bibinfo{pages}{6189} (\bibinfo{year}{1997}), \eprint{hep-th/9612146}.

\end{thebibliography}

\newpage

\newpage

\newpage

\newpage

\newpage

\newpage

\newpage

 \newpage
 
\end{document}